\documentclass[aps,prl,twocolumn,superscriptaddress,]{revtex4-2}
\usepackage{cancel}
\usepackage[normalem]{ulem}
\usepackage{bm}
\usepackage{amsmath}
\usepackage{amsthm}
\usepackage{natbib}
\usepackage{amssymb}
\usepackage{graphicx}
\usepackage{esint}

\makeatletter
\usepackage{times}
\usepackage{comment}
\usepackage{graphicx}
\usepackage{feynmf}
\usepackage{tabularx}
\usepackage{amsmath}
\usepackage{amstext}
\usepackage{amssymb}
\usepackage{xfrac}
\usepackage[colorlinks,citecolor=blue]{hyperref}
\usepackage{graphicx}
\usepackage{amsmath}
\usepackage{amstext}
\usepackage{amssymb}
\usepackage{amsfonts}
\usepackage{longtable,booktabs}
\usepackage{hyperref}
\usepackage{url}
\usepackage{subfigure}
\usepackage{dsfont}
\usepackage{booktabs}
\usepackage{amsbsy}
\usepackage{dcolumn}
\usepackage{amsthm}
\usepackage{bm}
\usepackage{esint}
\usepackage{multirow}
\usepackage{hyperref}
\usepackage{cleveref}
\usepackage{mathrsfs}
\usepackage{amsfonts}
\usepackage{amsbsy}
\usepackage{dcolumn}
\usepackage{bm}
\usepackage{multirow}
\usepackage{color}
\usepackage{extarrows}
\usepackage{datetime}
\usepackage{makecell}
\usepackage[super]{nth}
\hypersetup{
	colorlinks=magenta,
	linkcolor=blue,
	filecolor=magenta,
	urlcolor=magenta,
}

\newcommand{\tmmathbf}[1]{\ensuremath{\boldsymbol{#1}}}
\newcommand{\tmop}[1]{\ensuremath{\operatorname{#1}}}

\begin{document}
\title{Dissipative Superfluidity in a Molecular Bose-Einstein Condensate}
\author{Hongchao Li}
\thanks{These two authors contributed equally to this work.}
\affiliation{Department of Physics, The University of Tokyo, 7-3-1 Hongo, Tokyo 113-0033,
	Japan}
\email{lhc@cat.phys.s.u-tokyo.ac.jp}

\author{Xie-Hang Yu}
\thanks{These two authors contributed equally to this work.}
\affiliation{Max-Planck-Institut für Quantenoptik, Hans-Kopfermann-Straße 1, D-85748
	Garching, Germany}
\affiliation{Munich Center for Quantum Science and Technology, Schellingstraße
	4, 80799 München, Germany}
\email{xiehang.yu@mpq.mpg.de}

\author{Masaya Nakagawa}
\affiliation{Department of Physics, The University of Tokyo, 7-3-1 Hongo, Tokyo 113-0033,
	Japan}
\email{nakagawa@cat.phys.s.u-tokyo.ac.jp}

\author{Masahito Ueda}
\affiliation{Department of Physics, The University of Tokyo, 7-3-1 Hongo, Tokyo 113-0033,
	Japan}
\affiliation{RIKEN Center for Emergent Matter Science (CEMS), Wako, Saitama 351-0198,
	Japan}
\affiliation{Institute for Physics of Intelligence, The University of Tokyo, 7-3-1
	Hongo, Tokyo 113-0033, Japan}
\email{ueda@cat.phys.s.u-tokyo.ac.jp}

\date{\today}
\begin{abstract}
	Motivated by recent experimental realization of a Bose-Einstein condensate (BEC) of dipolar molecules, we develop superfluid transport theory for a dissipative BEC to show that a weak uniform two-body loss can induce phase rigidity, leading to superfluid transport of bosons even without repulsive interparticle interactions. 
	A generalized $f$-sum rule is shown to hold for a dissipative superfluid as a consequence of weak U(1) symmetry. We also demonstrate that dissipation enhances the stability of a molecular BEC with dipolar interactions. Possible experimental signature of dissipative superfluidity are discussed.
\end{abstract}
\maketitle
\emph{Introduction.---} The term superfluidity represents a collection of extraordinary phenomena such as zero viscosity, quantized vortices, and the nonclassical rotational inertia~\cite{Leggett2006,Coleman_2015,Schmitt2015,RevModPhys.29.205,RevModPhys.47.331,Ueda2010,RevModPhys.71.463,Pethick_Smith_2008}. 
A distinctive feature of superfluidity, as opposed to Bose-Einstein condensation, is the phase rigidity, which causes a free-energy penalty against phase twist \citep{Coleman_2015} and is usually attributed to interparticle interactions. 
While superfluidity and Bose-Einstein condensation often emerge simultaneously, they are neither necessary nor sufficient to each other~\cite{Leggett2006}. For example, free bosons show Bose-Einstein condensation but no superfluidity. The recent realization of a Bose-Einstein condensate (BEC) of heteronuclear molecules~\cite{Sebastian2023} raises a new question as to whether or not a dissipative BEC exhibits superfluidity because the system is intrinsically dissipative due to chemical reactions.
We address this problem and find that the phase rigidity is reinforced rather than weakened by dissipation.

Quantum gases of dipolar molecules, which serve as a platform to realize clean and controllable long-range interacting systems, have received considerable attention in the fields of many-body physics and quantum simulation \cite{Ni2008,Ni2010,Arthur2019,Liu2020,Bause2021,Gersema2021,rosenberg2022observation,Bause2023,Ospelkaus2010,Karman2018,Sebastian2016,Sebastian2023,Lin2023,Yan2020,Cairncross2021,Lam2022,Guo2022,jin2024boseeinstein}.  However, heteronuclear molecules inevitably suffer from two-body loss due to chemical reactions, which is particularly serious for bosonic molecules~\cite{takekoshi2014ultracold,guo2016creation,guo2018dipolar,Gersema2021,Bause2023,stevenson2023ultracold}. Recently, with the development of microwave shielding~\cite{Karman2018,Sebastian2023,Lin2023,anderegg2021observation,Schindewolf2022}, the first experimental realization of a BEC of heteronuclear molecules has been reported~\cite{Sebastian2023}. Thus, it is of fundamental importance to understand whether or not superfluidity exists in such dissipative BECs, since dissipation usually deteriorates the phase coherence of a superfluid~\cite{doi:10.1126/sciadv.1701513,Diehl2008,Tomita_2019,Keeling2011,dogra2019dissipation,Yamamoto2021,Mazza2023}.


In this Letter, we develop a theory of dissipative superfluidity of a molecular BEC in the presence of the dipole-dipole interaction and uniform two-body loss. In particular, we elucidate that dissipation can induce the phase rigidity and hence superfluidity even without interparticle interactions. 
We use the Schwinger-Keldysh theory of open quantum systems to show that the normal fluid density vanishes if we start from a BEC at absolute zero even in the presence of dissipation.  
This result implies that the non-condensate part (quantum depletion) fully participates in the superfluid transport in non-equilibrium dissipative dynamics. We also show that the quantum depletion can be induced by dissipation and is nonzero even without interactions. The underlying physics is that particle loss induces the correlations which play a role similar to effective repulsion between bosons.

Furthermore, we derive a generalized $f$-sum rule that is applicable to dissipative open quantum systems as a consequence of weak U(1) symmetry~\cite{buvca2012note} of the Lindbladian even in the absence of particle-number conservation. We study the stability of a dissipative molecular BEC by investigating elementary excitations and find that dissipation can effectively enhance the stability of the molecular BEC with the dipolar interaction. Lastly, we numerically verify our predictions by showing the nucleation of quantized vortices solely induced by two-body loss under typical experimental parameters \cite{Sebastian2023}. In comparison with Ref.~\cite{liu2022weakly}, we consider a molecular BEC with an anisotropic and long-range dipolar interaction and focus on transport properties of a dissipative superfluid.



\emph{Dissipation-induced phase rigidity.--- }We consider a three-dimensional
gas of bosonic molecules with an electric dipole-dipole interaction~\cite{Lahaye_2009,Chomaz_2023} described by the Hamiltonian
\begin{align}\label{eq:hamiltonians}
	H=&\int d^3\bm{r}\frac{1}{2m}\nabla a_{\bm{r}}^{\dagger}\nabla a_{\bm{r}}+\frac{U_{R}}{2}\int d^3\bm{r}(a_{\bm{r}}^{\dagger})^{2}(a_{\bm{r}})^{2}\nonumber\\
	&+c_{dd}\int d^3\bm{r}_1d^3\bm{r}_2a_{\bm{r}_1}^{\dagger}a_{\bm{r}_2}^{\dagger}\frac{1-3\cos^2\theta}{|\bm{r}_{12}|^3}a_{\bm{r}_2}a_{\bm{r}_1},
\end{align}
where $m$ is the mass of a single boson, $U_R>0$ is the strength of the contact interaction, $c_{dd}$ is that of the dipole-dipole interaction, $a_{\bm{\bm{r}}}^{(\dagger)}$ is the annihilation (creation) operator of a boson at position $\bm{r}$, $\bm{r}_{12}:=\bm{r}_1-\bm{r}_2$, $\cos\theta:=\hat{\bm{z}}\cdot\bm{r}_{12}/|\bm{r}_{12}|$ and we set $\hbar=1$. Here we assume that all the dipoles are polarized along the $z$-axis.
The dissipative dynamics induced by two-body loss of bosonic molecules can be described by the Lindblad equation \citep{10.1093/acprof:oso/9780199213900.001.0001}
\begin{equation}
	\frac{d\rho}{dt}=\mathcal{L}\rho=-i[H,\rho]-\frac{\gamma}{2}\int d^3\bm{r}(\{L_{\bm{r}}^{\dagger}L_{\bm{r}},\rho\}-2L_{\bm{r}}\rho L_{\bm{r}}^{\dagger}),\label{eq: Lindblad}
\end{equation}
where $\rho$ is the density matrix of the bosonic molecules, $\mathcal{L}$ is the Lindbladian, and the Lindblad operator $L_{\bm{r}}=a_{\bm{r}}^{2}$ describes a two-body loss at position $\bm{r}$ with loss rate $\gamma>0$. Here we ignore three-body loss which is suppressed by the enhanced microwave shielding~\cite{Sebastian2023}.

We first investigate the nature of superfluidity of a dissipative BEC by using an effective field theory. We begin with a BEC at absolute zero and switch on the two-body loss at infinite past. We consider the path-integral representation of Eq.~(\ref{eq: Lindblad}) on the Schwinger-Keldysh contour with the action given by \citep{Sieberer_2016}
\begin{equation}
	\begin{aligned}S= & \int_{-\infty}^{\infty}dt\biggl[\int d^3\bm{r}(\varphi^{*}_{+}i\partial_{t}\varphi_{+}-\varphi^{*}_{-}i\partial_{t}\varphi_{-})-H_{+}\\&+H_{-}
		+\frac{i\gamma}{2}\int d\bm{r}(\bar{L}_{\tmmathbf{r}+}L_{\tmmathbf{r}+}+\bar{L}_{\tmmathbf{r}-}L_{\tmmathbf{r}-}-2L_{\tmmathbf{r}+}\bar{L}_{\tmmathbf{r}-})\biggr],
	\end{aligned}
	\label{eq:Keldysh_action}
\end{equation}
where we use the subscripts $+$ and $-$ to label the forward and
backward contours and $\varphi_{\alpha}=\varphi_{\alpha}(\bm{r},t)$ ($\alpha=\pm$) describes a bosonic field. Here the Hamiltonian $H_\alpha$ is given by replacing $a_{\bm{r}}$ ($a_{\bm{r}}^\dag$) in Eq.~\eqref{eq:hamiltonians} with $\varphi_{\alpha}(\bm{r},t)$ ($\varphi^{*}_{\alpha}(\bm{r},t)$), $L_{\tmmathbf{r}\alpha}=\varphi_{\alpha}(\bm{r},t)^{2}$ and $\bar{L}_{\tmmathbf{r}\alpha}=\varphi_{\alpha}^{*}(\bm{r},t)^{2}$. We note that the action \eqref{eq:Keldysh_action} has a weak U(1) symmetry~\cite{buvca2012note}, i.e., invariance under $\varphi_{\alpha}\rightarrow \varphi_{\alpha}e^{i\theta}$, which keeps the density matrix block-diagonal in the particle-number representation, 
while the system does not conserve the particle number~\cite{Albert2014}.
With the action, we introduce the Keldysh partition function~\cite{Sieberer_2016}:
\begin{equation}
	Z=\tmop{Tr}\rho=\int D[\varphi_{-},\varphi^{*}_{-},\varphi_{+},\varphi^*_{+}]e^{iS}.
\end{equation}
We assume that the system undergoes Bose-Einstein condensation, in which the order parameters are determined by the saddle-point condition of the action. Then, we can decompose the bosonic fields
as \citep{Wen2007}
\begin{eqnarray}
	\varphi_{\alpha}(\bm{r},t) & = & \varphi_{0}(t)(1+\phi_{\alpha}(\tmmathbf{r},t))e^{i\theta_{\alpha}(\bm{r},t)},\nonumber \\
	\varphi^{*}_{\alpha}(\bm{r},t) & = & \varphi_{0}(t)(1+\phi_{\alpha}(\tmmathbf{r},t))e^{-i\theta_{\alpha}(\bm{r},t)},\label{eq:field_decomposation}
\end{eqnarray}
where $\varphi_{0}(t)$ is the saddle point of the action. The real fields $\phi_{\alpha}$ and $\theta_{\alpha}$ denote the amplitude and phase fluctuations of the bosonic field on the contour
$\alpha$. Physically, the amplitude field $\phi$ represents the Higgs mode and the phase field $\theta$ represents the Nambu-Goldstone (NG) mode. The saddle-point solution is given by $\varphi_{0}(t)=\sqrt{n_0(t)}\exp(-i\int_0^t dt' \mu(t'))$ with $n_0$ being the number density of condensate bosons and $\mu(t):=n_0(t)(U_R-8\pi c_{dd}/3)=n_0(t)U_R(1-\varepsilon_{dd})$ representing the time-dependent frequency of the phase of the condensate where $\varepsilon_{dd}:=8\pi c_{dd}/3U_R$. Substituting Eq.~(\ref{eq:field_decomposation}) in Eq. \eqref{eq:Keldysh_action}, we have the action $S(\phi_{+},\phi_{-},\theta_{+},\theta_{-})$ as
a function of the phase and amplitude fluctuations. 
Consequently, the Keldysh partition function is rewritten as
\begin{equation}
	Z=\int D[\phi_{+},\phi_{-},\theta_{+},\theta_{-}]e^{iS(\phi_{+},\phi_{-},\theta_{+},\theta_{-})}.\label{eq:Keldysh_partition_function}
\end{equation}

We now investigate the superfluid density. Since the contact interaction (dissipation) terms give rise to a real (imaginary) gap proportional to $U_R|\varphi_{0}|^2$ ($i\gamma|\varphi_{0}|^2$) for the amplitude mode, we integrate out the amplitude fields if either of them is nonzero, obtaining an effective action for the phase mode~\citep{PhysRevD.103.056020}
\begin{equation}
	S^{\tmop{eff}}=S_{+}-S_{-},S_{\alpha}=-\int dt d^{3}\bm{r}\frac{\varphi_{0}^{2}}{2m}(\nabla\tilde{\theta}_{\alpha})^{2},\label{eq:effective}
\end{equation}
where $\nabla\tilde{\theta}_{\alpha}:=\nabla\theta_{\alpha}+\bm{\psi}(\bm{r})$. Here the function $\bm{\psi}(\bm{r})$ satisfying $\nabla\cdot\bm{\psi}(\bm{r})=2\gamma m\varphi_{0}^{2}$
describes a loss of particles into an environment, as shown below. The derivation of the effective action is given in Supplemental Material \citep{SupplementaryMaterial}. The quadratic effective action~\eqref{eq:effective} exhibits the phase rigidity, leading to superfluidity. To find the superfluid component in an open system, we twist the phases on both the
forward and backward contours. These phase twists can be related to the superfluid velocity as \citep{Coleman_2015} $\bm{v}_{\alpha}=\alpha\nabla\theta_{\alpha}/m.$ The current density is given by the expectation value of the current operators on both contours, i.e., 
\begin{equation}
	\bm{j}=\frac{1}{2}\langle\bm{j}_{+}+\bm{j}_{-}\rangle=\frac{1}{2Z}\int D[\varphi_{+}]D[\varphi_{-}](\bm{j}_{+}+\bm{j}_{-})e^{iS},
\end{equation}
where $\bm{v}_+=-\bm{v}_-=\bm{v}_s$ and the current density on each contour is defined as $\bm{j}_{\alpha}=-\delta S/\delta(m\bm{v}_{\alpha})$. The superfluid density is defined as $\rho_s:=\delta\bm{j}/\delta\bm{v}_s$, which is given by the coefficient of the quadratic term of the superfluid velocity $\bm{v}_s$ in the effective action \eqref{eq:effective}. From the effective action,
we obtain the superfluid density
\begin{equation}
	\rho_s(t)=|\varphi_0(t)|^2=n_0(t).\label{eq:supercurrent}
\end{equation}
The continuity equation for a dissipative superfluid is given by
\begin{equation}
	\frac{dn}{dt}=-\nabla\cdot\bm{j}=-\nabla\cdot(\bm{j}_{s}+\bm{j}_{d}),
\end{equation}
where $\bm{j}_{s}=|\varphi_{0}|^{2}\bm{v}_{s}$ is the superfluid current induced by an external perturbation and $\bm{j}_{d}=|\varphi_{0}|^{2}\bm{\psi}/m$ represents the flow of particles to the environment, which results in the loss of particles. Therefore, the term involving $\bm{\psi}(\bm{r})$ in Eq. \eqref{eq:effective} is identified with the dissipative current induced by two-body loss~\citep{Sieberer_2016,PhysRevLett.93.160404}. In the long-time limit, this dissipative current induces the decay of the particle number as $n_0(t)\sim n_0(0)/(1+2\gamma n(0)t)$~\cite{Ce2022}. However, here we consider a short-time regime where the particle number is nearly constant during the observation. In this time scale, dissipation enhances rather than suppresses superfluidity, as shown below.

Here we point out that the phase rigidity and superfluidity can be induced by dissipation even without interactions. For a closed system composed of free bosons, the amplitude mode is gapless, leading to the breakdown of the effective action~\eqref{eq:effective} and the system does not show superfluidity. Therefore, repulsive interactions are necessary for superfluidity since they suppress density fluctuations and ensure a finite compressibility~\cite{RevModPhys.80.885}. By using effective field theory, we can show that compressibility of a dissipative BEC becomes finite due to two-body loss even in the absence of interactions~\cite{SupplementaryMaterial}, indicating the stability against external perturbations. The dipolar interaction does not essentially modify the above mechanism but only partially compensates the repulsive interaction, i.e., $U_R\to U_R(1-\varepsilon_{dd})$. In the dissipative superfluidity considered here, the two-body loss suppresses density fluctuations by eliminating particles predominantly from where the density is higher than the average. This suppression of density fluctuations leads to the phase rigidity and hence superfluidity. Similar mechanisms by which dissipation induces correlations akin to effective repulsive interactions have been discussed in the context of the quantum Zeno effect~\cite{Syassen2008,Garcia-Ripoll_2009,Yamamoto2019,Hongchao2023}. We note that not all kinds of dissipation can lead to the phase rigidity. Only the dissipation such as two-body and three-body losses leads to correlations between bosons and can generate phase rigidity. 

\emph{Quantum depletion.---}The above discussion only considers the quadratic terms in the effective action and is only valid for the condensate part. To further vindicate the superfluidity in a dissipative BEC, we now consider the quantum depletion and perform the Bogoliubov analysis by employing the mean-field approximation with $a_{0+}=a_{0-}=a_{0-}^\dagger=a_{0+}^\dagger\approx\sqrt{N}$ where $a_{0}$ represents the annihilation operator of a boson with zero momentum. Let us here consider a quasi-steady state where $\gamma n(0)T\ll1$ with $T$ being the observation time during which $n(t)$ can approximately be considered as a constant~\footnote{According to the experimental data in Ref. \cite{Sebastian2023}, we have $\gamma n(0)T\sim 10^{-4}\ll1$ for the observation time of the milisecond order, justifying the approximation we make. }. Under the quasi-steady-state approximation, we calculate the quantum depletion defined as $n_D:=\sum_{\bm{k}\neq0}\langle a_{\bm{k}}^{\dagger}a_{\bm{k}}\rangle/V$, where $V$ is the volume of the system.
To this end, we generalize the phase stiffness~\cite{Coleman_2015} to the case of open quantum systems by using $Q_{ab}=(-1/(2V))\partial^2 F / \partial v_a \partial v_b$, where $F[\bm{v}]=-i\log Z[\bm{v}]$ and $Z[\bm{v}]$ is the Keldysh partition function~(\ref{eq:Keldysh_partition_function}) perturbed by a velocity field $\bm{v}$~\cite{SupplementaryMaterial}. The phase stiffness is related to the superfluid density $n_{sD}$ in the quantum depletion as $\bar{Q}=mn_{sD}$~\cite{Coleman_2015}, where $\bar{Q}$ is the phase stiffness averaged over angles. As a result, we find that the normal fluid density vanishes and hence $n_{sD}=n_D$ if the dissipative dynamics starts from a BEC at absolute zero~\cite{SupplementaryMaterial}.
Together with Eq. \eqref{eq:supercurrent}, we arrive at the conclusion that all bosons contribute to the superfluid transport, i.e., $n_s=n$. The vanishing normal fluid component also indicates a nonzero Landau critical velocity.

We next discuss the quantum depletion density. In the weak-dissipation limit $U_{R}\gg\gamma$, it has no linear correction in $\gamma$: 
\begin{equation}\label{eq:depletion-1}
	n_{D}=\frac{(mnU_R)^{3/2}}{3\pi^{2}}\left(h_1+h_2\eta\left(\frac{\gamma}{U_{R}}\right)^{2}+O\left(\frac{\gamma}{U_R}\right)^3\right),
\end{equation}
where $\eta\simeq 0.76$, and $h_1(\varepsilon_{dd})$ and $h_2(\varepsilon_{dd})$ satisfy $h_1(0)=h_2(0)=1$ and $h_1(\varepsilon_{dd}),h_2(\varepsilon_{dd})>1$ for $\varepsilon_{dd}>0$. 
This quantum depletion can be expressed in terms of the complex scattering length \citep{PhysRevA.79.023614,PhysRevA.103.013724}
\begin{equation}
	a_{c}=a_{R}-ia_{I}:=\frac{m}{4\pi}(U_{R}-i\gamma).\label{eq:definition_ac}
\end{equation}
In the case of $\gamma=0$, Eq. \eqref{eq:depletion-1} reduces to the standard result~\cite{Ueda2010,Mathematical_method_SF_1968}:
$n_{D}=\frac{8}{3\sqrt{\pi}}(na_{R})^{3/2}$. In the weak-interaction limit $U_{R}\ll\gamma$,
the quantum depletion density is given by 
\begin{align}\label{eq:depletion-2}
	n_{D}=(\gamma n)^{3/2}&\frac{m^{3/2}}{24\pi^{5/2}}\Gamma\left(\frac{1}{4}\right)\nonumber\\
	&\times\left(1+\frac{6U_{R}}{\gamma}\frac{\Gamma(3/4)^{2}}{\Gamma(1/4)^{2}}+O\left[\left(\frac{U_{R}}{\gamma}\right)^{2}\right]\right),
\end{align}
which indicates that the dissipation leads to quantum depletion as in the case of repulsive interaction even without interactions between bosons. 
We note that the right-hand side of Eq.~(\ref{eq:depletion-2}) is irrelevant to the strength of the dipolar interaction since the quantum depletion is dominated by dissipation and the contribution of the dipolar interaction vanishes at the first order after averaged over angles. In terms of the complex scattering length in Eq. (\ref{eq:definition_ac}), the quantum depletion density for $U_{R}=0$ can be expressed as $n_{D}=\frac{1}{3\pi}\Gamma\left(\frac{1}{4}\right)(na_{I})^{3/2}$. By measuring the imaginary part of the complex scattering length from the scattering cross section \citep{PhysRevC.90.064004,PhysRevD.38.742,PhysRevA.78.023608}, we can verify this prediction. Equations \eqref{eq:depletion-1} and \eqref{eq:depletion-2} indicate that the two-body loss should be distinguished from thermal noise. They both increase the non-condensate part but only the latter increases the normal fluid density since the two-body loss induces an imaginary gap which gives rise to a phase rigidity. We note that our results \eqref{eq:depletion-1} and \eqref{eq:depletion-2} are derived from the space-averaged phase stiffness. In the presence of the dipolar interaction, the phase stiffness is spatially anisotropic~\cite{SupplementaryMaterial}. 

\

\emph{f-sum rule.--- }The $f$-sum rule plays a fundamental role in the superfluid transport theory~\cite{Ueda2010}. It is a consequence of the particle-number conservation and not expected to hold in dissipative superfluids. Here we show that weak U(1) symmetry leads to a generalized $f$-sum rule in the present case. We begin with the commutation relation:
\begin{equation}
	[\rho_{-\bm{k}},\mathcal{L}^{\dagger}(\rho_{\bm{k}})]=2i\varepsilon_{\bm{k}}\hat{N},\label{eq:f-sum-rule-exp}
\end{equation}
where $\rho_{-\bm{k}}:=\sum_{\bm{p}}a_{\bm{p}}^\dagger a_{\bm{p}-\bm{k}}$ is the density operator, $\hat{N}$ is the total particle-number operator, $\varepsilon_{\bm{k}}=|\bm{k}|^{2}/2m$ is the kinetic energy and $\mathcal{L}^{\dagger}$ represents the adjoint Lindbladian~\cite{10.1093/acprof:oso/9780199213900.001.0001}. Note that the right-hand side of Eq.~(\ref{eq:f-sum-rule-exp}) is independent of the strength $\gamma$ of dissipation. This implies that a generalized $f$-sum rule in open quantum systems holds as in closed systems. Physically, this is because the two-body loss only leads to uniform decrease in the particle-number density and hence does not generate a current inside the system. We use this identity to show that 
\begin{equation}
	\int\frac{d\omega}{2\pi\omega}\gamma^{L}(\bm{k},\omega,t_0)=N(t_0),\label{eq:Application_sum_rule}
\end{equation}
where $N(t_0)$ represents the total particle number at an arbitrary time $t_0$, and $\gamma^{L}(\bm{k},\omega,t_0)=\sum_{i,j}\frac{k_i k_j}{|\bm{k}|^2}\gamma^{i,j}(\bm{k},\omega,t_0)$ is the longitudinal component of the current-current correlation function $\gamma^{i,j}(\bm{k},\omega,t_0):=m\langle\int dt e^{i\omega t}[{j}_{\mathrm{t}}^{i}(\bm{k},t_0+t),{j}_{\mathrm{t}}^{j}(-\bm{k},t_0)]\rangle$
with
\begin{equation}
	{\bm{j}}_{\mathrm{t}}=\frac{i}{2m}\left(\nabla a_{\bm{r}}^{\dagger}a_{\bm{r}}-a_{\bm{r}}^{\dagger}\nabla a_{\bm{r}}\right)+{\bm{j}}_d(\bm{r})
\end{equation}
being the total current of the system \citep{SupplementaryMaterial,PhysRevLett.93.160404}. Here ${\bm{j}}_d(\bm{r})=\gamma\int \frac{d\bm{r}'}{2\pi}\frac{\bm{r}-\bm{r}'}{|\bm{r}-\bm{r}'|^3} a^{\dagger}_{\bm{r}'}a^{\dagger}_{\bm{r}'}a_{\bm{r}'}a_{\bm{r}'}$ represents a dissipative current. Equation \eqref{eq:Application_sum_rule} can be interpreted as a generalized
$f$-sum rule in open quantum systems. 
The normal fluid density corresponds to the transverse part of the current-current
correlation function. The difference between
longitudinal and transverse parts of the current-current correlation function
defines the superfluid
component~\cite{Mathematical_method_SF_1968}.

The generalized $f$-sum rule~(\ref{eq:Application_sum_rule}) implies that the longitudinal response is determined by the total particle number at a given time and is a direct consequence of the weak U(1) symmetry of the Lindbladian \eqref{eq: Lindblad} which holds even in the absence of particle-number conservation~\cite{Albert2014,Sieberer_2016}. The weak U(1) symmetry preserves the block diagonal form of the density matrix in the particle-number basis, and hence the $f$-sum rule holds in each particle-number sector, leading to the generalized $f$-sum rule which is general and exact.  

\emph{Excitation spectrum and stability of a molecular BEC.---}We now investigate the excitation spectrum of a dissipative BEC. 
With the mean-field and quasi-steady-state approximations, the action (\ref{eq:Keldysh_action}) is rewritten as
\begin{equation}
	S=\sum_{\bm{k},\omega}\Psi_{\tmmathbf{k},\omega}^{\dagger}\begin{pmatrix}h_{+}^{2\times2} & B\\
		B^{\dagger} & h_{-}^{2\times2}
	\end{pmatrix}\Psi_{\tmmathbf{k},\omega}\label{eq:Gaussian_action}
\end{equation}
where $\Psi_{\tmmathbf{k},\omega}=(a_{\tmmathbf{k},\omega,+},a_{\tmmathbf{-k},\omega,+}^{\dagger},a_{\tmmathbf{k},\omega,-},a_{\tmmathbf{-k},\omega,-}^{\dagger})^{T}$,
\begin{equation}
	h_{\pm}^{2\times2}=\mp\frac{1}{2}\begin{pmatrix}\varepsilon_{\bm{k},U}^{\mp}-\omega & \mp(\tilde{U}_R-i\gamma)n\\
		\mp(\tilde{U}_R+i\gamma)n & \varepsilon_{\bm{k},U}^{\mp}+\omega
	\end{pmatrix},
\end{equation}
and $B=\begin{pmatrix}0 & 0\\
	0 & -2i\gamma n
\end{pmatrix}$
with $\varepsilon_{\bm{k},U}^{\mp}:=\varepsilon_{\tmmathbf{k}}+\tilde{U}_Rn\mp2i\gamma n$, $\tilde{U}_R(\bm{k}):=U_R[1-\varepsilon_{dd}(1-3\cos^2\theta_{\bm{k}})]$, and $\cos\theta_{\bm{k}}=\bm{k}\cdot\hat{z}/|\bm{k}|$. The Green's functions of bosonic fields~\cite{Kamenev_2011} are given by the inverse
of the Gaussian matrix in Eq. (\ref{eq:Gaussian_action}) (see Supplemental Material for details~\citep{SupplementaryMaterial}). 

The excitation spectrum can be obtained from the poles of the Green's function determined by $\mathrm{det}\begin{pmatrix}h_{+}^{2\times2} & B\\
	B^{\dagger} & h_{-}^{2\times2}
\end{pmatrix}=0$:
\begin{equation}
	\begin{aligned}\omega_{1,2} & =-2i\gamma n\pm\sqrt{\varepsilon_{\tmmathbf{k}}(\varepsilon_{\tmmathbf{k}}+2\tilde{U}_{R}(\bm{k})n)-\gamma^{2}n^{2}},\\
		\omega_{3,4} & =2i\gamma n\pm\sqrt{\varepsilon_{\tmmathbf{k}}(\varepsilon_{\tmmathbf{k}}+2\tilde{U}_{R}(\bm{k})n)-\gamma^{2}n^{2}}.
	\end{aligned}
	\label{eq:complex_spectrum}
\end{equation}
The spectra $\omega_{1,2}$ originate from the poles of the retarded Green's function, which are also eigenfrequencies of the non-Hermitian Gross-Pitaevskii equation~\cite{liu2022weakly}. 
Similarly, $\omega_{3,4}$ are complex conjugate to $\omega_{1,2}$, representing the poles of the advanced Green's function~\footnote{Due to the nonunitary Bogoliubov transformation, the amplitude modes have mixed with the phase modes, leading to the difference between the spectrums of the NG mode and that of the bosonic operators.}. The complex spectrum is expected to be measured from the spectral function, which is defined as 
\begin{equation}
	A(\bm{k},\omega)  =\frac{i}{2\pi}(G^{>}(\bm{k},\omega)+G^{<}(\bm{k},\omega)),\label{eq:spectrum_function}
\end{equation}
where $G^{<}$ and $G^{>}$ represent the lesser and greater Green's function, respectively. Concrete analytic expressions of the spectral function and other plots under different limits and directions are given in the Supplemental Material \cite{SupplementaryMaterial}.

We investigate the dynamical stability of a dissipative BEC via the complex spectra in Eq. (\ref{eq:complex_spectrum}). The system is dynamically stable if and only if both of the two retarded (advanced) spectra $\omega_{1,2\,(3,4)}$ have negative (positive) imaginary parts~\cite{Ce2022}. We find that the system becomes stable when $\varepsilon_{dd}<1+\sqrt{3}\gamma/U_R$, and that the two-body loss enlarges the stable region. Physically, the enhanced stability due to dissipation can be understood as a consequence of the correlation induced by dissipation which plays a role similar to the repulsive interaction. A similar result for a dissipative attractive BEC has been obtained in Ref.~\cite{Ce2022}. This stability analysis sets an upper bound on the dipolar interaction strength in order for the molecular BEC to be stable. 
\begin{figure}
	\includegraphics[width=1.0\columnwidth]{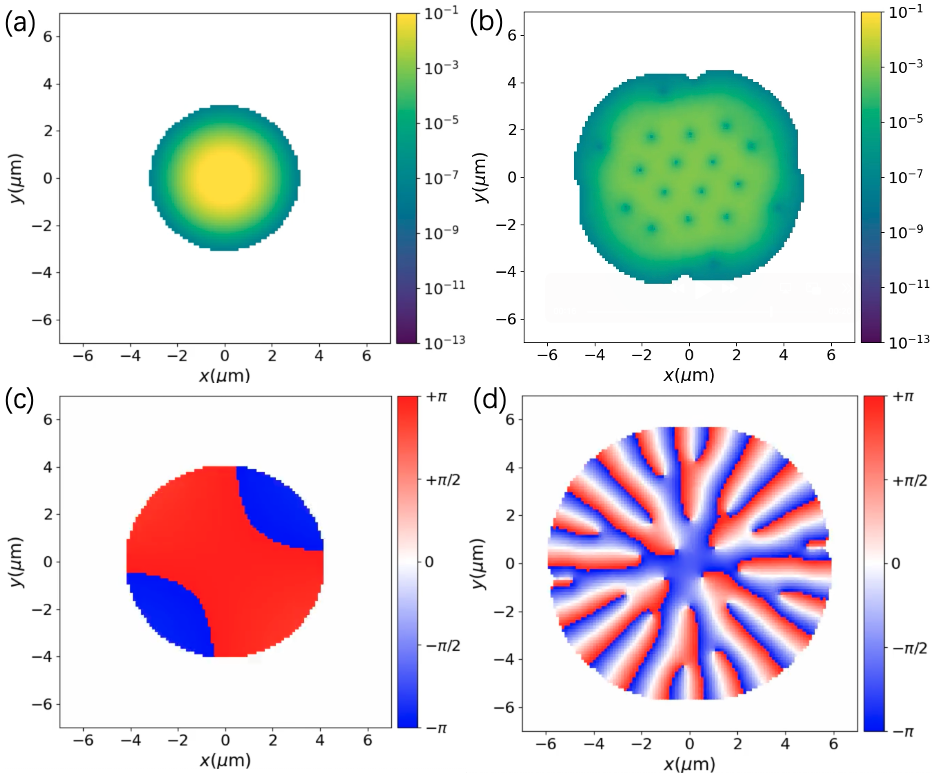}
	\caption{(a), (b): Density profiles of the order parameter without and with two-body loss at $t=200$ms. (c), (d): Phase profiles of the order parameter without and with two-body loss at $t=200$ms. We start with the ground state of a non-interacting BEC in a pancake-shaped harmonic trap at time $t=0$ and switch on the stirring of the BEC and two-body loss. Here the trap frequencies are $\omega_x=\omega_y=2\pi\times108\mathrm{Hz},\ \omega_z=100\omega_x$, the mass of a boson is $m=2.59\times10^{-22}\mathrm{g}$, and the dissipation rate is $\gamma=3\times10^{-13}\mathrm{cm}^3/\mathrm{s}$. We stir the BEC by introducing slight anisotropy in the trap frequencies ($\omega_x\rightarrow1.03\omega_x$ and $\omega_y\rightarrow1.09\omega_y$) and rotating the potential about the $z$ axis with frequency $\Omega=0.75\omega_x$.
	}
	
	\label{fig:stable}
\end{figure}

\emph{Experimental signatures of superfluidity.---}One of the important experimental evidences of superfluidity is the formation of quantized vortices by stirring the BEC with a high angular velocity~\cite{Ueda2010,Tsubota2002,Kasamatsu2003,Madison2001}. Here we simulate the dynamics of a non-interacting trapped BEC with two-body loss. We first prepare the ground state of free bosons in a pancake-shaped harmonic trap with trap frequencies $\omega_x=\omega_y=\omega_z/100$. Then we switch on an anisotropic trap, stirring, and two-body loss at time $t=0$. By simulating the dissipative Gross-Pitaevskii equation~\cite{liu2022weakly,Kasamatsu2003}, we obtain the dynamics of the density and phase profiles of the order parameter $\psi$ of the dissipative BEC as shown in Fig. \ref{fig:stable}. From the numerical results, we can see that quantized vortices are formed and enter the system in the dynamics in Fig. \ref{fig:stable}(b) compared with the non-dissipative case in Fig. \ref{fig:stable}(a), indicating that the superfluidity is induced solely by two-body loss which verifies our theoretical prediction. The superfluid density $\rho_{\mathrm{s}}$ can be also measured from the total angular momentum $\langle L\rangle$ of a quantum gas under a synthetic gauge field ~\cite{PhysRevLett.104.030401,Chen2018}:
\begin{equation}
	\frac{\rho_{\mathrm{s}}}{\rho_{\mathrm{t}}}=1-\lim_{\omega_{\mathrm{eff}}\to0}\frac{\langle L\rangle}{I_{\mathrm{cl}}\omega_{\mathrm{eff}}},
\end{equation}
where $\omega_{\mathrm{eff}}$ is an effective angular momentum induced by two Laguerre-Gauss beams~\cite{PhysRevLett.104.030401}, $\rho_{\mathrm{t}}$ is the total boson density and $I_{\mathrm{cl}}$ represents the classical moment of inertia of the gas. 

\emph{Conclusion.---}In this Letter, we have investigated the
transport property of a dissipative molecular BEC with two-body loss and
shown that superfluidity can be induced by dissipation without recourse to interparticle interaction both numerically and analytically. We have demonstrated that the normal fluid density in the dissipative BEC vanishes, i.e., all bosons contribute to superfluid transport in the presence of dissipation if we start from the absolute zero in the remote past. We also find that the quantum depletion is induced by dissipation, indicating that dissipation plays a role similar to repulsive interaction. We have further investigated the spectrum and the stability of a dissipative molecular BEC, which can be experimentally tested through the angle-resolved measurement of the spectral function~\citep{Brown2020,PhysRevB.97.125117}. 

\emph{Acknowledgements.---}We are grateful to Zongping Gong, Yukai Lu, Waseem S. Bakr, Michikazu Kobayashi and Masaya Kunimi for fruitful discussion. H. L. is
supported by Forefront Physics and Mathematics Program to Drive Transformation
(FoPM), a World-leading Innovative Graduate Study (WINGS) Program,
the University of Tokyo. M.N. is supported by JSPS KAKENHI Grant No.
JP20K14383 and No.~JP24K16989. M.U. is supported by JSPS KAKENHI Grant No. JP22H01152
and the CREST program “Quantum Frontiers” of JST (Grand No. JPMJCR23I1).

\bibliographystyle{apsrev4-2}
\bibliography{MyNewCollection}

\begin{thebibliography}{75}%
\makeatletter
\providecommand \@ifxundefined [1]{%
 \@ifx{#1\undefined}
}%
\providecommand \@ifnum [1]{%
 \ifnum #1\expandafter \@firstoftwo
 \else \expandafter \@secondoftwo
 \fi
}%
\providecommand \@ifx [1]{%
 \ifx #1\expandafter \@firstoftwo
 \else \expandafter \@secondoftwo
 \fi
}%
\providecommand \natexlab [1]{#1}%
\providecommand \enquote  [1]{``#1''}%
\providecommand \bibnamefont  [1]{#1}%
\providecommand \bibfnamefont [1]{#1}%
\providecommand \citenamefont [1]{#1}%
\providecommand \href@noop [0]{\@secondoftwo}%
\providecommand \href [0]{\begingroup \@sanitize@url \@href}%
\providecommand \@href[1]{\@@startlink{#1}\@@href}%
\providecommand \@@href[1]{\endgroup#1\@@endlink}%
\providecommand \@sanitize@url [0]{\catcode `\\12\catcode `\$12\catcode
  `\&12\catcode `\#12\catcode `\^12\catcode `\_12\catcode `\%12\relax}%
\providecommand \@@startlink[1]{}%
\providecommand \@@endlink[0]{}%
\providecommand \url  [0]{\begingroup\@sanitize@url \@url }%
\providecommand \@url [1]{\endgroup\@href {#1}{\urlprefix }}%
\providecommand \urlprefix  [0]{URL }%
\providecommand \Eprint [0]{\href }%
\providecommand \doibase [0]{https://doi.org/}%
\providecommand \selectlanguage [0]{\@gobble}%
\providecommand \bibinfo  [0]{\@secondoftwo}%
\providecommand \bibfield  [0]{\@secondoftwo}%
\providecommand \translation [1]{[#1]}%
\providecommand \BibitemOpen [0]{}%
\providecommand \bibitemStop [0]{}%
\providecommand \bibitemNoStop [0]{.\EOS\space}%
\providecommand \EOS [0]{\spacefactor3000\relax}%
\providecommand \BibitemShut  [1]{\csname bibitem#1\endcsname}%
\let\auto@bib@innerbib\@empty
\bibitem [{\citenamefont {Leggett}(2006)}]{Leggett2006}%
  \BibitemOpen
  \bibfield  {author} {\bibinfo {author} {\bibfnamefont {A.~J.}\ \bibnamefont
  {Leggett}},\ }\href
  {https://doi.org/10.1093/acprof:oso/9780198526438.001.0001} {\emph {\bibinfo
  {title} {{Quantum Liquids: Bose condensation and Cooper pairing in
  condensed-matter systems}}}}\ (\bibinfo  {publisher} {Oxford University
  Press, Oxford, UK},\ \bibinfo {year} {2006})\BibitemShut {NoStop}%
\bibitem [{\citenamefont {Coleman}(2015)}]{Coleman_2015}%
  \BibitemOpen
  \bibfield  {author} {\bibinfo {author} {\bibfnamefont {P.}~\bibnamefont
  {Coleman}},\ }\href@noop {} {\emph {\bibinfo {title} {Introduction to
  Many-Body Physics}}}\ (\bibinfo  {publisher} {Cambridge University Press,
  Cambridge, UK},\ \bibinfo {year} {2015})\BibitemShut {NoStop}%
\bibitem [{\citenamefont {Schmitt}(2015)}]{Schmitt2015}%
  \BibitemOpen
  \bibfield  {author} {\bibinfo {author} {\bibfnamefont {A.}~\bibnamefont
  {Schmitt}},\ }\href {https://doi.org/10.1007/978-3-319-07947-9} {\emph
  {\bibinfo {title} {Introduction to Superfluidity}}},\ Vol.\ \bibinfo {volume}
  {888}\ (\bibinfo  {publisher} {Springer International Publishing, New York,
  US},\ \bibinfo {year} {2015})\BibitemShut {NoStop}%
\bibitem [{\citenamefont {Feynman}(1957)}]{RevModPhys.29.205}%
  \BibitemOpen
  \bibfield  {author} {\bibinfo {author} {\bibfnamefont {R.~P.}\ \bibnamefont
  {Feynman}},\ }\href {https://doi.org/10.1103/RevModPhys.29.205} {\bibfield
  {journal} {\bibinfo  {journal} {Rev. Mod. Phys.}\ }\textbf {\bibinfo {volume}
  {29}},\ \bibinfo {pages} {205} (\bibinfo {year} {1957})}\BibitemShut
  {NoStop}%
\bibitem [{\citenamefont {Leggett}(1975)}]{RevModPhys.47.331}%
  \BibitemOpen
  \bibfield  {author} {\bibinfo {author} {\bibfnamefont {A.~J.}\ \bibnamefont
  {Leggett}},\ }\href {https://doi.org/10.1103/RevModPhys.47.331} {\bibfield
  {journal} {\bibinfo  {journal} {Rev. Mod. Phys.}\ }\textbf {\bibinfo {volume}
  {47}},\ \bibinfo {pages} {331} (\bibinfo {year} {1975})}\BibitemShut
  {NoStop}%
\bibitem [{\citenamefont {Ueda}(2010)}]{Ueda2010}%
  \BibitemOpen
  \bibfield  {author} {\bibinfo {author} {\bibfnamefont {M.}~\bibnamefont
  {Ueda}},\ }\href {https://doi.org/10.1142/7216} {\emph {\bibinfo {title}
  {Fundamentals and New Frontiers of Bose-Einstein Condensation}}}\ (\bibinfo
  {publisher} {WORLD SCIENTIFIC, Singapore},\ \bibinfo {year}
  {2010})\BibitemShut {NoStop}%
\bibitem [{\citenamefont {Dalfovo}\ \emph {et~al.}(1999)\citenamefont
  {Dalfovo}, \citenamefont {Giorgini}, \citenamefont {Pitaevskii},\ and\
  \citenamefont {Stringari}}]{RevModPhys.71.463}%
  \BibitemOpen
  \bibfield  {author} {\bibinfo {author} {\bibfnamefont {F.}~\bibnamefont
  {Dalfovo}}, \bibinfo {author} {\bibfnamefont {S.}~\bibnamefont {Giorgini}},
  \bibinfo {author} {\bibfnamefont {L.~P.}\ \bibnamefont {Pitaevskii}},\ and\
  \bibinfo {author} {\bibfnamefont {S.}~\bibnamefont {Stringari}},\ }\href
  {https://doi.org/10.1103/RevModPhys.71.463} {\bibfield  {journal} {\bibinfo
  {journal} {Rev. Mod. Phys.}\ }\textbf {\bibinfo {volume} {71}},\ \bibinfo
  {pages} {463} (\bibinfo {year} {1999})}\BibitemShut {NoStop}%
\bibitem [{\citenamefont {Pethick}\ and\ \citenamefont
  {Smith}(2008)}]{Pethick_Smith_2008}%
  \BibitemOpen
  \bibfield  {author} {\bibinfo {author} {\bibfnamefont {C.~J.}\ \bibnamefont
  {Pethick}}\ and\ \bibinfo {author} {\bibfnamefont {H.}~\bibnamefont
  {Smith}},\ }\href@noop {} {\emph {\bibinfo {title} {Bose-Einstein
  Condensation in Dilute Gases}}},\ \bibinfo {edition} {2nd}\ ed.\ (\bibinfo
  {publisher} {Cambridge University Press, Cambridge UK},\ \bibinfo {year}
  {2008})\BibitemShut {NoStop}%
\bibitem [{\citenamefont {Bigagli}\ \emph {et~al.}(2024)\citenamefont
  {Bigagli}, \citenamefont {Yuan}, \citenamefont {Zhang}, \citenamefont
  {Bulatovic}, \citenamefont {Karman}, \citenamefont {Stevenson},\ and\
  \citenamefont {Will}}]{Sebastian2023}%
  \BibitemOpen
  \bibfield  {author} {\bibinfo {author} {\bibfnamefont {N.}~\bibnamefont
  {Bigagli}}, \bibinfo {author} {\bibfnamefont {W.}~\bibnamefont {Yuan}},
  \bibinfo {author} {\bibfnamefont {S.}~\bibnamefont {Zhang}}, \bibinfo
  {author} {\bibfnamefont {B.}~\bibnamefont {Bulatovic}}, \bibinfo {author}
  {\bibfnamefont {T.}~\bibnamefont {Karman}}, \bibinfo {author} {\bibfnamefont
  {I.}~\bibnamefont {Stevenson}},\ and\ \bibinfo {author} {\bibfnamefont
  {S.}~\bibnamefont {Will}},\ }\href
  {https://doi.org/10.1038/s41586-024-07492-z} {\bibfield  {journal} {\bibinfo
  {journal} {Nature}\ }\textbf {\bibinfo {volume} {631}},\ \bibinfo {pages}
  {289} (\bibinfo {year} {2024})}\BibitemShut {NoStop}%
\bibitem [{\citenamefont {Ni}\ \emph {et~al.}(2008)\citenamefont {Ni},
  \citenamefont {Ospelkaus}, \citenamefont {de~Miranda}, \citenamefont {Pe'er},
  \citenamefont {Neyenhuis}, \citenamefont {Zirbel}, \citenamefont
  {Kotochigova}, \citenamefont {Julienne}, \citenamefont {Jin},\ and\
  \citenamefont {Ye}}]{Ni2008}%
  \BibitemOpen
  \bibfield  {author} {\bibinfo {author} {\bibfnamefont {K.-K.}\ \bibnamefont
  {Ni}}, \bibinfo {author} {\bibfnamefont {S.}~\bibnamefont {Ospelkaus}},
  \bibinfo {author} {\bibfnamefont {M.~H.~G.}\ \bibnamefont {de~Miranda}},
  \bibinfo {author} {\bibfnamefont {A.}~\bibnamefont {Pe'er}}, \bibinfo
  {author} {\bibfnamefont {B.}~\bibnamefont {Neyenhuis}}, \bibinfo {author}
  {\bibfnamefont {J.~J.}\ \bibnamefont {Zirbel}}, \bibinfo {author}
  {\bibfnamefont {S.}~\bibnamefont {Kotochigova}}, \bibinfo {author}
  {\bibfnamefont {P.~S.}\ \bibnamefont {Julienne}}, \bibinfo {author}
  {\bibfnamefont {D.~S.}\ \bibnamefont {Jin}},\ and\ \bibinfo {author}
  {\bibfnamefont {J.}~\bibnamefont {Ye}},\ }\href
  {https://doi.org/10.1126/science.1163861} {\bibfield  {journal} {\bibinfo
  {journal} {Science}\ }\textbf {\bibinfo {volume} {322}},\ \bibinfo {pages}
  {231} (\bibinfo {year} {2008})}\BibitemShut {NoStop}%
\bibitem [{\citenamefont {Ni}\ \emph {et~al.}(2010)\citenamefont {Ni},
  \citenamefont {Ospelkaus}, \citenamefont {Wang}, \citenamefont
  {Qu{\'e}m{\'e}ner}, \citenamefont {Neyenhuis}, \citenamefont {de~Miranda},
  \citenamefont {Bohn}, \citenamefont {Ye},\ and\ \citenamefont
  {Jin}}]{Ni2010}%
  \BibitemOpen
  \bibfield  {author} {\bibinfo {author} {\bibfnamefont {K.~K.}\ \bibnamefont
  {Ni}}, \bibinfo {author} {\bibfnamefont {S.}~\bibnamefont {Ospelkaus}},
  \bibinfo {author} {\bibfnamefont {D.}~\bibnamefont {Wang}}, \bibinfo {author}
  {\bibfnamefont {G.}~\bibnamefont {Qu{\'e}m{\'e}ner}}, \bibinfo {author}
  {\bibfnamefont {B.}~\bibnamefont {Neyenhuis}}, \bibinfo {author}
  {\bibfnamefont {M.~H.~G.}\ \bibnamefont {de~Miranda}}, \bibinfo {author}
  {\bibfnamefont {J.~L.}\ \bibnamefont {Bohn}}, \bibinfo {author}
  {\bibfnamefont {J.}~\bibnamefont {Ye}},\ and\ \bibinfo {author}
  {\bibfnamefont {D.~S.}\ \bibnamefont {Jin}},\ }\href
  {https://doi.org/10.1038/nature08953} {\bibfield  {journal} {\bibinfo
  {journal} {Nature}\ }\textbf {\bibinfo {volume} {464}},\ \bibinfo {pages}
  {1324} (\bibinfo {year} {2010})}\BibitemShut {NoStop}%
\bibitem [{\citenamefont {Christianen}\ \emph {et~al.}(2019)\citenamefont
  {Christianen}, \citenamefont {Zwierlein}, \citenamefont {Groenenboom},\ and\
  \citenamefont {Karman}}]{Arthur2019}%
  \BibitemOpen
  \bibfield  {author} {\bibinfo {author} {\bibfnamefont {A.}~\bibnamefont
  {Christianen}}, \bibinfo {author} {\bibfnamefont {M.~W.}\ \bibnamefont
  {Zwierlein}}, \bibinfo {author} {\bibfnamefont {G.~C.}\ \bibnamefont
  {Groenenboom}},\ and\ \bibinfo {author} {\bibfnamefont {T.}~\bibnamefont
  {Karman}},\ }\href {https://doi.org/10.1103/PhysRevLett.123.123402}
  {\bibfield  {journal} {\bibinfo  {journal} {Phys. Rev. Lett.}\ }\textbf
  {\bibinfo {volume} {123}},\ \bibinfo {pages} {123402} (\bibinfo {year}
  {2019})}\BibitemShut {NoStop}%
\bibitem [{\citenamefont {Liu}\ \emph {et~al.}(2020)\citenamefont {Liu},
  \citenamefont {Hu}, \citenamefont {Nichols}, \citenamefont {Grimes},
  \citenamefont {Karman}, \citenamefont {Guo},\ and\ \citenamefont
  {Ni}}]{Liu2020}%
  \BibitemOpen
  \bibfield  {author} {\bibinfo {author} {\bibfnamefont {Y.}~\bibnamefont
  {Liu}}, \bibinfo {author} {\bibfnamefont {M.-G.}\ \bibnamefont {Hu}},
  \bibinfo {author} {\bibfnamefont {M.~A.}\ \bibnamefont {Nichols}}, \bibinfo
  {author} {\bibfnamefont {D.~D.}\ \bibnamefont {Grimes}}, \bibinfo {author}
  {\bibfnamefont {T.}~\bibnamefont {Karman}}, \bibinfo {author} {\bibfnamefont
  {H.}~\bibnamefont {Guo}},\ and\ \bibinfo {author} {\bibfnamefont {K.-K.}\
  \bibnamefont {Ni}},\ }\href {https://doi.org/10.1038/s41567-020-0968-8}
  {\bibfield  {journal} {\bibinfo  {journal} {Nature Physics}\ }\textbf
  {\bibinfo {volume} {16}},\ \bibinfo {pages} {1132} (\bibinfo {year}
  {2020})}\BibitemShut {NoStop}%
\bibitem [{\citenamefont {Bause}\ \emph {et~al.}(2021)\citenamefont {Bause},
  \citenamefont {Schindewolf}, \citenamefont {Tao}, \citenamefont {Duda},
  \citenamefont {Chen}, \citenamefont {Qu\'em\'ener}, \citenamefont {Karman},
  \citenamefont {Christianen}, \citenamefont {Bloch},\ and\ \citenamefont
  {Luo}}]{Bause2021}%
  \BibitemOpen
  \bibfield  {author} {\bibinfo {author} {\bibfnamefont {R.}~\bibnamefont
  {Bause}}, \bibinfo {author} {\bibfnamefont {A.}~\bibnamefont {Schindewolf}},
  \bibinfo {author} {\bibfnamefont {R.}~\bibnamefont {Tao}}, \bibinfo {author}
  {\bibfnamefont {M.}~\bibnamefont {Duda}}, \bibinfo {author} {\bibfnamefont
  {X.-Y.}\ \bibnamefont {Chen}}, \bibinfo {author} {\bibfnamefont
  {G.}~\bibnamefont {Qu\'em\'ener}}, \bibinfo {author} {\bibfnamefont
  {T.}~\bibnamefont {Karman}}, \bibinfo {author} {\bibfnamefont
  {A.}~\bibnamefont {Christianen}}, \bibinfo {author} {\bibfnamefont
  {I.}~\bibnamefont {Bloch}},\ and\ \bibinfo {author} {\bibfnamefont {X.-Y.}\
  \bibnamefont {Luo}},\ }\href
  {https://doi.org/10.1103/PhysRevResearch.3.033013} {\bibfield  {journal}
  {\bibinfo  {journal} {Phys. Rev. Res.}\ }\textbf {\bibinfo {volume} {3}},\
  \bibinfo {pages} {033013} (\bibinfo {year} {2021})}\BibitemShut {NoStop}%
\bibitem [{\citenamefont {Gersema}\ \emph {et~al.}(2021)\citenamefont
  {Gersema}, \citenamefont {Voges}, \citenamefont {Meyer~zum Alten~Borgloh},
  \citenamefont {Koch}, \citenamefont {Hartmann}, \citenamefont {Zenesini},
  \citenamefont {Ospelkaus}, \citenamefont {Lin}, \citenamefont {He},\ and\
  \citenamefont {Wang}}]{Gersema2021}%
  \BibitemOpen
  \bibfield  {author} {\bibinfo {author} {\bibfnamefont {P.}~\bibnamefont
  {Gersema}}, \bibinfo {author} {\bibfnamefont {K.~K.}\ \bibnamefont {Voges}},
  \bibinfo {author} {\bibfnamefont {M.}~\bibnamefont {Meyer~zum
  Alten~Borgloh}}, \bibinfo {author} {\bibfnamefont {L.}~\bibnamefont {Koch}},
  \bibinfo {author} {\bibfnamefont {T.}~\bibnamefont {Hartmann}}, \bibinfo
  {author} {\bibfnamefont {A.}~\bibnamefont {Zenesini}}, \bibinfo {author}
  {\bibfnamefont {S.}~\bibnamefont {Ospelkaus}}, \bibinfo {author}
  {\bibfnamefont {J.}~\bibnamefont {Lin}}, \bibinfo {author} {\bibfnamefont
  {J.}~\bibnamefont {He}},\ and\ \bibinfo {author} {\bibfnamefont
  {D.}~\bibnamefont {Wang}},\ }\href
  {https://doi.org/10.1103/PhysRevLett.127.163401} {\bibfield  {journal}
  {\bibinfo  {journal} {Phys. Rev. Lett.}\ }\textbf {\bibinfo {volume} {127}},\
  \bibinfo {pages} {163401} (\bibinfo {year} {2021})}\BibitemShut {NoStop}%
\bibitem [{\citenamefont {Rosenberg}\ \emph {et~al.}(2022)\citenamefont
  {Rosenberg}, \citenamefont {Christakis}, \citenamefont {Guardado-Sanchez},
  \citenamefont {Yan},\ and\ \citenamefont {Bakr}}]{rosenberg2022observation}%
  \BibitemOpen
  \bibfield  {author} {\bibinfo {author} {\bibfnamefont {J.~S.}\ \bibnamefont
  {Rosenberg}}, \bibinfo {author} {\bibfnamefont {L.}~\bibnamefont
  {Christakis}}, \bibinfo {author} {\bibfnamefont {E.}~\bibnamefont
  {Guardado-Sanchez}}, \bibinfo {author} {\bibfnamefont {Z.~Z.}\ \bibnamefont
  {Yan}},\ and\ \bibinfo {author} {\bibfnamefont {W.~S.}\ \bibnamefont
  {Bakr}},\ }\href {https://doi.org/10.1038/s41567-022-01695-9} {\bibfield
  {journal} {\bibinfo  {journal} {Nature Physics}\ }\textbf {\bibinfo {volume}
  {18}},\ \bibinfo {pages} {1062} (\bibinfo {year} {2022})}\BibitemShut
  {NoStop}%
\bibitem [{\citenamefont {Bause}\ \emph {et~al.}(2023)\citenamefont {Bause},
  \citenamefont {Christianen}, \citenamefont {Schindewolf}, \citenamefont
  {Bloch},\ and\ \citenamefont {Luo}}]{Bause2023}%
  \BibitemOpen
  \bibfield  {author} {\bibinfo {author} {\bibfnamefont {R.}~\bibnamefont
  {Bause}}, \bibinfo {author} {\bibfnamefont {A.}~\bibnamefont {Christianen}},
  \bibinfo {author} {\bibfnamefont {A.}~\bibnamefont {Schindewolf}}, \bibinfo
  {author} {\bibfnamefont {I.}~\bibnamefont {Bloch}},\ and\ \bibinfo {author}
  {\bibfnamefont {X.-Y.}\ \bibnamefont {Luo}},\ }\href
  {https://doi.org/10.1021/acs.jpca.2c08095} {\bibfield  {journal} {\bibinfo
  {journal} {The Journal of Physical Chemistry A}\ }\textbf {\bibinfo {volume}
  {127}},\ \bibinfo {pages} {729} (\bibinfo {year} {2023})}\BibitemShut
  {NoStop}%
\bibitem [{\citenamefont {Ospelkaus}\ \emph {et~al.}(2010)\citenamefont
  {Ospelkaus}, \citenamefont {Ni}, \citenamefont {Wang}, \citenamefont
  {de~Miranda}, \citenamefont {Neyenhuis}, \citenamefont {Qu{\'e}m{\'e}ner},
  \citenamefont {Julienne}, \citenamefont {Bohn}, \citenamefont {Jin},\ and\
  \citenamefont {Ye}}]{Ospelkaus2010}%
  \BibitemOpen
  \bibfield  {author} {\bibinfo {author} {\bibfnamefont {S.}~\bibnamefont
  {Ospelkaus}}, \bibinfo {author} {\bibfnamefont {K.~K.}\ \bibnamefont {Ni}},
  \bibinfo {author} {\bibfnamefont {D.}~\bibnamefont {Wang}}, \bibinfo {author}
  {\bibfnamefont {M.~H.~G.}\ \bibnamefont {de~Miranda}}, \bibinfo {author}
  {\bibfnamefont {B.}~\bibnamefont {Neyenhuis}}, \bibinfo {author}
  {\bibfnamefont {G.}~\bibnamefont {Qu{\'e}m{\'e}ner}}, \bibinfo {author}
  {\bibfnamefont {P.~S.}\ \bibnamefont {Julienne}}, \bibinfo {author}
  {\bibfnamefont {J.~L.}\ \bibnamefont {Bohn}}, \bibinfo {author}
  {\bibfnamefont {D.~S.}\ \bibnamefont {Jin}},\ and\ \bibinfo {author}
  {\bibfnamefont {J.}~\bibnamefont {Ye}},\ }\href
  {https://doi.org/10.1126/science.1184121} {\bibfield  {journal} {\bibinfo
  {journal} {Science}\ }\textbf {\bibinfo {volume} {327}},\ \bibinfo {pages}
  {853} (\bibinfo {year} {2010})}\BibitemShut {NoStop}%
\bibitem [{\citenamefont {Karman}\ and\ \citenamefont
  {Hutson}(2018)}]{Karman2018}%
  \BibitemOpen
  \bibfield  {author} {\bibinfo {author} {\bibfnamefont {T.}~\bibnamefont
  {Karman}}\ and\ \bibinfo {author} {\bibfnamefont {J.~M.}\ \bibnamefont
  {Hutson}},\ }\href {https://doi.org/10.1103/PhysRevLett.121.163401}
  {\bibfield  {journal} {\bibinfo  {journal} {Phys. Rev. Lett.}\ }\textbf
  {\bibinfo {volume} {121}},\ \bibinfo {pages} {163401} (\bibinfo {year}
  {2018})}\BibitemShut {NoStop}%
\bibitem [{\citenamefont {Will}\ \emph {et~al.}(2016)\citenamefont {Will},
  \citenamefont {Park}, \citenamefont {Yan}, \citenamefont {Loh},\ and\
  \citenamefont {Zwierlein}}]{Sebastian2016}%
  \BibitemOpen
  \bibfield  {author} {\bibinfo {author} {\bibfnamefont {S.~A.}\ \bibnamefont
  {Will}}, \bibinfo {author} {\bibfnamefont {J.~W.}\ \bibnamefont {Park}},
  \bibinfo {author} {\bibfnamefont {Z.~Z.}\ \bibnamefont {Yan}}, \bibinfo
  {author} {\bibfnamefont {H.}~\bibnamefont {Loh}},\ and\ \bibinfo {author}
  {\bibfnamefont {M.~W.}\ \bibnamefont {Zwierlein}},\ }\href
  {https://doi.org/10.1103/PhysRevLett.116.225306} {\bibfield  {journal}
  {\bibinfo  {journal} {Phys. Rev. Lett.}\ }\textbf {\bibinfo {volume} {116}},\
  \bibinfo {pages} {225306} (\bibinfo {year} {2016})}\BibitemShut {NoStop}%
\bibitem [{\citenamefont {Lin}\ \emph {et~al.}(2023)\citenamefont {Lin},
  \citenamefont {Chen}, \citenamefont {Jin}, \citenamefont {Shi}, \citenamefont
  {Deng}, \citenamefont {Zhang}, \citenamefont {Qu\'em\'ener}, \citenamefont
  {Shi}, \citenamefont {Yi},\ and\ \citenamefont {Wang}}]{Lin2023}%
  \BibitemOpen
  \bibfield  {author} {\bibinfo {author} {\bibfnamefont {J.}~\bibnamefont
  {Lin}}, \bibinfo {author} {\bibfnamefont {G.}~\bibnamefont {Chen}}, \bibinfo
  {author} {\bibfnamefont {M.}~\bibnamefont {Jin}}, \bibinfo {author}
  {\bibfnamefont {Z.}~\bibnamefont {Shi}}, \bibinfo {author} {\bibfnamefont
  {F.}~\bibnamefont {Deng}}, \bibinfo {author} {\bibfnamefont {W.}~\bibnamefont
  {Zhang}}, \bibinfo {author} {\bibfnamefont {G.}~\bibnamefont {Qu\'em\'ener}},
  \bibinfo {author} {\bibfnamefont {T.}~\bibnamefont {Shi}}, \bibinfo {author}
  {\bibfnamefont {S.}~\bibnamefont {Yi}},\ and\ \bibinfo {author}
  {\bibfnamefont {D.}~\bibnamefont {Wang}},\ }\href
  {https://doi.org/10.1103/PhysRevX.13.031032} {\bibfield  {journal} {\bibinfo
  {journal} {Phys. Rev. X}\ }\textbf {\bibinfo {volume} {13}},\ \bibinfo
  {pages} {031032} (\bibinfo {year} {2023})}\BibitemShut {NoStop}%
\bibitem [{\citenamefont {Yan}\ \emph {et~al.}(2020)\citenamefont {Yan},
  \citenamefont {Park}, \citenamefont {Ni}, \citenamefont {Loh}, \citenamefont
  {Will}, \citenamefont {Karman},\ and\ \citenamefont {Zwierlein}}]{Yan2020}%
  \BibitemOpen
  \bibfield  {author} {\bibinfo {author} {\bibfnamefont {Z.~Z.}\ \bibnamefont
  {Yan}}, \bibinfo {author} {\bibfnamefont {J.~W.}\ \bibnamefont {Park}},
  \bibinfo {author} {\bibfnamefont {Y.}~\bibnamefont {Ni}}, \bibinfo {author}
  {\bibfnamefont {H.}~\bibnamefont {Loh}}, \bibinfo {author} {\bibfnamefont
  {S.}~\bibnamefont {Will}}, \bibinfo {author} {\bibfnamefont {T.}~\bibnamefont
  {Karman}},\ and\ \bibinfo {author} {\bibfnamefont {M.}~\bibnamefont
  {Zwierlein}},\ }\href {https://doi.org/10.1103/PhysRevLett.125.063401}
  {\bibfield  {journal} {\bibinfo  {journal} {Phys. Rev. Lett.}\ }\textbf
  {\bibinfo {volume} {125}},\ \bibinfo {pages} {063401} (\bibinfo {year}
  {2020})}\BibitemShut {NoStop}%
\bibitem [{\citenamefont {Cairncross}\ \emph {et~al.}(2021)\citenamefont
  {Cairncross}, \citenamefont {Zhang}, \citenamefont {Picard}, \citenamefont
  {Yu}, \citenamefont {Wang},\ and\ \citenamefont {Ni}}]{Cairncross2021}%
  \BibitemOpen
  \bibfield  {author} {\bibinfo {author} {\bibfnamefont {W.~B.}\ \bibnamefont
  {Cairncross}}, \bibinfo {author} {\bibfnamefont {J.~T.}\ \bibnamefont
  {Zhang}}, \bibinfo {author} {\bibfnamefont {L.~R.~B.}\ \bibnamefont
  {Picard}}, \bibinfo {author} {\bibfnamefont {Y.}~\bibnamefont {Yu}}, \bibinfo
  {author} {\bibfnamefont {K.}~\bibnamefont {Wang}},\ and\ \bibinfo {author}
  {\bibfnamefont {K.-K.}\ \bibnamefont {Ni}},\ }\href
  {https://doi.org/10.1103/PhysRevLett.126.123402} {\bibfield  {journal}
  {\bibinfo  {journal} {Phys. Rev. Lett.}\ }\textbf {\bibinfo {volume} {126}},\
  \bibinfo {pages} {123402} (\bibinfo {year} {2021})}\BibitemShut {NoStop}%
\bibitem [{\citenamefont {Lam}\ \emph {et~al.}(2022)\citenamefont {Lam},
  \citenamefont {Bigagli}, \citenamefont {Warner}, \citenamefont {Yuan},
  \citenamefont {Zhang}, \citenamefont {Tiemann}, \citenamefont {Stevenson},\
  and\ \citenamefont {Will}}]{Lam2022}%
  \BibitemOpen
  \bibfield  {author} {\bibinfo {author} {\bibfnamefont {A.~Z.}\ \bibnamefont
  {Lam}}, \bibinfo {author} {\bibfnamefont {N.}~\bibnamefont {Bigagli}},
  \bibinfo {author} {\bibfnamefont {C.}~\bibnamefont {Warner}}, \bibinfo
  {author} {\bibfnamefont {W.}~\bibnamefont {Yuan}}, \bibinfo {author}
  {\bibfnamefont {S.}~\bibnamefont {Zhang}}, \bibinfo {author} {\bibfnamefont
  {E.}~\bibnamefont {Tiemann}}, \bibinfo {author} {\bibfnamefont
  {I.}~\bibnamefont {Stevenson}},\ and\ \bibinfo {author} {\bibfnamefont
  {S.}~\bibnamefont {Will}},\ }\href
  {https://doi.org/10.1103/PhysRevResearch.4.L022019} {\bibfield  {journal}
  {\bibinfo  {journal} {Phys. Rev. Res.}\ }\textbf {\bibinfo {volume} {4}},\
  \bibinfo {pages} {L022019} (\bibinfo {year} {2022})}\BibitemShut {NoStop}%
\bibitem [{\citenamefont {Guo}\ \emph {et~al.}(2022)\citenamefont {Guo},
  \citenamefont {Jia}, \citenamefont {Zhu}, \citenamefont {Li}, \citenamefont
  {Hutson},\ and\ \citenamefont {Wang}}]{Guo2022}%
  \BibitemOpen
  \bibfield  {author} {\bibinfo {author} {\bibfnamefont {Z.}~\bibnamefont
  {Guo}}, \bibinfo {author} {\bibfnamefont {F.}~\bibnamefont {Jia}}, \bibinfo
  {author} {\bibfnamefont {B.}~\bibnamefont {Zhu}}, \bibinfo {author}
  {\bibfnamefont {L.}~\bibnamefont {Li}}, \bibinfo {author} {\bibfnamefont
  {J.~M.}\ \bibnamefont {Hutson}},\ and\ \bibinfo {author} {\bibfnamefont
  {D.}~\bibnamefont {Wang}},\ }\href
  {https://doi.org/10.1103/PhysRevA.105.023313} {\bibfield  {journal} {\bibinfo
   {journal} {Phys. Rev. A}\ }\textbf {\bibinfo {volume} {105}},\ \bibinfo
  {pages} {023313} (\bibinfo {year} {2022})}\BibitemShut {NoStop}%
\bibitem [{\citenamefont {Jin}\ \emph {et~al.}(2025)\citenamefont {Jin},
  \citenamefont {Deng}, \citenamefont {Yi},\ and\ \citenamefont
  {Shi}}]{jin2024boseeinstein}%
  \BibitemOpen
  \bibfield  {author} {\bibinfo {author} {\bibfnamefont {W.-J.}\ \bibnamefont
  {Jin}}, \bibinfo {author} {\bibfnamefont {F.}~\bibnamefont {Deng}}, \bibinfo
  {author} {\bibfnamefont {S.}~\bibnamefont {Yi}},\ and\ \bibinfo {author}
  {\bibfnamefont {T.}~\bibnamefont {Shi}},\ }\href
  {https://doi.org/10.1103/b8y9-yvz9} {\bibfield  {journal} {\bibinfo
  {journal} {Phys. Rev. Lett.}\ }\textbf {\bibinfo {volume} {134}},\ \bibinfo
  {pages} {233003} (\bibinfo {year} {2025})}\BibitemShut {NoStop}%
\bibitem [{\citenamefont {Takekoshi}\ \emph {et~al.}(2014)\citenamefont
  {Takekoshi}, \citenamefont {Reichs{\"o}llner}, \citenamefont {Schindewolf},
  \citenamefont {Hutson}, \citenamefont {Le~Sueur}, \citenamefont {Dulieu},
  \citenamefont {Ferlaino}, \citenamefont {Grimm},\ and\ \citenamefont
  {N{\"a}gerl}}]{takekoshi2014ultracold}%
  \BibitemOpen
  \bibfield  {author} {\bibinfo {author} {\bibfnamefont {T.}~\bibnamefont
  {Takekoshi}}, \bibinfo {author} {\bibfnamefont {L.}~\bibnamefont
  {Reichs{\"o}llner}}, \bibinfo {author} {\bibfnamefont {A.}~\bibnamefont
  {Schindewolf}}, \bibinfo {author} {\bibfnamefont {J.~M.}\ \bibnamefont
  {Hutson}}, \bibinfo {author} {\bibfnamefont {C.~R.}\ \bibnamefont
  {Le~Sueur}}, \bibinfo {author} {\bibfnamefont {O.}~\bibnamefont {Dulieu}},
  \bibinfo {author} {\bibfnamefont {F.}~\bibnamefont {Ferlaino}}, \bibinfo
  {author} {\bibfnamefont {R.}~\bibnamefont {Grimm}},\ and\ \bibinfo {author}
  {\bibfnamefont {H.-C.}\ \bibnamefont {N{\"a}gerl}},\ }\href
  {https://doi.org/10.1103/PhysRevLett.113.205301} {\bibfield  {journal}
  {\bibinfo  {journal} {Physical review letters}\ }\textbf {\bibinfo {volume}
  {113}},\ \bibinfo {pages} {205301} (\bibinfo {year} {2014})}\BibitemShut
  {NoStop}%
\bibitem [{\citenamefont {Guo}\ \emph {et~al.}(2016)\citenamefont {Guo},
  \citenamefont {Zhu}, \citenamefont {Lu}, \citenamefont {Ye}, \citenamefont
  {Wang}, \citenamefont {Vexiau}, \citenamefont {Bouloufa-Maafa}, \citenamefont
  {Qu{\'e}m{\'e}ner}, \citenamefont {Dulieu},\ and\ \citenamefont
  {Wang}}]{guo2016creation}%
  \BibitemOpen
  \bibfield  {author} {\bibinfo {author} {\bibfnamefont {M.}~\bibnamefont
  {Guo}}, \bibinfo {author} {\bibfnamefont {B.}~\bibnamefont {Zhu}}, \bibinfo
  {author} {\bibfnamefont {B.}~\bibnamefont {Lu}}, \bibinfo {author}
  {\bibfnamefont {X.}~\bibnamefont {Ye}}, \bibinfo {author} {\bibfnamefont
  {F.}~\bibnamefont {Wang}}, \bibinfo {author} {\bibfnamefont {R.}~\bibnamefont
  {Vexiau}}, \bibinfo {author} {\bibfnamefont {N.}~\bibnamefont
  {Bouloufa-Maafa}}, \bibinfo {author} {\bibfnamefont {G.}~\bibnamefont
  {Qu{\'e}m{\'e}ner}}, \bibinfo {author} {\bibfnamefont {O.}~\bibnamefont
  {Dulieu}},\ and\ \bibinfo {author} {\bibfnamefont {D.}~\bibnamefont {Wang}},\
  }\href {https://doi.org/10.1103/PhysRevLett.116.205303} {\bibfield  {journal}
  {\bibinfo  {journal} {Physical review letters}\ }\textbf {\bibinfo {volume}
  {116}},\ \bibinfo {pages} {205303} (\bibinfo {year} {2016})}\BibitemShut
  {NoStop}%
\bibitem [{\citenamefont {Guo}\ \emph {et~al.}(2018)\citenamefont {Guo},
  \citenamefont {Ye}, \citenamefont {He}, \citenamefont
  {Gonz{\'a}lez-Mart{\'\i}nez}, \citenamefont {Vexiau}, \citenamefont
  {Qu{\'e}m{\'e}ner},\ and\ \citenamefont {Wang}}]{guo2018dipolar}%
  \BibitemOpen
  \bibfield  {author} {\bibinfo {author} {\bibfnamefont {M.}~\bibnamefont
  {Guo}}, \bibinfo {author} {\bibfnamefont {X.}~\bibnamefont {Ye}}, \bibinfo
  {author} {\bibfnamefont {J.}~\bibnamefont {He}}, \bibinfo {author}
  {\bibfnamefont {M.~L.}\ \bibnamefont {Gonz{\'a}lez-Mart{\'\i}nez}}, \bibinfo
  {author} {\bibfnamefont {R.}~\bibnamefont {Vexiau}}, \bibinfo {author}
  {\bibfnamefont {G.}~\bibnamefont {Qu{\'e}m{\'e}ner}},\ and\ \bibinfo {author}
  {\bibfnamefont {D.}~\bibnamefont {Wang}},\ }\href
  {https://doi.org/10.1103/PhysRevX.8.041044} {\bibfield  {journal} {\bibinfo
  {journal} {Physical Review X}\ }\textbf {\bibinfo {volume} {8}},\ \bibinfo
  {pages} {041044} (\bibinfo {year} {2018})}\BibitemShut {NoStop}%
\bibitem [{\citenamefont {Stevenson}\ \emph {et~al.}(2023)\citenamefont
  {Stevenson}, \citenamefont {Lam}, \citenamefont {Bigagli}, \citenamefont
  {Warner}, \citenamefont {Yuan}, \citenamefont {Zhang},\ and\ \citenamefont
  {Will}}]{stevenson2023ultracold}%
  \BibitemOpen
  \bibfield  {author} {\bibinfo {author} {\bibfnamefont {I.}~\bibnamefont
  {Stevenson}}, \bibinfo {author} {\bibfnamefont {A.~Z.}\ \bibnamefont {Lam}},
  \bibinfo {author} {\bibfnamefont {N.}~\bibnamefont {Bigagli}}, \bibinfo
  {author} {\bibfnamefont {C.}~\bibnamefont {Warner}}, \bibinfo {author}
  {\bibfnamefont {W.}~\bibnamefont {Yuan}}, \bibinfo {author} {\bibfnamefont
  {S.}~\bibnamefont {Zhang}},\ and\ \bibinfo {author} {\bibfnamefont
  {S.}~\bibnamefont {Will}},\ }\href
  {https://doi.org/10.1103/PhysRevLett.130.113002} {\bibfield  {journal}
  {\bibinfo  {journal} {Physical Review Letters}\ }\textbf {\bibinfo {volume}
  {130}},\ \bibinfo {pages} {113002} (\bibinfo {year} {2023})}\BibitemShut
  {NoStop}%
\bibitem [{\citenamefont {Anderegg}\ \emph {et~al.}(2021)\citenamefont
  {Anderegg}, \citenamefont {Burchesky}, \citenamefont {Bao}, \citenamefont
  {Yu}, \citenamefont {Karman}, \citenamefont {Chae}, \citenamefont {Ni},
  \citenamefont {Ketterle},\ and\ \citenamefont
  {Doyle}}]{anderegg2021observation}%
  \BibitemOpen
  \bibfield  {author} {\bibinfo {author} {\bibfnamefont {L.}~\bibnamefont
  {Anderegg}}, \bibinfo {author} {\bibfnamefont {S.}~\bibnamefont {Burchesky}},
  \bibinfo {author} {\bibfnamefont {Y.}~\bibnamefont {Bao}}, \bibinfo {author}
  {\bibfnamefont {S.~S.}\ \bibnamefont {Yu}}, \bibinfo {author} {\bibfnamefont
  {T.}~\bibnamefont {Karman}}, \bibinfo {author} {\bibfnamefont
  {E.}~\bibnamefont {Chae}}, \bibinfo {author} {\bibfnamefont {K.-K.}\
  \bibnamefont {Ni}}, \bibinfo {author} {\bibfnamefont {W.}~\bibnamefont
  {Ketterle}},\ and\ \bibinfo {author} {\bibfnamefont {J.~M.}\ \bibnamefont
  {Doyle}},\ }\href@noop {} {\bibfield  {journal} {\bibinfo  {journal}
  {Science}\ }\textbf {\bibinfo {volume} {373}},\ \bibinfo {pages} {779}
  (\bibinfo {year} {2021})}\BibitemShut {NoStop}%
\bibitem [{\citenamefont {Schindewolf}\ \emph {et~al.}(2022)\citenamefont
  {Schindewolf}, \citenamefont {Bause}, \citenamefont {Chen}, \citenamefont
  {Duda}, \citenamefont {Karman}, \citenamefont {Bloch},\ and\ \citenamefont
  {Luo}}]{Schindewolf2022}%
  \BibitemOpen
  \bibfield  {author} {\bibinfo {author} {\bibfnamefont {A.}~\bibnamefont
  {Schindewolf}}, \bibinfo {author} {\bibfnamefont {R.}~\bibnamefont {Bause}},
  \bibinfo {author} {\bibfnamefont {X.-Y.}\ \bibnamefont {Chen}}, \bibinfo
  {author} {\bibfnamefont {M.}~\bibnamefont {Duda}}, \bibinfo {author}
  {\bibfnamefont {T.}~\bibnamefont {Karman}}, \bibinfo {author} {\bibfnamefont
  {I.}~\bibnamefont {Bloch}},\ and\ \bibinfo {author} {\bibfnamefont {X.-Y.}\
  \bibnamefont {Luo}},\ }\href {https://doi.org/10.1038/s41586-022-04900-0}
  {\bibfield  {journal} {\bibinfo  {journal} {Nature}\ }\textbf {\bibinfo
  {volume} {607}},\ \bibinfo {pages} {677} (\bibinfo {year}
  {2022})}\BibitemShut {NoStop}%
\bibitem [{\citenamefont {Tomita}\ \emph {et~al.}(2017)\citenamefont {Tomita},
  \citenamefont {Nakajima}, \citenamefont {Danshita}, \citenamefont {Takasu},\
  and\ \citenamefont {Takahashi}}]{doi:10.1126/sciadv.1701513}%
  \BibitemOpen
  \bibfield  {author} {\bibinfo {author} {\bibfnamefont {T.}~\bibnamefont
  {Tomita}}, \bibinfo {author} {\bibfnamefont {S.}~\bibnamefont {Nakajima}},
  \bibinfo {author} {\bibfnamefont {I.}~\bibnamefont {Danshita}}, \bibinfo
  {author} {\bibfnamefont {Y.}~\bibnamefont {Takasu}},\ and\ \bibinfo {author}
  {\bibfnamefont {Y.}~\bibnamefont {Takahashi}},\ }\href
  {https://doi.org/10.1126/sciadv.1701513} {\bibfield  {journal} {\bibinfo
  {journal} {Science Advances}\ }\textbf {\bibinfo {volume} {3}},\ \bibinfo
  {pages} {e1701513} (\bibinfo {year} {2017})}\BibitemShut {NoStop}%
\bibitem [{\citenamefont {Diehl}\ \emph {et~al.}(2008)\citenamefont {Diehl},
  \citenamefont {Micheli}, \citenamefont {Kantian}, \citenamefont {Kraus},
  \citenamefont {B{\"u}chler},\ and\ \citenamefont {Zoller}}]{Diehl2008}%
  \BibitemOpen
  \bibfield  {author} {\bibinfo {author} {\bibfnamefont {S.}~\bibnamefont
  {Diehl}}, \bibinfo {author} {\bibfnamefont {A.}~\bibnamefont {Micheli}},
  \bibinfo {author} {\bibfnamefont {A.}~\bibnamefont {Kantian}}, \bibinfo
  {author} {\bibfnamefont {B.}~\bibnamefont {Kraus}}, \bibinfo {author}
  {\bibfnamefont {H.~P.}\ \bibnamefont {B{\"u}chler}},\ and\ \bibinfo {author}
  {\bibfnamefont {P.}~\bibnamefont {Zoller}},\ }\href
  {https://doi.org/10.1038/nphys1073} {\bibfield  {journal} {\bibinfo
  {journal} {Nature Physics}\ }\textbf {\bibinfo {volume} {4}},\ \bibinfo
  {pages} {878} (\bibinfo {year} {2008})}\BibitemShut {NoStop}%
\bibitem [{\citenamefont {Tomita}\ \emph {et~al.}(2019)\citenamefont {Tomita},
  \citenamefont {Nakajima}, \citenamefont {Takasu},\ and\ \citenamefont
  {Takahashi}}]{Tomita_2019}%
  \BibitemOpen
  \bibfield  {author} {\bibinfo {author} {\bibfnamefont {T.}~\bibnamefont
  {Tomita}}, \bibinfo {author} {\bibfnamefont {S.}~\bibnamefont {Nakajima}},
  \bibinfo {author} {\bibfnamefont {Y.}~\bibnamefont {Takasu}},\ and\ \bibinfo
  {author} {\bibfnamefont {Y.}~\bibnamefont {Takahashi}},\ }\href
  {https://doi.org/10.1103/PhysRevA.99.031601} {\bibfield  {journal} {\bibinfo
  {journal} {Phys. Rev. A}\ }\textbf {\bibinfo {volume} {99}},\ \bibinfo
  {pages} {031601} (\bibinfo {year} {2019})}\BibitemShut {NoStop}%
\bibitem [{\citenamefont {Keeling}(2011)}]{Keeling2011}%
  \BibitemOpen
  \bibfield  {author} {\bibinfo {author} {\bibfnamefont {J.}~\bibnamefont
  {Keeling}},\ }\href {https://doi.org/10.1103/PhysRevLett.107.080402}
  {\bibfield  {journal} {\bibinfo  {journal} {Phys. Rev. Lett.}\ }\textbf
  {\bibinfo {volume} {107}},\ \bibinfo {pages} {080402} (\bibinfo {year}
  {2011})}\BibitemShut {NoStop}%
\bibitem [{\citenamefont {Dogra}\ \emph {et~al.}(2019)\citenamefont {Dogra},
  \citenamefont {Landini}, \citenamefont {Kroeger}, \citenamefont {Hruby},
  \citenamefont {Donner},\ and\ \citenamefont
  {Esslinger}}]{dogra2019dissipation}%
  \BibitemOpen
  \bibfield  {author} {\bibinfo {author} {\bibfnamefont {N.}~\bibnamefont
  {Dogra}}, \bibinfo {author} {\bibfnamefont {M.}~\bibnamefont {Landini}},
  \bibinfo {author} {\bibfnamefont {K.}~\bibnamefont {Kroeger}}, \bibinfo
  {author} {\bibfnamefont {L.}~\bibnamefont {Hruby}}, \bibinfo {author}
  {\bibfnamefont {T.}~\bibnamefont {Donner}},\ and\ \bibinfo {author}
  {\bibfnamefont {T.}~\bibnamefont {Esslinger}},\ }\href
  {https://doi.org/10.1126/science.aaw4465} {\bibfield  {journal} {\bibinfo
  {journal} {Science}\ }\textbf {\bibinfo {volume} {366}},\ \bibinfo {pages}
  {1496} (\bibinfo {year} {2019})}\BibitemShut {NoStop}%
\bibitem [{\citenamefont {Yamamoto}\ \emph {et~al.}(2021)\citenamefont
  {Yamamoto}, \citenamefont {Nakagawa}, \citenamefont {Tsuji}, \citenamefont
  {Ueda},\ and\ \citenamefont {Kawakami}}]{Yamamoto2021}%
  \BibitemOpen
  \bibfield  {author} {\bibinfo {author} {\bibfnamefont {K.}~\bibnamefont
  {Yamamoto}}, \bibinfo {author} {\bibfnamefont {M.}~\bibnamefont {Nakagawa}},
  \bibinfo {author} {\bibfnamefont {N.}~\bibnamefont {Tsuji}}, \bibinfo
  {author} {\bibfnamefont {M.}~\bibnamefont {Ueda}},\ and\ \bibinfo {author}
  {\bibfnamefont {N.}~\bibnamefont {Kawakami}},\ }\href
  {https://doi.org/10.1103/PhysRevLett.127.055301} {\bibfield  {journal}
  {\bibinfo  {journal} {Phys. Rev. Lett.}\ }\textbf {\bibinfo {volume} {127}},\
  \bibinfo {pages} {055301} (\bibinfo {year} {2021})}\BibitemShut {NoStop}%
\bibitem [{\citenamefont {Mazza}\ and\ \citenamefont
  {Schir\`o}(2023)}]{Mazza2023}%
  \BibitemOpen
  \bibfield  {author} {\bibinfo {author} {\bibfnamefont {G.}~\bibnamefont
  {Mazza}}\ and\ \bibinfo {author} {\bibfnamefont {M.}~\bibnamefont
  {Schir\`o}},\ }\href {https://doi.org/10.1103/PhysRevA.107.L051301}
  {\bibfield  {journal} {\bibinfo  {journal} {Phys. Rev. A}\ }\textbf {\bibinfo
  {volume} {107}},\ \bibinfo {pages} {L051301} (\bibinfo {year}
  {2023})}\BibitemShut {NoStop}%
\bibitem [{\citenamefont {Bu{\v{c}}a}\ and\ \citenamefont
  {Prosen}(2012)}]{buvca2012note}%
  \BibitemOpen
  \bibfield  {author} {\bibinfo {author} {\bibfnamefont {B.}~\bibnamefont
  {Bu{\v{c}}a}}\ and\ \bibinfo {author} {\bibfnamefont {T.}~\bibnamefont
  {Prosen}},\ }\href {https://doi.org/10.1088/1367-2630/14/7/073007} {\bibfield
   {journal} {\bibinfo  {journal} {New Journal of Physics}\ }\textbf {\bibinfo
  {volume} {14}},\ \bibinfo {pages} {073007} (\bibinfo {year}
  {2012})}\BibitemShut {NoStop}%
\bibitem [{\citenamefont {Liu}\ \emph {et~al.}(2024)\citenamefont {Liu},
  \citenamefont {Shi},\ and\ \citenamefont {Wang}}]{liu2022weakly}%
  \BibitemOpen
  \bibfield  {author} {\bibinfo {author} {\bibfnamefont {C.}~\bibnamefont
  {Liu}}, \bibinfo {author} {\bibfnamefont {Z.}~\bibnamefont {Shi}},\ and\
  \bibinfo {author} {\bibfnamefont {C.}~\bibnamefont {Wang}},\ }\href
  {https://doi.org/10.21468/SciPostPhys.16.5.116} {\bibfield  {journal}
  {\bibinfo  {journal} {SciPost Physics}\ }\textbf {\bibinfo {volume} {16}},\
  \bibinfo {pages} {116} (\bibinfo {year} {2024})}\BibitemShut {NoStop}%
\bibitem [{\citenamefont {Lahaye}\ \emph {et~al.}(2009)\citenamefont {Lahaye},
  \citenamefont {Menotti}, \citenamefont {Santos}, \citenamefont {Lewenstein},\
  and\ \citenamefont {Pfau}}]{Lahaye_2009}%
  \BibitemOpen
  \bibfield  {author} {\bibinfo {author} {\bibfnamefont {T.}~\bibnamefont
  {Lahaye}}, \bibinfo {author} {\bibfnamefont {C.}~\bibnamefont {Menotti}},
  \bibinfo {author} {\bibfnamefont {L.}~\bibnamefont {Santos}}, \bibinfo
  {author} {\bibfnamefont {M.}~\bibnamefont {Lewenstein}},\ and\ \bibinfo
  {author} {\bibfnamefont {T.}~\bibnamefont {Pfau}},\ }\href
  {https://doi.org/10.1088/0034-4885/72/12/126401} {\bibfield  {journal}
  {\bibinfo  {journal} {Reports on Progress in Physics}\ }\textbf {\bibinfo
  {volume} {72}},\ \bibinfo {pages} {126401} (\bibinfo {year}
  {2009})}\BibitemShut {NoStop}%
\bibitem [{\citenamefont {Chomaz}\ \emph {et~al.}(2022)\citenamefont {Chomaz},
  \citenamefont {Ferrier-Barbut}, \citenamefont {Ferlaino}, \citenamefont
  {Laburthe-Tolra}, \citenamefont {Lev},\ and\ \citenamefont
  {Pfau}}]{Chomaz_2023}%
  \BibitemOpen
  \bibfield  {author} {\bibinfo {author} {\bibfnamefont {L.}~\bibnamefont
  {Chomaz}}, \bibinfo {author} {\bibfnamefont {I.}~\bibnamefont
  {Ferrier-Barbut}}, \bibinfo {author} {\bibfnamefont {F.}~\bibnamefont
  {Ferlaino}}, \bibinfo {author} {\bibfnamefont {B.}~\bibnamefont
  {Laburthe-Tolra}}, \bibinfo {author} {\bibfnamefont {B.~L.}\ \bibnamefont
  {Lev}},\ and\ \bibinfo {author} {\bibfnamefont {T.}~\bibnamefont {Pfau}},\
  }\href {https://doi.org/10.1088/1361-6633/aca814} {\bibfield  {journal}
  {\bibinfo  {journal} {Reports on Progress in Physics}\ }\textbf {\bibinfo
  {volume} {86}},\ \bibinfo {pages} {026401} (\bibinfo {year}
  {2022})}\BibitemShut {NoStop}%
\bibitem [{\citenamefont {Breuer}\ and\ \citenamefont
  {Petruccione}(2007)}]{10.1093/acprof:oso/9780199213900.001.0001}%
  \BibitemOpen
  \bibfield  {author} {\bibinfo {author} {\bibfnamefont {H.-P.}\ \bibnamefont
  {Breuer}}\ and\ \bibinfo {author} {\bibfnamefont {F.}~\bibnamefont
  {Petruccione}},\ }\href
  {https://doi.org/10.1093/acprof:oso/9780199213900.001.0001} {\emph {\bibinfo
  {title} {{The Theory of Open Quantum Systems}}}}\ (\bibinfo  {publisher}
  {Oxford University Press, Oxford, UK},\ \bibinfo {year} {2007})\BibitemShut
  {NoStop}%
\bibitem [{\citenamefont {Sieberer}\ \emph {et~al.}(2016)\citenamefont
  {Sieberer}, \citenamefont {Buchhold},\ and\ \citenamefont
  {Diehl}}]{Sieberer_2016}%
  \BibitemOpen
  \bibfield  {author} {\bibinfo {author} {\bibfnamefont {L.~M.}\ \bibnamefont
  {Sieberer}}, \bibinfo {author} {\bibfnamefont {M.}~\bibnamefont {Buchhold}},\
  and\ \bibinfo {author} {\bibfnamefont {S.}~\bibnamefont {Diehl}},\ }\href
  {https://doi.org/10.1088/0034-4885/79/9/096001} {\bibfield  {journal}
  {\bibinfo  {journal} {Reports on Progress in Physics}\ }\textbf {\bibinfo
  {volume} {79}},\ \bibinfo {pages} {096001} (\bibinfo {year}
  {2016})}\BibitemShut {NoStop}%
\bibitem [{\citenamefont {Albert}\ and\ \citenamefont
  {Jiang}(2014)}]{Albert2014}%
  \BibitemOpen
  \bibfield  {author} {\bibinfo {author} {\bibfnamefont {V.~V.}\ \bibnamefont
  {Albert}}\ and\ \bibinfo {author} {\bibfnamefont {L.}~\bibnamefont {Jiang}},\
  }\href {https://doi.org/10.1103/PhysRevA.89.022118} {\bibfield  {journal}
  {\bibinfo  {journal} {Phys. Rev. A}\ }\textbf {\bibinfo {volume} {89}},\
  \bibinfo {pages} {022118} (\bibinfo {year} {2014})}\BibitemShut {NoStop}%
\bibitem [{\citenamefont {Wen}(2007)}]{Wen2007}%
  \BibitemOpen
  \bibfield  {author} {\bibinfo {author} {\bibfnamefont {X.-G.}\ \bibnamefont
  {Wen}},\ }\href {https://doi.org/10.1093/acprof:oso/9780199227259.001.0001}
  {\emph {\bibinfo {title} {{Quantum Field Theory of Many-Body Systems: From
  the Origin of Sound to an Origin of Light and Electrons}}}}\ (\bibinfo
  {publisher} {Oxford University Press, Oxford, UK},\ \bibinfo {year}
  {2007})\BibitemShut {NoStop}%
\bibitem [{\citenamefont {Hongo}\ \emph {et~al.}(2021)\citenamefont {Hongo},
  \citenamefont {Kim}, \citenamefont {Noumi},\ and\ \citenamefont
  {Ota}}]{PhysRevD.103.056020}%
  \BibitemOpen
  \bibfield  {author} {\bibinfo {author} {\bibfnamefont {M.}~\bibnamefont
  {Hongo}}, \bibinfo {author} {\bibfnamefont {S.}~\bibnamefont {Kim}}, \bibinfo
  {author} {\bibfnamefont {T.}~\bibnamefont {Noumi}},\ and\ \bibinfo {author}
  {\bibfnamefont {A.}~\bibnamefont {Ota}},\ }\href
  {https://doi.org/10.1103/PhysRevD.103.056020} {\bibfield  {journal} {\bibinfo
   {journal} {Phys. Rev. D}\ }\textbf {\bibinfo {volume} {103}},\ \bibinfo
  {pages} {056020} (\bibinfo {year} {2021})}\BibitemShut {NoStop}%
\bibitem [{Sup()}]{SupplementaryMaterial}%
  \BibitemOpen
  \href@noop {} {\bibinfo {title} {See supplemental material for
  details.}}\BibitemShut {Stop}%
\bibitem [{\citenamefont {Gebauer}\ and\ \citenamefont
  {Car}(2004)}]{PhysRevLett.93.160404}%
  \BibitemOpen
  \bibfield  {author} {\bibinfo {author} {\bibfnamefont {R.}~\bibnamefont
  {Gebauer}}\ and\ \bibinfo {author} {\bibfnamefont {R.}~\bibnamefont {Car}},\
  }\href {https://doi.org/10.1103/PhysRevLett.93.160404} {\bibfield  {journal}
  {\bibinfo  {journal} {Phys. Rev. Lett.}\ }\textbf {\bibinfo {volume} {93}},\
  \bibinfo {pages} {160404} (\bibinfo {year} {2004})}\BibitemShut {NoStop}%
\bibitem [{\citenamefont {Wang}\ \emph {et~al.}(2022)\citenamefont {Wang},
  \citenamefont {Liu},\ and\ \citenamefont {Shi}}]{Ce2022}%
  \BibitemOpen
  \bibfield  {author} {\bibinfo {author} {\bibfnamefont {C.}~\bibnamefont
  {Wang}}, \bibinfo {author} {\bibfnamefont {C.}~\bibnamefont {Liu}},\ and\
  \bibinfo {author} {\bibfnamefont {Z.-Y.}\ \bibnamefont {Shi}},\ }\href
  {https://doi.org/10.1103/PhysRevLett.129.203401} {\bibfield  {journal}
  {\bibinfo  {journal} {Phys. Rev. Lett.}\ }\textbf {\bibinfo {volume} {129}},\
  \bibinfo {pages} {203401} (\bibinfo {year} {2022})}\BibitemShut {NoStop}%
\bibitem [{\citenamefont {Bloch}\ \emph {et~al.}(2008)\citenamefont {Bloch},
  \citenamefont {Dalibard},\ and\ \citenamefont {Zwerger}}]{RevModPhys.80.885}%
  \BibitemOpen
  \bibfield  {author} {\bibinfo {author} {\bibfnamefont {I.}~\bibnamefont
  {Bloch}}, \bibinfo {author} {\bibfnamefont {J.}~\bibnamefont {Dalibard}},\
  and\ \bibinfo {author} {\bibfnamefont {W.}~\bibnamefont {Zwerger}},\ }\href
  {https://doi.org/10.1103/RevModPhys.80.885} {\bibfield  {journal} {\bibinfo
  {journal} {Rev. Mod. Phys.}\ }\textbf {\bibinfo {volume} {80}},\ \bibinfo
  {pages} {885} (\bibinfo {year} {2008})}\BibitemShut {NoStop}%
\bibitem [{\citenamefont {Syassen}\ \emph {et~al.}(2008)\citenamefont
  {Syassen}, \citenamefont {Bauer}, \citenamefont {Lettner}, \citenamefont
  {Volz}, \citenamefont {Dietze}, \citenamefont {García-Ripoll}, \citenamefont
  {Cirac}, \citenamefont {Rempe},\ and\ \citenamefont {Dürr}}]{Syassen2008}%
  \BibitemOpen
  \bibfield  {author} {\bibinfo {author} {\bibfnamefont {N.}~\bibnamefont
  {Syassen}}, \bibinfo {author} {\bibfnamefont {D.~M.}\ \bibnamefont {Bauer}},
  \bibinfo {author} {\bibfnamefont {M.}~\bibnamefont {Lettner}}, \bibinfo
  {author} {\bibfnamefont {T.}~\bibnamefont {Volz}}, \bibinfo {author}
  {\bibfnamefont {D.}~\bibnamefont {Dietze}}, \bibinfo {author} {\bibfnamefont
  {J.~J.}\ \bibnamefont {García-Ripoll}}, \bibinfo {author} {\bibfnamefont
  {J.~I.}\ \bibnamefont {Cirac}}, \bibinfo {author} {\bibfnamefont
  {G.}~\bibnamefont {Rempe}},\ and\ \bibinfo {author} {\bibfnamefont
  {S.}~\bibnamefont {Dürr}},\ }\href {https://doi.org/10.1126/science.1155309}
  {\bibfield  {journal} {\bibinfo  {journal} {Science}\ }\textbf {\bibinfo
  {volume} {320}},\ \bibinfo {pages} {1329} (\bibinfo {year}
  {2008})}\BibitemShut {NoStop}%
\bibitem [{\citenamefont {García-Ripoll}\ \emph {et~al.}(2009)\citenamefont
  {García-Ripoll}, \citenamefont {Dürr}, \citenamefont {Syassen},
  \citenamefont {Bauer}, \citenamefont {Lettner}, \citenamefont {Rempe},\ and\
  \citenamefont {Cirac}}]{Garcia-Ripoll_2009}%
  \BibitemOpen
  \bibfield  {author} {\bibinfo {author} {\bibfnamefont {J.~J.}\ \bibnamefont
  {García-Ripoll}}, \bibinfo {author} {\bibfnamefont {S.}~\bibnamefont
  {Dürr}}, \bibinfo {author} {\bibfnamefont {N.}~\bibnamefont {Syassen}},
  \bibinfo {author} {\bibfnamefont {D.~M.}\ \bibnamefont {Bauer}}, \bibinfo
  {author} {\bibfnamefont {M.}~\bibnamefont {Lettner}}, \bibinfo {author}
  {\bibfnamefont {G.}~\bibnamefont {Rempe}},\ and\ \bibinfo {author}
  {\bibfnamefont {J.~I.}\ \bibnamefont {Cirac}},\ }\href
  {https://doi.org/10.1088/1367-2630/11/1/013053} {\bibfield  {journal}
  {\bibinfo  {journal} {New Journal of Physics}\ }\textbf {\bibinfo {volume}
  {11}},\ \bibinfo {pages} {013053} (\bibinfo {year} {2009})}\BibitemShut
  {NoStop}%
\bibitem [{\citenamefont {Yamamoto}\ \emph {et~al.}(2019)\citenamefont
  {Yamamoto}, \citenamefont {Nakagawa}, \citenamefont {Adachi}, \citenamefont
  {Takasan}, \citenamefont {Ueda},\ and\ \citenamefont
  {Kawakami}}]{Yamamoto2019}%
  \BibitemOpen
  \bibfield  {author} {\bibinfo {author} {\bibfnamefont {K.}~\bibnamefont
  {Yamamoto}}, \bibinfo {author} {\bibfnamefont {M.}~\bibnamefont {Nakagawa}},
  \bibinfo {author} {\bibfnamefont {K.}~\bibnamefont {Adachi}}, \bibinfo
  {author} {\bibfnamefont {K.}~\bibnamefont {Takasan}}, \bibinfo {author}
  {\bibfnamefont {M.}~\bibnamefont {Ueda}},\ and\ \bibinfo {author}
  {\bibfnamefont {N.}~\bibnamefont {Kawakami}},\ }\href
  {https://doi.org/10.1103/PhysRevLett.123.123601} {\bibfield  {journal}
  {\bibinfo  {journal} {Phys. Rev. Lett.}\ }\textbf {\bibinfo {volume} {123}},\
  \bibinfo {pages} {123601} (\bibinfo {year} {2019})}\BibitemShut {NoStop}%
\bibitem [{\citenamefont {Li}\ \emph {et~al.}(2023)\citenamefont {Li},
  \citenamefont {Yu}, \citenamefont {Nakagawa},\ and\ \citenamefont
  {Ueda}}]{Hongchao2023}%
  \BibitemOpen
  \bibfield  {author} {\bibinfo {author} {\bibfnamefont {H.}~\bibnamefont
  {Li}}, \bibinfo {author} {\bibfnamefont {X.-H.}\ \bibnamefont {Yu}}, \bibinfo
  {author} {\bibfnamefont {M.}~\bibnamefont {Nakagawa}},\ and\ \bibinfo
  {author} {\bibfnamefont {M.}~\bibnamefont {Ueda}},\ }\href
  {https://doi.org/10.1103/PhysRevLett.131.216001} {\bibfield  {journal}
  {\bibinfo  {journal} {Phys. Rev. Lett.}\ }\textbf {\bibinfo {volume} {131}},\
  \bibinfo {pages} {216001} (\bibinfo {year} {2023})}\BibitemShut {NoStop}%
\bibitem [{Note1()}]{Note1}%
  \BibitemOpen
  \bibinfo {note} {According to the experimental data in Ref. \cite
  {Sebastian2023}, we have $\gamma n(0)T\sim 10^{-4}\ll 1$ for the observation
  time of the milisecond order, justifying the approximation we
  make.}\BibitemShut {Stop}%
\bibitem [{\citenamefont {D\"urr}\ \emph {et~al.}(2009)\citenamefont {D\"urr},
  \citenamefont {Garc\'{\i}a-Ripoll}, \citenamefont {Syassen}, \citenamefont
  {Bauer}, \citenamefont {Lettner}, \citenamefont {Cirac},\ and\ \citenamefont
  {Rempe}}]{PhysRevA.79.023614}%
  \BibitemOpen
  \bibfield  {author} {\bibinfo {author} {\bibfnamefont {S.}~\bibnamefont
  {D\"urr}}, \bibinfo {author} {\bibfnamefont {J.~J.}\ \bibnamefont
  {Garc\'{\i}a-Ripoll}}, \bibinfo {author} {\bibfnamefont {N.}~\bibnamefont
  {Syassen}}, \bibinfo {author} {\bibfnamefont {D.~M.}\ \bibnamefont {Bauer}},
  \bibinfo {author} {\bibfnamefont {M.}~\bibnamefont {Lettner}}, \bibinfo
  {author} {\bibfnamefont {J.~I.}\ \bibnamefont {Cirac}},\ and\ \bibinfo
  {author} {\bibfnamefont {G.}~\bibnamefont {Rempe}},\ }\href
  {https://doi.org/10.1103/PhysRevA.79.023614} {\bibfield  {journal} {\bibinfo
  {journal} {Phys. Rev. A}\ }\textbf {\bibinfo {volume} {79}},\ \bibinfo
  {pages} {023614} (\bibinfo {year} {2009})}\BibitemShut {NoStop}%
\bibitem [{\citenamefont {Iskin}(2021)}]{PhysRevA.103.013724}%
  \BibitemOpen
  \bibfield  {author} {\bibinfo {author} {\bibfnamefont {M.}~\bibnamefont
  {Iskin}},\ }\href {https://doi.org/10.1103/PhysRevA.103.013724} {\bibfield
  {journal} {\bibinfo  {journal} {Phys. Rev. A}\ }\textbf {\bibinfo {volume}
  {103}},\ \bibinfo {pages} {013724} (\bibinfo {year} {2021})}\BibitemShut
  {NoStop}%
\bibitem [{\citenamefont {Clark}\ and\ \citenamefont
  {Derrick}(1968)}]{Mathematical_method_SF_1968}%
  \BibitemOpen
  \bibfield  {author} {\bibinfo {author} {\bibfnamefont {R.~C.}\ \bibnamefont
  {Clark}}\ and\ \bibinfo {author} {\bibfnamefont {G.~H.}\ \bibnamefont
  {Derrick}},\ }\href {https://doi.org/10.1007/978-1-4899-6435-9} {\emph
  {\bibinfo {title} {Mathematical Methods in Solid State and Superfluid
  Theory}}},\ \bibinfo {edition} {1st}\ ed.,\ edited by\ \bibinfo {editor}
  {\bibfnamefont {R.~C.}\ \bibnamefont {Clark}}\ and\ \bibinfo {editor}
  {\bibfnamefont {G.~H.}\ \bibnamefont {Derrick}}\ (\bibinfo  {publisher}
  {Springer US},\ \bibinfo {year} {1968})\BibitemShut {NoStop}%
\bibitem [{\citenamefont {Huber}\ \emph {et~al.}(2014)\citenamefont {Huber},
  \citenamefont {Arif}, \citenamefont {Chen}, \citenamefont {Gentile},
  \citenamefont {Hussey}, \citenamefont {Black}, \citenamefont {Pushin},
  \citenamefont {Shahi}, \citenamefont {Wietfeldt},\ and\ \citenamefont
  {Yang}}]{PhysRevC.90.064004}%
  \BibitemOpen
  \bibfield  {author} {\bibinfo {author} {\bibfnamefont {M.~G.}\ \bibnamefont
  {Huber}}, \bibinfo {author} {\bibfnamefont {M.}~\bibnamefont {Arif}},
  \bibinfo {author} {\bibfnamefont {W.~C.}\ \bibnamefont {Chen}}, \bibinfo
  {author} {\bibfnamefont {T.~R.}\ \bibnamefont {Gentile}}, \bibinfo {author}
  {\bibfnamefont {D.~S.}\ \bibnamefont {Hussey}}, \bibinfo {author}
  {\bibfnamefont {T.~C.}\ \bibnamefont {Black}}, \bibinfo {author}
  {\bibfnamefont {D.~A.}\ \bibnamefont {Pushin}}, \bibinfo {author}
  {\bibfnamefont {C.~B.}\ \bibnamefont {Shahi}}, \bibinfo {author}
  {\bibfnamefont {F.~E.}\ \bibnamefont {Wietfeldt}},\ and\ \bibinfo {author}
  {\bibfnamefont {L.}~\bibnamefont {Yang}},\ }\href
  {https://doi.org/10.1103/PhysRevC.90.064004} {\bibfield  {journal} {\bibinfo
  {journal} {Phys. Rev. C}\ }\textbf {\bibinfo {volume} {90}},\ \bibinfo
  {pages} {064004} (\bibinfo {year} {2014})}\BibitemShut {NoStop}%
\bibitem [{\citenamefont {Mutchler}\ \emph {et~al.}(1988)\citenamefont
  {Mutchler}, \citenamefont {Clement}, \citenamefont {Kruk}, \citenamefont
  {Moss}, \citenamefont {Hungerford}, \citenamefont {Kishimoto}, \citenamefont
  {Mayes}, \citenamefont {Pinsky}, \citenamefont {Tang}, \citenamefont {Xue},
  \citenamefont {Bassalleck}, \citenamefont {Armstrong}, \citenamefont
  {Hartman}, \citenamefont {Hicks}, \citenamefont {Lewis}, \citenamefont
  {Lochstet}, \citenamefont {Smith}, \citenamefont {Lowenstein}, \citenamefont
  {Poth}, \citenamefont {von Witsch},\ and\ \citenamefont
  {Furic}}]{PhysRevD.38.742}%
  \BibitemOpen
  \bibfield  {author} {\bibinfo {author} {\bibfnamefont {G.~S.}\ \bibnamefont
  {Mutchler}}, \bibinfo {author} {\bibfnamefont {J.}~\bibnamefont {Clement}},
  \bibinfo {author} {\bibfnamefont {J.}~\bibnamefont {Kruk}}, \bibinfo {author}
  {\bibfnamefont {R.}~\bibnamefont {Moss}}, \bibinfo {author} {\bibfnamefont
  {E.}~\bibnamefont {Hungerford}}, \bibinfo {author} {\bibfnamefont
  {T.}~\bibnamefont {Kishimoto}}, \bibinfo {author} {\bibfnamefont
  {B.}~\bibnamefont {Mayes}}, \bibinfo {author} {\bibfnamefont
  {L.}~\bibnamefont {Pinsky}}, \bibinfo {author} {\bibfnamefont
  {L.}~\bibnamefont {Tang}}, \bibinfo {author} {\bibfnamefont {Y.}~\bibnamefont
  {Xue}}, \bibinfo {author} {\bibfnamefont {B.}~\bibnamefont {Bassalleck}},
  \bibinfo {author} {\bibfnamefont {T.}~\bibnamefont {Armstrong}}, \bibinfo
  {author} {\bibfnamefont {K.}~\bibnamefont {Hartman}}, \bibinfo {author}
  {\bibfnamefont {A.}~\bibnamefont {Hicks}}, \bibinfo {author} {\bibfnamefont
  {R.}~\bibnamefont {Lewis}}, \bibinfo {author} {\bibfnamefont
  {W.}~\bibnamefont {Lochstet}}, \bibinfo {author} {\bibfnamefont {G.~A.}\
  \bibnamefont {Smith}}, \bibinfo {author} {\bibfnamefont {D.}~\bibnamefont
  {Lowenstein}}, \bibinfo {author} {\bibfnamefont {H.}~\bibnamefont {Poth}},
  \bibinfo {author} {\bibfnamefont {W.}~\bibnamefont {von Witsch}},\ and\
  \bibinfo {author} {\bibfnamefont {M.}~\bibnamefont {Furic}},\ }\href
  {https://doi.org/10.1103/PhysRevD.38.742} {\bibfield  {journal} {\bibinfo
  {journal} {Phys. Rev. D}\ }\textbf {\bibinfo {volume} {38}},\ \bibinfo
  {pages} {742} (\bibinfo {year} {1988})}\BibitemShut {NoStop}%
\bibitem [{\citenamefont {Li}\ \emph {et~al.}(2008)\citenamefont {Li},
  \citenamefont {Zhang}, \citenamefont {Li}, \citenamefont {Chen},\ and\
  \citenamefont {Liu}}]{PhysRevA.78.023608}%
  \BibitemOpen
  \bibfield  {author} {\bibinfo {author} {\bibfnamefont {B.}~\bibnamefont
  {Li}}, \bibinfo {author} {\bibfnamefont {X.-F.}\ \bibnamefont {Zhang}},
  \bibinfo {author} {\bibfnamefont {Y.-Q.}\ \bibnamefont {Li}}, \bibinfo
  {author} {\bibfnamefont {Y.}~\bibnamefont {Chen}},\ and\ \bibinfo {author}
  {\bibfnamefont {W.~M.}\ \bibnamefont {Liu}},\ }\href
  {https://doi.org/10.1103/PhysRevA.78.023608} {\bibfield  {journal} {\bibinfo
  {journal} {Phys. Rev. A}\ }\textbf {\bibinfo {volume} {78}},\ \bibinfo
  {pages} {023608} (\bibinfo {year} {2008})}\BibitemShut {NoStop}%
\bibitem [{\citenamefont {Kamenev}(2011)}]{Kamenev_2011}%
  \BibitemOpen
  \bibfield  {author} {\bibinfo {author} {\bibfnamefont {A.}~\bibnamefont
  {Kamenev}},\ }\href@noop {} {\emph {\bibinfo {title} {Field Theory of
  Non-Equilibrium Systems}}}\ (\bibinfo  {publisher} {Cambridge University
  Press, Cambridge, UK},\ \bibinfo {year} {2011})\BibitemShut {NoStop}%
\bibitem [{Note2()}]{Note2}%
  \BibitemOpen
  \bibinfo {note} {Due to the nonunitary Bogoliubov transformation, the
  amplitude modes have mixed with the phase modes, leading to the difference
  between the spectrums of the NG mode and that of the bosonic
  operators.}\BibitemShut {Stop}%
\bibitem [{\citenamefont {Tsubota}\ \emph {et~al.}(2002)\citenamefont
  {Tsubota}, \citenamefont {Kasamatsu},\ and\ \citenamefont
  {Ueda}}]{Tsubota2002}%
  \BibitemOpen
  \bibfield  {author} {\bibinfo {author} {\bibfnamefont {M.}~\bibnamefont
  {Tsubota}}, \bibinfo {author} {\bibfnamefont {K.}~\bibnamefont {Kasamatsu}},\
  and\ \bibinfo {author} {\bibfnamefont {M.}~\bibnamefont {Ueda}},\ }\href
  {https://doi.org/10.1103/PhysRevA.65.023603} {\bibfield  {journal} {\bibinfo
  {journal} {Phys. Rev. A}\ }\textbf {\bibinfo {volume} {65}},\ \bibinfo
  {pages} {023603} (\bibinfo {year} {2002})}\BibitemShut {NoStop}%
\bibitem [{\citenamefont {Kasamatsu}\ \emph {et~al.}(2003)\citenamefont
  {Kasamatsu}, \citenamefont {Tsubota},\ and\ \citenamefont
  {Ueda}}]{Kasamatsu2003}%
  \BibitemOpen
  \bibfield  {author} {\bibinfo {author} {\bibfnamefont {K.}~\bibnamefont
  {Kasamatsu}}, \bibinfo {author} {\bibfnamefont {M.}~\bibnamefont {Tsubota}},\
  and\ \bibinfo {author} {\bibfnamefont {M.}~\bibnamefont {Ueda}},\ }\href
  {https://doi.org/10.1103/PhysRevA.67.033610} {\bibfield  {journal} {\bibinfo
  {journal} {Phys. Rev. A}\ }\textbf {\bibinfo {volume} {67}},\ \bibinfo
  {pages} {033610} (\bibinfo {year} {2003})}\BibitemShut {NoStop}%
\bibitem [{\citenamefont {Madison}\ \emph {et~al.}(2001)\citenamefont
  {Madison}, \citenamefont {Chevy}, \citenamefont {Bretin},\ and\ \citenamefont
  {Dalibard}}]{Madison2001}%
  \BibitemOpen
  \bibfield  {author} {\bibinfo {author} {\bibfnamefont {K.~W.}\ \bibnamefont
  {Madison}}, \bibinfo {author} {\bibfnamefont {F.}~\bibnamefont {Chevy}},
  \bibinfo {author} {\bibfnamefont {V.}~\bibnamefont {Bretin}},\ and\ \bibinfo
  {author} {\bibfnamefont {J.}~\bibnamefont {Dalibard}},\ }\href
  {https://doi.org/10.1103/PhysRevLett.86.4443} {\bibfield  {journal} {\bibinfo
   {journal} {Phys. Rev. Lett.}\ }\textbf {\bibinfo {volume} {86}},\ \bibinfo
  {pages} {4443} (\bibinfo {year} {2001})}\BibitemShut {NoStop}%
\bibitem [{\citenamefont {Cooper}\ and\ \citenamefont
  {Hadzibabic}(2010)}]{PhysRevLett.104.030401}%
  \BibitemOpen
  \bibfield  {author} {\bibinfo {author} {\bibfnamefont {N.~R.}\ \bibnamefont
  {Cooper}}\ and\ \bibinfo {author} {\bibfnamefont {Z.}~\bibnamefont
  {Hadzibabic}},\ }\href {https://doi.org/10.1103/PhysRevLett.104.030401}
  {\bibfield  {journal} {\bibinfo  {journal} {Phys. Rev. Lett.}\ }\textbf
  {\bibinfo {volume} {104}},\ \bibinfo {pages} {030401} (\bibinfo {year}
  {2010})}\BibitemShut {NoStop}%
\bibitem [{\citenamefont {Chen}\ \emph {et~al.}(2018)\citenamefont {Chen},
  \citenamefont {Liu}, \citenamefont {Tsai}, \citenamefont {Chiu},
  \citenamefont {Kawaguchi}, \citenamefont {Yip}, \citenamefont {Chang},\ and\
  \citenamefont {Lin}}]{Chen2018}%
  \BibitemOpen
  \bibfield  {author} {\bibinfo {author} {\bibfnamefont {P.-K.}\ \bibnamefont
  {Chen}}, \bibinfo {author} {\bibfnamefont {L.-R.}\ \bibnamefont {Liu}},
  \bibinfo {author} {\bibfnamefont {M.-J.}\ \bibnamefont {Tsai}}, \bibinfo
  {author} {\bibfnamefont {N.-C.}\ \bibnamefont {Chiu}}, \bibinfo {author}
  {\bibfnamefont {Y.}~\bibnamefont {Kawaguchi}}, \bibinfo {author}
  {\bibfnamefont {S.-K.}\ \bibnamefont {Yip}}, \bibinfo {author} {\bibfnamefont
  {M.-S.}\ \bibnamefont {Chang}},\ and\ \bibinfo {author} {\bibfnamefont
  {Y.-J.}\ \bibnamefont {Lin}},\ }\href
  {https://doi.org/10.1103/PhysRevLett.121.250401} {\bibfield  {journal}
  {\bibinfo  {journal} {Phys. Rev. Lett.}\ }\textbf {\bibinfo {volume} {121}},\
  \bibinfo {pages} {250401} (\bibinfo {year} {2018})}\BibitemShut {NoStop}%
\bibitem [{\citenamefont {Brown}\ \emph {et~al.}(2020)\citenamefont {Brown},
  \citenamefont {Guardado-Sanchez}, \citenamefont {Spar}, \citenamefont
  {Huang}, \citenamefont {Devereaux},\ and\ \citenamefont {Bakr}}]{Brown2020}%
  \BibitemOpen
  \bibfield  {author} {\bibinfo {author} {\bibfnamefont {P.~T.}\ \bibnamefont
  {Brown}}, \bibinfo {author} {\bibfnamefont {E.}~\bibnamefont
  {Guardado-Sanchez}}, \bibinfo {author} {\bibfnamefont {B.~M.}\ \bibnamefont
  {Spar}}, \bibinfo {author} {\bibfnamefont {E.~W.}\ \bibnamefont {Huang}},
  \bibinfo {author} {\bibfnamefont {T.~P.}\ \bibnamefont {Devereaux}},\ and\
  \bibinfo {author} {\bibfnamefont {W.~S.}\ \bibnamefont {Bakr}},\ }\href
  {https://doi.org/10.1038/s41567-019-0696-0} {\bibfield  {journal} {\bibinfo
  {journal} {Nature Physics}\ }\textbf {\bibinfo {volume} {16}},\ \bibinfo
  {pages} {26} (\bibinfo {year} {2020})}\BibitemShut {NoStop}%
\bibitem [{\citenamefont {Bohrdt}\ \emph {et~al.}(2018)\citenamefont {Bohrdt},
  \citenamefont {Greif}, \citenamefont {Demler}, \citenamefont {Knap},\ and\
  \citenamefont {Grusdt}}]{PhysRevB.97.125117}%
  \BibitemOpen
  \bibfield  {author} {\bibinfo {author} {\bibfnamefont {A.}~\bibnamefont
  {Bohrdt}}, \bibinfo {author} {\bibfnamefont {D.}~\bibnamefont {Greif}},
  \bibinfo {author} {\bibfnamefont {E.}~\bibnamefont {Demler}}, \bibinfo
  {author} {\bibfnamefont {M.}~\bibnamefont {Knap}},\ and\ \bibinfo {author}
  {\bibfnamefont {F.}~\bibnamefont {Grusdt}},\ }\href
  {https://doi.org/10.1103/PhysRevB.97.125117} {\bibfield  {journal} {\bibinfo
  {journal} {Phys. Rev. B}\ }\textbf {\bibinfo {volume} {97}},\ \bibinfo
  {pages} {125117} (\bibinfo {year} {2018})}\BibitemShut {NoStop}%
\bibitem [{\citenamefont {Scarlatella}\ \emph {et~al.}(2019)\citenamefont
  {Scarlatella}, \citenamefont {Clerk},\ and\ \citenamefont
  {Schiro}}]{Scarlatella2019}%
  \BibitemOpen
  \bibfield  {author} {\bibinfo {author} {\bibfnamefont {O.}~\bibnamefont
  {Scarlatella}}, \bibinfo {author} {\bibfnamefont {A.~A.}\ \bibnamefont
  {Clerk}},\ and\ \bibinfo {author} {\bibfnamefont {M.}~\bibnamefont
  {Schiro}},\ }\href {https://doi.org/10.1088/1367-2630/ab0ce9} {\bibfield
  {journal} {\bibinfo  {journal} {New Journal of Physics}\ }\textbf {\bibinfo
  {volume} {21}},\ \bibinfo {pages} {043040} (\bibinfo {year}
  {2019})}\BibitemShut {NoStop}%
\bibitem [{\citenamefont {Bismut}\ \emph {et~al.}(2012)\citenamefont {Bismut},
  \citenamefont {Laburthe-Tolra}, \citenamefont {Mar\'echal}, \citenamefont
  {Pedri}, \citenamefont {Gorceix},\ and\ \citenamefont {Vernac}}]{Bismut2012}%
  \BibitemOpen
  \bibfield  {author} {\bibinfo {author} {\bibfnamefont {G.}~\bibnamefont
  {Bismut}}, \bibinfo {author} {\bibfnamefont {B.}~\bibnamefont
  {Laburthe-Tolra}}, \bibinfo {author} {\bibfnamefont {E.}~\bibnamefont
  {Mar\'echal}}, \bibinfo {author} {\bibfnamefont {P.}~\bibnamefont {Pedri}},
  \bibinfo {author} {\bibfnamefont {O.}~\bibnamefont {Gorceix}},\ and\ \bibinfo
  {author} {\bibfnamefont {L.}~\bibnamefont {Vernac}},\ }\href
  {https://doi.org/10.1103/PhysRevLett.109.155302} {\bibfield  {journal}
  {\bibinfo  {journal} {Phys. Rev. Lett.}\ }\textbf {\bibinfo {volume} {109}},\
  \bibinfo {pages} {155302} (\bibinfo {year} {2012})}\BibitemShut {NoStop}%
\bibitem [{\citenamefont {Boudjem\^aa}\ and\ \citenamefont
  {Shlyapnikov}(2013)}]{Boudjem2013}%
  \BibitemOpen
  \bibfield  {author} {\bibinfo {author} {\bibfnamefont {A.}~\bibnamefont
  {Boudjem\^aa}}\ and\ \bibinfo {author} {\bibfnamefont {G.~V.}\ \bibnamefont
  {Shlyapnikov}},\ }\href {https://doi.org/10.1103/PhysRevA.87.025601}
  {\bibfield  {journal} {\bibinfo  {journal} {Phys. Rev. A}\ }\textbf {\bibinfo
  {volume} {87}},\ \bibinfo {pages} {025601} (\bibinfo {year}
  {2013})}\BibitemShut {NoStop}%
\end{thebibliography}%

\clearpage{}

\onecolumngrid
\appendix
\renewcommand{\thefigure}{S\arabic{figure}}
\setcounter{figure}{0} 
\renewcommand{\thepage}{S\arabic{page}}
\setcounter{page}{1} 
\renewcommand{\theequation}{S.\arabic{equation}}
\setcounter{equation}{0} 
\renewcommand{\thesection}{S\arabic{section}}
\setcounter{section}{0}

\begin{center}
	\large{Supplemental Material for}\\
	\textbf{``Dissipative Superfluidity in a Molecular Bose-Einstein Condensate"}
\end{center}

\tableofcontents{}

\section{Derivation of The Effective Action}

\subsection{Effective Action in Closed Quantum Systems}

We derive the effective Keldysh action for a dissipative Bose-Einstein condensate (BEC) that is used in the main text. To set the basis for discussions, we first review the theory of superfluidity
in a closed bosonic system \citep{Wen2007}. Let $\varphi_+$ and $\varphi_-$ be the bosonic field for the forward and backward paths, respectively. Then the action of the system is given by  
\begin{equation}
	S=\int dtd^{3}\bm{r}[i\varphi_{+}^{\ast}(\bm{r},t)\partial_{t}\varphi_{+}(\bm{r},t)-H_{+}-i\varphi_{-}^{\ast}(\bm{r},t)\partial_{t}\varphi_{-}(\bm{r},t)+H_{-}],\label{eq:closed_action}
\end{equation}
where 
\begin{equation}
	H_{\pm}=\frac{1}{2m}(\nabla\varphi_{\pm}^{\ast})\cdot(\nabla\varphi_{\pm})-\mu|\varphi_{\pm}|^{2}+\frac{U}{2}|\varphi_{\pm}|^{4}.
\end{equation}
Let us introduce the retarded and advanced fields~\cite{PhysRevD.103.056020} defined by
\begin{equation}\label{eq: retarded_frame}
	\varphi_{R}=\frac{1}{2}(\varphi_{+}+\varphi_{-}),\varphi_{A}=\varphi_{+}-\varphi_{-},
\end{equation}
which correspond to the classical and quantum fields in Refs. \cite{Sieberer_2016,Kamenev_2011}. In terms of them, the action~(\ref{eq:closed_action}) can be expressed as 
\begin{eqnarray}
	S & = & \int dtd^{3}\bm{r}\left[i\varphi_{R}^{\ast}\partial_{t}\varphi_{A}+i\varphi_{A}^{\ast}\partial_{t}\varphi_{R}-\frac{1}{2m}(\nabla\varphi_{R}^{\ast})\cdot(\nabla\varphi_{A})-\frac{1}{2m}(\nabla\varphi_{A}^{\ast})\cdot(\nabla\varphi_{R})\right]\nonumber \\
	&  & +\int dtd^{3}\bm{r}\left[\mu|\varphi_{+}|^2-\mu|\varphi_{-}|^2-\frac{U}{2}|\varphi_{+}|^{4}+\frac{U}{2}|\varphi_{-}|^{4}\right].
\end{eqnarray}

Let us decompose the bosonic fields
$\varphi_{+},\varphi_{-}$ around a U(1)-symmetry-broken ground state $\varphi_0$ as \citep{PhysRevD.103.056020}
\begin{eqnarray}
	\varphi_{+} & = & \varphi_{0}(1+\phi_{+})e^{i\theta_{+}},\label{eq:phi+}\\
	\varphi_{-} & = & \varphi_{0}(1+\phi_{-})e^{i\theta_{-}},\label{eq:phi-}
\end{eqnarray}
where $\phi_{\alpha}$ and $\theta_{\alpha}$ represent the Higgs mode and the Nambu-Goldstone mode on the contour $\alpha=\pm$. At the first step, we consider a uniform field solution $\varphi_0$. The value of $\varphi_R$ is determined from the mean-field equation $\delta S/\delta\varphi_{A}=0$ and setting $\varphi_{A}=0$. The mean-field equation gives
\begin{equation}
	(i\partial_{t}+\mu-U|\varphi_{R}|^{2}-U|\varphi_{A}|^{2})\varphi_{R}=0.
\end{equation}
The stationary solution to this equation is 
\begin{eqnarray*}
	\varphi_{R} & = & \sqrt{n_0},\\
	\mu & = & n_0U,\\
	\varphi_{A} & = & 0,
\end{eqnarray*}
where $n_0$ is the particle-number density of the condensate. Thus, we have $\varphi_0=\sqrt{n_0}$. By expanding the action \eqref{eq:closed_action} up to the second order in $\phi_{\pm}$ and $\theta_{\pm}$, we have 
\begin{equation}
	S=S_{+}-S_{-},\label{eq:total_action}
\end{equation}
where 
\begin{equation}\label{eq:alpha_action}
	S_{\alpha}=\int dtd^{3}\bm{r}\left[-\varphi_{0}^{2}\partial_{t}\theta_{\alpha}-\frac{1}{2m}\varphi_{0}^{2}(\nabla\theta_{\alpha})^{2}-2\varphi_{0}^{2}\phi_{\alpha}\partial_{t}\theta_{\alpha}-\frac{1}{2m}\varphi_{0}^{2}(\nabla\phi_{\alpha})^{2}-2U\varphi_{0}^{4}\phi_{\alpha}^{2}\right].
\end{equation}
By substituting the retarded and advanced fields in Eq. \eqref{eq: retarded_frame} into Eqs. \eqref{eq:total_action} and \eqref{eq:alpha_action}, we
obtain 
\begin{eqnarray}
	S & = & \int dtd^{3}\bm{r}\left[-\varphi_{0}^{2}\partial_{t}\theta_{A}-\frac{1}{m}\varphi_{0}^{2}(\nabla\theta_{R})(\nabla\theta_{A})-2\varphi_{0}^{2}\phi_{R}\partial_{t}\theta_{A}-2\varphi_{0}^{2}\phi_{A}\partial_{t}\theta_{R}-\frac{1}{m}\varphi_{0}^{2}(\nabla\phi_{R})(\nabla\phi_{A})\right]\nonumber \\
	&  & +\int dtd^{3}\bm{r}[-4U\varphi_{0}^{4}\phi_{R}\phi_{A}]\nonumber \\
	& = & \int dtd^{3}\bm{r}\left[-\frac{1}{m}\varphi_{0}^{2}(\nabla\theta_{R})(\nabla\theta_{A})-2\varphi_{0}^{2}\left(\begin{array}{cc}
		\phi_{R} & \phi_{A}\end{array}\right)\left(\begin{array}{c}
		\partial_{t}\theta_{A}\\
		\partial_{t}\theta_{R}
	\end{array}\right)\right]\nonumber \\
	&  & +\int dtd^{3}\bm{r}\left[-\left(\begin{array}{cc}
		\phi_{R} & \phi_{A}\end{array}\right)\left(\begin{array}{cc}
		0 & -\frac{\varphi_{0}^{2}}{2m}\nabla^{2}+2U\varphi_{0}^{4}\\
		-\frac{\varphi_{0}^{2}}{2m}\nabla^{2}+2U\varphi_{0}^{4} & 0
	\end{array}\right)\left(\begin{array}{c}
		\phi_{R}\\
		\phi_{A}
	\end{array}\right)\right].
\end{eqnarray}
Eliminating the amplitude modes $\phi_{R}$ and $\phi_{A}$ through path integration, we obtain
\begin{eqnarray}
	S^{\text{eff}} & = & \int dtd^{3}\bm{r}\left[-\frac{1}{m}\varphi_{0}^{2}(\nabla\theta_{R})(\nabla\theta_{A})\right]\nonumber \\
	&  & +\varphi_{0}^{4}\int dtd^{3}\bm{r}\left[\left(\begin{array}{cc}
		\partial_{t}\theta_{A} & \partial_{t}\theta_{R}\end{array}\right)\left(\begin{array}{cc}
		0 & \left(-\frac{\varphi_{0}^{2}}{2m}\nabla^{2}+2U\varphi_{0}^{4}\right)^{-1}\\
		\left(-\frac{\varphi_{0}^{2}}{2m}\nabla^{2}+2U\varphi_{0}^{4}\right)^{-1} & 0
	\end{array}\right)\left(\begin{array}{c}
		\partial_{t}\theta_{A}\\
		\partial_{t}\theta_{R}
	\end{array}\right)\right].\nonumber \\
	& = & \int dtd^{3}\bm{r}\left[-\frac{1}{m}\varphi_{0}^{2}(\nabla\theta_{R})(\nabla\theta_{A})+\partial_{t}\theta_{R}\frac{2\varphi_{0}^{4}}{-\frac{\varphi_{0}^{2}}{2m}\nabla^{2}+2U\varphi_{0}^{4}}\partial_{t}\theta_{A}\right].
\end{eqnarray}
Rewriting this in terms of quantities belonging to the forward and backward contours, we obtain 
\begin{equation}
	S^{\text{eff}}=S_{+}^{\text{eff}}-S_{-}^{\text{eff}},
\end{equation}
where 
\begin{equation}
	S_{\alpha}^{\text{eff}}=\int dtd^{3}\bm{r}\left[-\frac{1}{2m}\varphi_{0}^{2}(\nabla\theta_{\alpha})(\nabla\theta_{\alpha})+\partial_{t}\theta_{\alpha}\frac{\varphi_{0}^{2}}{-\frac{1}{2m}\nabla^{2}+2U\varphi_{0}^{2}}\partial_{t}\theta_{\alpha}\right].\label{eq:action_2}
\end{equation}
Applying the Fourier transformation to this action and requiring $S_{\alpha}^{\text{eff}}=0$ which corresponds to the poles of the propagator, we obtain the excitation spectrum:
\begin{equation}
	\omega=\sqrt{\frac{k^{2}}{2m}\left(\frac{k^{2}}{2m}+2Un_0\right)}.
\end{equation}
Setting $\partial_{t}\theta_{+}=\partial_{t}\theta_{-}=0$ and performing a Wick rotation, we obtain
\begin{equation}
	S=\int d\tau d^{3}\bm{r}\frac{1}{2m}\varphi_{0}^{2}(\nabla\theta)^{2}.
\end{equation}
This positive semidefinite term expresses the phase rigidity. We note that the last term in Eq. \eqref{eq:action_2} diverges for free bosons. Thus the present analysis should apply only for interacting systems with $U\neq0$.

\subsection{Effective Action in Open Quantum Systems without a Dipolar Interaction}

We turn to the effective action in a Lindblad system, in which case the action becomes
\begin{equation}
	S=\int dtd^{3}\bm{r}[i\varphi_{+}^{\ast}(\bm{r},t)\partial_{t}\varphi_{+}(\bm{r},t)-H_{+}-i\varphi_{-}^{\ast}(\bm{r},t)\partial_{t}\varphi_{-}(\bm{r},t)+H_{-}-i\gamma\varphi_{-}^{\ast}(\bm{r},t)^{2}\varphi_{+}(\bm{r},t)^{2}],
\end{equation}
where 
\begin{equation}
	H_{\pm}=\frac{1}{2m}(\nabla\varphi_{\pm}^{\ast})\cdot(\nabla\varphi_{\pm})+\frac{U_{\pm}}{2}|\varphi_{\pm}|^{4}
\end{equation}
with $U_{\pm}=U_{R}\mp i\gamma$. The uniform mean-field equation $\delta S/\delta\varphi_{A}=0$ can be expressed in terms of the retarded and advanced fields in Eq. (\ref{eq: retarded_frame}) as
\begin{equation}
	(i\partial_{t}-(U_{R}-i\gamma)|\varphi_{R}|^{2}-(U_{R}-i\gamma)|\varphi_{A}|^{2})\varphi_{R}=0.
\end{equation}
The solution of this equation is 
\begin{eqnarray*}
	\varphi_{R}(t) & = & \frac{\varphi_{R}(0)}{\sqrt{1+2\gamma\varphi_{R}(0)^{2}t}}\exp\left(-i\int_0^t\mu(t')dt'\right)=\sqrt{\frac{n_0(0)}{1+2\gamma n_0(0)t}}\exp\left(-i\int_0^t\mu(t')dt'\right),\\
	\mu(t) & = & n_0(t)U_{R},\\
	\varphi_{A} & = & 0,
\end{eqnarray*}
where $n_0(t)=n_0(0)/(1+2\gamma n_0(0)t)$ is the number density of the condensate. Hence, the mean-field solution is given by $\varphi_{0}=\varphi_{R}(t)$, which is a function of time $t$.

The bosonic fields can be decomposed into fields of $\theta_{\pm}$
and $\phi_{\pm}$ as in Eqs. (\ref{eq:phi+}) and (\ref{eq:phi-}), and the expansion up to the second order of them gives
\begin{equation}\label{eq:total_action_1}
	S=S_{+}-S_{-}+\int dtd^{3}\bm{r}[(-4i\gamma)|\varphi_0|^{4}\phi_{+}\phi_{-}+2\gamma|\varphi_0|^{4}(\theta_{+}-\theta_{-})+4\gamma|\varphi_0|^{4}(\phi_{+}+\phi_{-})(\theta_{+}-\theta_{-})],
\end{equation}
where 
\begin{equation}\label{eq:contour_action}
	S_{\alpha}=\int dtd^{3}\bm{r}\left[-|\varphi_0|^{2}\partial_{t}\theta_{\alpha}-\frac{1}{2m}|\varphi_0|^{2}(\nabla\theta_{\alpha})^{2}-2|\varphi_0|^{2}\phi_{\alpha}\partial_{t}\theta_{\alpha}-\frac{1}{2m}|\varphi_0|^{2}(\nabla\phi_{\alpha})^{2}-2(U_{R}-i\alpha\gamma)|\varphi_0|^{4}\phi_{\alpha}^{2}\right].
\end{equation}
Substituting Eq. \eqref{eq: retarded_frame} into Eqs. \eqref{eq:total_action_1} and \eqref{eq:contour_action}, we have
\begin{eqnarray}
	S & = & \int dtd^{3}\bm{r}\left[-|\varphi_0|^{2}\partial_{t}\theta_{A}-\frac{1}{m}|\varphi_0|^{2}(\nabla\theta_{R})(\nabla\theta_{A})-2|\varphi_0|^{2}\phi_{R}\partial_{t}\theta_{A}-2|\varphi_0|^{2}\phi_{A}\partial_{t}\theta_{R}-\frac{1}{m}|\varphi_0|^{2}(\nabla\phi_{R})(\nabla\phi_{A})\right]\nonumber \\
	&  & +\int dtd^{3}\bm{r}\left[-4U_{R}|\varphi_0|^{4}\phi_{R}\phi_{A}+2i\gamma|\varphi_0|^{4}\left(2\phi_{R}^{2}+\frac{1}{2}\phi_{A}^{2}\right)+2\gamma|\varphi_0|^{4}\theta_{A}+8\gamma|\varphi_0|^{4}\phi_{R}\theta_{A}\right]\nonumber \\
	& = & \int dtd^{3}\bm{r}\left[-|\varphi_0|^{2}\partial_{t}\theta_{A}-\frac{1}{m}|\varphi_0|^{2}(\nabla\theta_{R})(\nabla\theta_{A})-2|\varphi_0|^{2}\left(\begin{array}{cc}
		\phi_{R} & \phi_{A}\end{array}\right)\left(\begin{array}{c}
		\partial_{t}\theta_{A}-4\gamma|\varphi_0|^{2}\theta_{A}\\
		\partial_{t}\theta_{R}
	\end{array}\right)\right]\nonumber \\
	&  & +\int dtd^{3}\bm{r}\left[-\left(\begin{array}{cc}
		\phi_{R} & \phi_{A}\end{array}\right)\left(\begin{array}{cc}
		-4i\gamma|\varphi_0|^{4} & -\frac{|\varphi_0|^{2}}{2m}\nabla^{2}+2U_{R}|\varphi_0|^{4}\\
		-\frac{|\varphi_0|^{2}}{2m}\nabla^{2}+2U_{R}|\varphi_0|^{4} & -i\gamma|\varphi_0|^{4}
	\end{array}\right)\left(\begin{array}{c}
		\phi_{R}\\
		\phi_{A}
	\end{array}\right)+2\gamma|\varphi_0|^{4}\theta_{A}\right].
\end{eqnarray}
Integrating out the amplitude fields $\phi_{R,A}$, we obtain the effective action as 
\begin{eqnarray}
	S^{\text{eff}} & = & \int dtd^{3}\bm{r}\left[-|\varphi_0|^{2}\partial_{t}\theta_{A}-\frac{1}{m}|\varphi_0|^{2}(\nabla\theta_{R})(\nabla\theta_{A})\right]\nonumber \\ \label{eq: effective}
	&  & +\int dtd^{3}\bm{r}\left[|\varphi_0|^{4}\left(\begin{array}{cc}
		\partial_{t}\theta_{A}-4\gamma|\varphi_0|^{2}\theta_{A} & \partial_{t}\theta_{R}\end{array}\right)A^{-1}\left(\begin{array}{c}
		\partial_{t}\theta_{A}-4\gamma|\varphi_0|^{2}\theta_{A}\\
		\partial_{t}\theta_{R}
	\end{array}\right)+2\gamma|\varphi_0|^{4}\theta_{A}\right],
\end{eqnarray}
where 
\begin{equation}
	A=\left(\begin{array}{cc}
		-4i\gamma|\varphi_0|^4 & -\frac{|\varphi_0|^{2}}{2m}\nabla^{2}+2U_{R}|\varphi_0|^{4}\\
		-\frac{|\varphi_0|^{2}}{2m}\nabla^{2}+2U_{R}|\varphi_0|^{4} & -i\gamma|\varphi_0|^4
	\end{array}\right).
\end{equation}
We note that $\det(A)\neq0$ holds for an arbitrary interaction strength and an arbitrary dissipation strength unless they both vanish. Hence, the effective field theory derived here is valid even without the interaction. Furthermore, if we impose the conditions $\partial_{t}\theta_{+}=\partial_{t}\theta_{-}=0$,
we obtain 
\begin{eqnarray}\label{eq: effective_action_1}
	S^{\text{eff}} & = & \int dtd^{3}\bm{r}\left[-\frac{1}{m}|\varphi_0|^{2}(\nabla\theta_{R})(\nabla\theta_{A})+16\gamma^{2}|\varphi_0|^{4}\theta_{A}^{2}\frac{|\varphi_0|^{4}(-i\gamma|\varphi_0|^{4})}{-4\gamma^{2}|\varphi_0|^{8}-\left(-\frac{|\varphi_0|^{2}}{2m}\nabla^{2}+2U_{R}|\varphi_0|^{4}\right)^{2}}+2\gamma|\varphi_0|^{4}\theta_{A}\right]\nonumber \\
	& \simeq & \int dtd^{3}\bm{r}\left[-\frac{1}{m}|\varphi_0|^{2}(\nabla\theta_{R})(\nabla\theta_{A})+4i\gamma^{2}|\varphi_0|^{4}\theta_{A}^{2}\frac{\gamma}{\gamma^{2}+U_{R}^{2}}+2\gamma|\varphi_0|^{4}\theta_{A}\right],
\end{eqnarray}
where we ignore the contribution from $\nabla^{2}$ since we are interested in the long-wavelength limit. We can see that the second term on the right-hand side of Eq. \eqref{eq: effective_action_1} 
includes the second-order terms in $\theta_A$, which can be neglected in the following calculation since we only consider the lowest-order perturbation around the mean-field solution. Therefore, the effective action becomes
\begin{eqnarray}
	S^{\text{eff}} & = & -\int dt d^{3}\bm{r}\left[\frac{1}{m}|\varphi_0|^{2}(\nabla\theta_{R})(\nabla\theta_{A})-2\gamma|\varphi_0|^{4}\theta_{A}\right]\nonumber \\
	& = & -\int dt d^{3}\bm{r}\left[\frac{1}{2m}|\varphi_0|^{2}(\nabla\theta_{+})(\nabla\theta_{+})-\frac{1}{2m}|\varphi_0|^{2}(\nabla\theta_{-})(\nabla\theta_{-})-2\gamma|\varphi_0|^{4}(\theta_{+}-\theta_{-})\right]\nonumber \\
	& = & -\int dt d^{3}\bm{r}|\varphi_0|^{2}\left[\frac{1}{2m}(\nabla\theta_{+}+\bm{\psi}(\bm{r}))(\nabla\theta_{+}+\bm{\psi}(\bm{r}))-\frac{1}{2m}(\nabla\theta_{-}+\bm{\psi}(\bm{r}))(\nabla\theta_{-}+\bm{\psi}(\bm{r}))\right],\label{action-superfluid}
\end{eqnarray}
with $\nabla\cdot\bm{\psi}=2m\gamma|\varphi_0|^{2}$. 
Here the vector field $\bm{\psi}(\bm{r})$ gives the dissipative current, which describes the loss of particles into the environment \citep{PhysRevLett.93.160404}.
Since the twist of the phase of a condensate can be related to the superfluid velocity as $\bm{v}_{s\alpha}=\alpha\nabla\theta_{\alpha}/m$ \citep{Coleman_2015},
the current density is determined as 
\begin{eqnarray}
	\langle\bm{j}\rangle & = & \frac{1}{2}\langle\bm{j}_{+}+\bm{j}_{-}\rangle =-\frac{1}{2}\left(\frac{\delta S^{\text{eff}}}{m\delta\bm{v}_{s+}}+\frac{\delta S^{\text{eff}}}{m\delta\bm{v}_{s-}}\right)|_{\bm{v}_{s+}=-\bm{v}_{s-}=\bm{v}_{s}}\nonumber \\
	& = & \frac{|\varphi_0|^{2}}{m}(m\bm{v}_s+\bm{\psi}),\label{eq:super_current_action}
\end{eqnarray}
where $\bm{j}_{\alpha}:=-\delta S^{\text{eff}}/\delta(m\bm{v}_{\alpha}) $. Equation \eqref{eq:super_current_action} gives $\rho_{s}=|\varphi_0|^{2}=n_0(t)$ by the definition of the superfluid density $\rho_{s}=\partial\langle\bm{j}\rangle/\partial\bm{v}_{s}$. We therefore find that the system shows superfluid transport if the dissipative current is subtracted from the total current.

Let us now introduce the current-density operator as $\hat{\bm{j}_c}=\frac{1}{2}\sum_{\bm{k}}\bm{k}(a_{\bm{k}+}^{\dagger}a_{\bm{k}+}+a_{\bm{k}-}^{\dagger}a_{\bm{k}-})$
and we calculate the expectation value of the current as 
\begin{equation}
	\bm{j}_c=\langle\hat{\bm{j}_c}\rangle=\frac{1}{Z}\int D[\varphi_{+}]D[\varphi_{-}]\bm{j}_c e^{iS}.
\end{equation}
This is the ``classical'' current in the open quantum systems defined in~Ref. \cite{Sieberer_2016}. However, here we have a  decay of the particle-number density, which can be seen from the continuous equation as 
\begin{equation}
	\frac{dn}{dt}=-\nabla\cdot(\bm{j}_{c}+\bm{j}_{d}),
\end{equation}
where $\bm{j}=\bm{j}_{c}+\bm{j}_{d}$ is the total current, $\bm{j}_{c}$
is the nondissipative current density and $\bm{j}_{d}$ is the dissipative current density due to the loss of particles. This dissipative current density obeys the following equation of continuity: 
\begin{equation}
	\nabla\cdot\bm{j}_{d}=-\frac{dn_0(t)}{dt}=\frac{2\gamma n_0(0)^{2}}{(1+2\gamma n_0(0)t)^{2}}.
\end{equation}
One immediately recognizes that $\bm{j}_c=\bm{j}_s=|\varphi_0|^2\bm{v}_s$ and $\bm{j}_d=|\varphi_0|^2\bm{\psi}/m$.
Nevertheless, the current $\bm{j}_d$ will not influence the
superfluid density since $\partial j_{\alpha}/\partial v_{\beta}=\rho_s\delta_{\alpha\beta}$.
This is also the case in the above field-theoretic calculations.

To investigate the stability of the bosonic gas against density fluctuations, we consider the compressibility of the gas subject to dissipation. In the closed quantum systems, the compressibility is defined as
\begin{equation}
	\kappa = - \frac{1}{V} \left( \frac{\partial V}{\partial p} \right)_T,
\end{equation}
which is equivalent to
\begin{equation}\label{eq: compressibility}
	\kappa = \frac{1}{n^2} \frac{\partial n}{\partial \mu} .
\end{equation}
However, since the thermodynamic relations do not hold in nonequilibrium open systems, we define the compressibility from the response $\delta n(x)$ of the density distribution $n(x)$ to the applied external potential $\delta\nu(x)$
\begin{equation}
	\delta n (x) = \int d x' \kappa (x, x') \delta \nu (x'),
\end{equation}
where $x:=(\bm{r},t)$. The response function $\kappa (x, x')$ can be calculated with the Schwinger-Keldysh
field theory as
\begin{equation}
	\kappa (x, x') = i (\langle n_+ (x) n_+ (x') \rangle - \langle n_- (x) n_-
	(x') \rangle) .
\end{equation}
Since the Schwinger-Keldysh action can be rewritten as
\begin{equation}
	S = \int d^4 x [n_+ (x) \partial_t \theta_+ (x) - n_- (x) \partial_t
	\theta_- (x)] + S_{\text{int}} [n, \theta, \nabla \theta],
\end{equation}
where $S_{\mathrm{int}}[n,\theta,\nabla\theta]$ does not include $\partial_t \theta_\alpha$, we can rewrite the response function as
\begin{equation}
	\kappa (x, x') = i \left( \left\langle \frac{\delta S^{\text{eff}}}{\delta \partial_t
		\theta_+ (x)} \frac{\delta S^{\text{eff}}}{\delta \partial_t \theta_+ (x')} \right\rangle
	- \left\langle \frac{\delta S^{\text{eff}}}{\delta \partial_t \theta_- (x)} \frac{\delta
		S^{\text{eff}}}{\delta \partial_t \theta_- (x')} \right\rangle \right) .
\end{equation}
Performing the Fourier transformation, we have
\begin{equation}
	\delta n (k) = \kappa (k) \delta \nu (k),
\end{equation}
where
\begin{equation}
	\kappa (k) = \frac{i}{\omega^2} \left( \left\langle \frac{\delta S^{\text{eff}}}{\delta
		\theta_+ (k)} \frac{\delta S^{\text{eff}}}{\delta
		\theta_+ (-k)} \right\rangle - \left\langle \frac{\delta S^{\text{eff}}}{\delta
		\theta_- (k)} \frac{\delta S^{\text{eff}}}{\delta
		\theta_- (-k)} \right\rangle \right) = \frac{i}{\omega^2} \left\langle \frac{\delta
		S^{\text{eff}}}{\delta \theta_R (k)} \frac{\delta S^{\text{eff}}}{\delta \theta_A (-k)} \right\rangle,
\end{equation}
and $k:=(\bm{k},\omega)$. Therefore, the compressibility in Eq. \eqref{eq: compressibility} is given by
\begin{equation}
	\kappa = \frac{1}{n^2} \lim_{k \rightarrow 0} \kappa
	(k) .
\end{equation}
From Eq. \eqref{eq: effective}, we obtain the compressibility as
\begin{equation}
	\kappa = \frac{U_R}{(\gamma^2 + U_R^2) n^2} .
\end{equation}
In the closed quantum systems without loss, the compressibility diverges as $U_R\to0$, indicating the instability of free bosonic systems against the external perturbation. However, in the open quantum system, the compressibility vanishes and thus remains finite if $U_R \to 0$ and $\gamma >0$. This result indicates that the density fluctuations are suppressed due to dissipation, ensuring the stability of the bosonic gas. We note that the vanishing compressibility does not indicate the infinite speed of sound since
\begin{equation}
	\kappa \neq - \frac{1}{V} \left( \frac{\partial V}{\partial p} \right)_T,
\end{equation}
i.e., the compressibility which is defined from the response function of the nonequilibrium system does not obey the thermodynamic relation. Therefore, the compressibility in the case of $\gamma\neq0$ should be distinguished from the conventional one in a closed quantum system and the two limits $U_R\to0$ and $\gamma\to0$ cannot be exchanged.

\subsection{Effective Action in Open Quantum Systems with a Dipolar Interaction}
We now move on to the molecular BEC system and consider an effective field theory of a dipolar Bose
gas. Here we take into account the dipole-dipole interaction, which takes the form of 
\begin{equation}
	V(\bm{r}-\bm{r}')=c_{dd}\frac{1-3\cos^{2}\theta_{r-r'}}{|\bm{r}-\bm{r}'|^{3}}a_{\bm{r}}^{\dagger}a_{\bm{r}'}^{\dagger}a_{\bm{r}'}a_{\bm{r}},
\end{equation}
where we assume that the dipole moments are polarized in the $z$ direction and $\cos\theta_{r-r'}=(\bm{r}-\bm{r}')\cdot\hat{\bm{z}}/|\bm{r}-\bm{r}'|$.
By adding this interaction, the
action becomes 
\begin{equation}
	S=\int dtd^{3}\bm{r}[i\varphi_{+}^{\ast}(\bm{r},t)\partial_{t}\varphi_{+}(\bm{r},t)-H_{+}-i\varphi_{-}^{\ast}(\bm{r},t)\partial_{t}\varphi_{-}(\bm{r},t)+H_{-}-i\gamma\varphi_{-}^{\ast}(\bm{r},t)^{2}\varphi_{+}(\bm{r},t)^{2}],
\end{equation}
where 
\begin{equation}
	H_{\pm}=\frac{1}{2m}(\nabla\varphi_{\pm}^{\ast})\cdot(\nabla\varphi_{\pm})+\frac{U_{\pm}}{2}|\varphi_{\pm}|^{4}+c_{dd}\int d^{3}\bm{r}'\frac{1-3\cos^{2}\theta_{r-r'}}{|\bm{r}-\bm{r}'|^{3}}|\varphi_{\pm}(\bm{r},t)|^{2}|\varphi_{\pm}(\bm{r}',t)|^{2}.
\end{equation}
We consider the uniform mean-field solution by $\delta S/\delta\varphi_{A}^{\ast}=0$
which gives
\begin{equation}
	\left(i\partial_{t}-(U_{R}-i\gamma)|\varphi_{R}|^{2}(\bm{r},t)-c_{dd}\int d^{3}\bm{r}'\frac{1-3\cos^{2}\theta_{r-r'}}{|\bm{r}-\bm{r}'|^{3}}|\varphi_{R}(\bm{r}',t)|^{2}\right)\varphi_{R}(\bm{r},t)=0.
\end{equation}
Here we assume $\varphi_{A}=0$ as before since in the saddle-point solution the field is the same on the forward and backward contours. This equation can be solved as
\begin{eqnarray}
	\varphi_{R}(t) & = & \frac{\varphi_{R}(0)}{\sqrt{1+2\gamma\varphi_{R}(0)^{2}t}}\exp\left(-i\int_0^t\mu(t')dt'\right)=\sqrt{\frac{n_0(0)}{1+2\gamma n_0(0)t}}\exp\left(-i\int_0^t\mu(t')dt'\right),\nonumber \\
	\mu(t) & = & |\varphi_{R}(t)|^{2}\left(U_{R}+2c_{dd}\int d^{3}\bm{r}'\frac{1-3\cos^{2}\theta_{r-r'}}{|\bm{r}-\bm{r}'|^{3}}\right)\nonumber \\
	& = & n_0(t)\left(U_{R}-\frac{8\pi}{3}c_{dd}\right).
\end{eqnarray}
Here we regularize the integral over $\bm{r}'$ as~\cite{Lahaye_2009}
\begin{eqnarray}
	\int d^{3}\bm{r}'\frac{1-3\cos^{2}\theta_{r-r'}}{|\bm{r}-\bm{r}'|^{3}} & = & \int d^{3}\bm{r}'\frac{1-3\cos^{2}\theta_{r'}}{|\bm{r}'|^{3}}\nonumber \\
	& = & \lim_{\bm{k}\rightarrow0}\int d^{3}\bm{r}'\frac{1-3\cos^{2}\theta_{r'}}{|\bm{r}'|^{3}}e^{i\bm{k}\cdot\bm{r}'}\nonumber \\
	& = & -\lim_{\bm{k}\rightarrow0}4\pi\int\frac{dr'}{r'}j_{2}(kr')\nonumber \\
	& = & -\frac{4\pi}{3},
\end{eqnarray}
where $j_2(x)$ is the Bessel function of the second kind. We can see that the effect of the dipolar interaction is just to partially cancel the repulsive interaction. By defining 
\begin{equation}
	\varepsilon_{dd}:=\frac{8\pi c_{dd}}{3U_{R}},
\end{equation}
we have $\mu=n_0(t)(1-\varepsilon_{dd})U_{R}$. By adding the dipolar interaction to the action \eqref{eq:contour_action} on each contour, we obtain 
\begin{equation}
	S_{\alpha}\rightarrow S_{\alpha}+\int dtd^{3}\bm{r}\left[\frac{8\pi}{3}c_{dd}|\varphi_{0}|^{2}(1+\phi_{\alpha}(\bm{r},t))^{2}+c_{dd}\int d^{3}\bm{r}'\frac{1-3\cos^{2}\theta_{r-r'}}{|\bm{r}-\bm{r}'|^{3}}|\varphi_{0}|^{4}(1+\phi_{\alpha}(\bm{r},t))^{2}(1+\phi_{\alpha}(\bm{r}',t))^{2}\right].
\end{equation}
After simplification, the modified action becomes 
\begin{equation}
	S_{\alpha}\rightarrow S_{\alpha}-4c_{dd}\int dt d^{3}\bm{r}d^{3}\bm{r}'\frac{1-3\cos^{2}\theta_{r-r'}}{|\bm{r}-\bm{r}'|^{3}}|\varphi_{0}|^{4}\phi_{+}(\bm{r},t)\phi_{-}(\bm{r}',t).
\end{equation}
By substituting the modified action on each contour into the total
action \eqref{eq:total_action_1}, we have
\begin{eqnarray}
	S & = & \int dtd^{3}\bm{r}\left[-|\varphi_{0}|^{2}\partial_{t}\theta_{A}-\frac{1}{m}|\varphi_{0}|^{2}(\nabla\theta_{R})(\nabla\theta_{A})-2|\varphi_{0}|^{2}\phi_{R}\partial_{t}\theta_{A}-2|\varphi_{0}|^{2}\phi_{A}\partial_{t}\theta_{R}-\frac{1}{m}|\varphi_{0}|^{2}(\nabla\phi_{R})(\nabla\phi_{A})\right]\nonumber \\
	&  & +\int dtd^{3}\bm{r}\left[-4U_{R}|\varphi_{0}|^{4}\phi_{R}\phi_{A}+2i\gamma\left(2\phi_{R}^{2}+\frac{1}{2}\phi_{A}^{2}\right)+2\gamma|\varphi_{0}|^{4}\theta_{A}+8\gamma|\varphi_{0}|^{4}\phi_{R}\theta_{A}\right]\nonumber \\
	&  & -4c_{dd}\int dtd^{3}\bm{r}d^{3}\bm{r}'\frac{1-3\cos^{2}\theta_{r-r'}}{|\bm{r}-\bm{r}'|^{3}}|\varphi_{0}|^{4}(\phi_{R}(\bm{r},t)\phi_{A}(\bm{r}',t)+\phi_{A}(\bm{r},t)\phi_{R}(\bm{r'},t))\nonumber \\
	& = & \int dtd^{3}\bm{r}\left[-|\varphi_{0}|^{2}\partial_{t}\theta_{A}-\frac{1}{m}|\varphi_{0}|^{2}(\nabla\theta_{R})(\nabla\theta_{A})-2|\varphi_{0}|^{2}\left(\begin{array}{cc}
		\phi_{R} & \phi_{A}\end{array}\right)\left(\begin{array}{c}
		\partial_{t}\theta_{A}-4\gamma|\varphi_{0}|^{2}\theta_{A}\\
		\partial_{t}\theta_{R}
	\end{array}\right)\right]\nonumber \\
	&  & -\int dtd^{3}\bm{r}\left[\left(\begin{array}{cc}
		\phi_{R} & \phi_{A} \end{array}\right)\left(\begin{array}{cc}
		-4i\gamma|\varphi_{0}|^{4} & -\frac{|\varphi_{0}|^{2}}{2m}\nabla^{2}+2U_{R}|\varphi_{0}|^{4}\\
		-\frac{|\varphi_{0}|^{2}}{2m}\nabla^{2}+2U_{R}|\varphi_{0}|^{4} & -i\gamma|\varphi_{0}|^{4}
	\end{array}\right)\left(\begin{array}{c}
		\phi_{R}\\
		\phi_{A}
	\end{array}\right)-2\gamma|\varphi_{0}|^{4}\theta_{A}\right]\nonumber \\
	&  & -\int dtd^{3}\bm{r}d^{3}\bm{r}'\left(\begin{array}{cc}
		\phi_{R} & \phi_{A}\end{array}\right)(\bm{r},t)\left(\begin{array}{cc}
		0 & 4c_{dd}\frac{1-3\cos^{2}\theta_{r-r'}}{|\bm{r}-\bm{r}'|^{3}}|\varphi_{0}|^{4}\\
		4c_{dd}\frac{1-3\cos^{2}\theta_{r-r'}}{|\bm{r}-\bm{r}'|^{3}}|\varphi_{0}|^{4} & 0
	\end{array}\right)\left(\begin{array}{c}
		\phi_{R}\\
		\phi_{A}
	\end{array}\right)(\bm{r'},t).
\end{eqnarray}
Then we integrate out the Higgs mode to obtain the effective field
theory for the NG mode. By the Fourier transformation, the action becomes
\begin{eqnarray*}
	S^{\text{eff}} & = & \int dtd^{3}\bm{r}\left[-|\varphi_{0}|^{2}\partial_{t}\theta_{A}-\frac{1}{m}|\varphi_{0}|^{2}(\nabla\theta_{R})(\nabla\theta_{A})-2|\varphi_{0}|^{2}\left(\begin{array}{cc}
		\phi_{R} & \phi_{A}\end{array}\right)\left(\begin{array}{c}
		\partial_{t}\theta_{A}-4\gamma|\varphi_{0}|^{2}\theta_{A}\\
		\partial_{t}\theta_{R}
	\end{array}\right)+2\gamma|\varphi_{0}|^{4}\theta_{A}\right]\\
	&  & +\int dtd^{3}\bm{k}\left[-\left(\begin{array}{cc}
		\phi_{R} & \phi_{A}\end{array}\right)(\bm{k},t)\left(\begin{array}{cc}
		-4i\gamma|\varphi_{0}|^{4} & \frac{|\varphi_{0}|^{2}k^{2}}{2m}+2(U_{R}+V_{dd}(\bm{k}))|\varphi_{0}|^{4}\\
		\frac{|\varphi_{0}|^{2}k^{2}}{2m}+2(U_{R}+V_{dd}(\bm{k}))|\varphi_{0}|^{4} & -i\gamma|\varphi_{0}|^{4}
	\end{array}\right)\left(\begin{array}{c}
		\phi_{R}\\
		\phi_{A}
	\end{array}\right)(-\bm{k},t)\right],
\end{eqnarray*}
where we define 
\begin{equation}
	V_{dd}(\bm{k}):=2c_{dd}\int d^{3}\bm{r}\frac{1-3\cos^{2}\theta_{r}}{r^{3}}e^{i\bm{k}\cdot\bm{r}}=-\frac{8\pi}{3}c_{dd}(1-3\cos^{2}\theta_{\bm{k}})
\end{equation}
with $\theta_{\bm{k}}$ being the angle between the momentum
$\bm{k}$ and the $z$-axis. We integrate out the Higgs mode in the
momentum space and set $\partial_{t}\theta_{A}=\partial_{t}\theta_{R}=0$,
obtaining the effective action as 
\begin{eqnarray}
	S^{\text{eff}} & = & \int dtd^{3}\bm{r}\left[-\frac{1}{m}|\varphi_{0}|^{2}(\nabla\theta_{R})(\nabla\theta_{A})+2\gamma|\varphi_{0}|^{4}\theta_{A}\right]+\int dtd^3\bm{k}\left[16\gamma^{2}|\varphi_{0}|^{4}\theta_{A}^{2}\frac{|\varphi_{0}|^{4}(-i\gamma|\varphi_{0}|^{4})}{-4\gamma^{2}|\varphi_{0}|^{8}-\left(\frac{|\varphi_{0}|^{2}k^{2}}{2m}+2(U_{R}+V_{dd}(\bm{k}))|\varphi_{0}|^{4}\right)^{2}}\right]\nonumber \\ \label{eq: effective_41}
	& \simeq & \int dtd^{3}\bm{r}\left[-\frac{1}{m}|\varphi_{0}|^{2}(\nabla\theta_{R})(\nabla\theta_{A})+2\gamma|\varphi_{0}|^{4}\theta_{A}\right]+\int dtd^3\bm{k}\left[4i\gamma^{2}|\varphi_{0}|^{4}\theta_{A}^{2}\frac{\gamma}{\gamma^{2}+(U_{R}+V_{dd}(\bm{k}))^{2}}\right].
\end{eqnarray}
In the second equality we also take the long-wavelength limit. Since the last term in Eq. \eqref{eq: effective_41} is of the second order in $\theta_A$, we can ignore it and arrive at the same
conclusion as before. Here, the dipolar interaction only plays
a role in modifying the strength of repulsive interaction, and therefore we
can understand the physics behind superfluidity in the same
way as the system only with contact interaction unless $\varepsilon_{dd}$ is so large that the system becomes unstable. We discuss the critical point of $\varepsilon_{dd}$ in Sec. \ref{sec: stability} (see Eq. (\ref{eq:stability_result})).

\section{$f$-sum Rule}
In this section, we generalize the $f$-sum rule to an open quantum system where the number of particles is not conserved, and discuss its relation to the weak U$(1)$ symmetry. We first consider the following quantity 
\begin{equation}
	\langle[\rho_{-\bm{k}},\mathcal{L}^{\dagger}(\rho_{\bm{k}})]\rangle=\text{Tr}[\rho(t)[\rho_{-\bm{k}},\mathcal{L}^{\dagger}(\rho_{\bm{k}})]],
\end{equation}
where $\rho(t)$ is the density matrix of the system at time $t$ which obeys the Lindblad master equation and
\begin{eqnarray}
	\rho_{\bm{k}}&:=&\sum_{\bm{p}}a_{\bm{p}}^{\dagger}a_{\bm{p}+\bm{k}},\nonumber \\ \mathcal{L}^{\dagger}(O)&:=&-i[O,H]-\frac{\gamma}{2}\int d^3\bm{r}\{O,L_{\bm{r}}^{\dagger}L_{\bm{r}}\}+\gamma\int d^3\bm{r}L_{\bm{r}}^{\dagger}OL_{\bm{r}}.
\end{eqnarray}
Since the Hamiltonian is given by
\begin{eqnarray}
	H & = & \sum_{\bm{k}}\varepsilon_{\bm{k}}a_{\bm{k}}^{\dagger}a_{\bm{k}}+\frac{U_{R}}{2}\int d^3\bm{r}(a_{\bm{r}}^{\dagger})^{2}(a_{\bm{r}})^{2}\nonumber \\
	& = & \sum_{\bm{k}}\varepsilon_{\bm{k}}a_{\bm{k}}^{\dagger}a_{\bm{k}}+\frac{U_{R}}{2V}\sum_{\bm{k},\bm{q},\bm{p}}a_{\bm{k}}^{\dagger}a_{\bm{p}}^{\dagger}a_{\bm{p}-\bm{q}}a_{\bm{k}+\bm{q}},
\end{eqnarray}
with $\varepsilon_{\bm{k}}=\bm{k}^2/2m$ being the kinetic energy, a straightforward calculation \citep{Ueda2010} gives
\begin{equation}
	i\left[{\rho_{\bm{k}}},H\right]=i\sum_{\bm{p}}(\varepsilon_{\bm{p}+\bm{k}}-\varepsilon_{\bm{p}})a_{\bm{p}}^{\dagger}a_{\bm{p}+\bm{k}},
\end{equation}
where the right-hand side originates from the kinetic-energy term in the Hamiltonian. The interaction terms do not contribute to the commutator since they only involve the density-density interaction. Then we calculate the following terms in $\mathcal{L}^{\dagger}(\rho_{\bm{k}})$.
We can see 
\begin{equation}
	-\frac{\gamma}{2}\int d^3\bm{r}\{O,L_{\bm{r}}^{\dagger}L_{\bm{r}}\}+\gamma\int d^3\bm{r}L_{\bm{r}}^{\dagger}OL_{\bm{r}}=-\frac{\gamma}{2}\int d^3\bm{r}[O,L_{\bm{r}}^{\dagger}]L_{\bm{r}}+\frac{\gamma}{2}\int d^3\bm{r}L_{\bm{r}}^{\dagger}[O,L_{\bm{r}}].
\end{equation}
Substituting $L_{\bm{r}}=a_{\bm{r}}^{2}$ and performing the Fourier transformation,
we obtain 
\begin{eqnarray}
	&  & -\frac{\gamma}{2}\sum_{\bm{p},\bm{q},\bm{l}}[\rho_{\bm{k}},a_{\bm{p}+\bm{l}}^{\dagger}a_{\bm{q}-\bm{l}}^{\dagger}]a_{\bm{q}}a_{\bm{p}}+\frac{\gamma}{2}\sum_{\bm{p},\bm{q},\bm{l}}a_{\bm{p}+\bm{l}}^{\dagger}a_{\bm{q}-\bm{l}}^{\dagger}[\rho_{\bm{k}},a_{\bm{q}}a_{\bm{p}}]\nonumber \\
	& = & -\gamma\sum_{\bm{p},\bm{q},\bm{l}}(a_{\bm{l}}^{\dagger}a_{\bm{p}}^{\dagger}a_{\bm{l}+\bm{k}-\bm{q}}a_{\bm{p}+\bm{q}}+a_{\bm{l}}^{\dagger}a_{\bm{p}}^{\dagger}a_{\bm{p}-\bm{q}}{a_{\bm{l}+\bm{k}+\bm{q}}})\nonumber \\
	& = & -2\gamma\left(\sum_{\bm{q}}\rho_{\bm{k}-\bm{q}}\rho_{\bm{q}}-\sum_{\bm{q}}\rho_{\bm{k}}\right).
\end{eqnarray}
By utilizing the fact that $[\rho_{\bm{k}},\rho_{\bm{p}}]=0$ for
arbitrary $\bm{k}$ and $\bm{p}$, we have 
\begin{eqnarray}
	[\rho_{-\bm{k}},\mathcal{L}^{\dagger}(\rho_{\bm{k}})] & = & \left[\rho_{-\bm{k}},-i\sum_{\bm{p}}(\varepsilon_{\bm{p}+\bm{k}}-\varepsilon_{\bm{p}})a_{\bm{p}}^{\dagger}a_{\bm{p}+\bm{k}}-2\gamma\left(\sum_{\bm{q}}\rho_{\bm{k}-\bm{q}}\rho_{\bm{q}}-\sum_{\bm{q}}\rho_{\bm{k}}\right)\right]\nonumber \\
	& = & i\sum_{\bm{p}}(\varepsilon_{\bm{p}+\bm{k}}+\varepsilon_{\bm{p}-\bm{k}}-2\varepsilon_{\bm{p}})a_{\bm{p}}^{\dagger}a_{\bm{p}}\nonumber \\
	& = &  2i\varepsilon_{\bm{k}}\hat{N}, \label{eq: f-sum-1}
\end{eqnarray}
where $\hat{N}$ is the total particle-number operator. Therefore, 
\begin{equation}
	\langle[\rho_{-\bm{k}},\mathcal{L}^{\dagger}(\rho_{\bm{k}})]\rangle=2i\varepsilon_{\bm{k}}N, \label{eq:commutation}
\end{equation}
where $N$ is the number of particles corresponding to the current state. We can see that dissipation does not change the expression of $[\rho_{-\bm{k}},\mathcal{L}^{\dagger}(\rho_{\bm{k}})]$ since the two-body loss only causes a uniform decrease in the density and hence does not generate current inside the system. Actually, this commutation relation originates from the weak U$(1)$ symmetry of the Lindblad equation. One can verify that other systems obeying weak U$(1)$ symmetry, like systems with $n$-body loss, also satisfy Eq. (\ref{eq:commutation}). Below, we show how to use the eigensystem of the Lindbladian to represent $\langle[\rho_{-\bm{k}},\mathcal{L}^{\dagger}(\rho_{\bm{k}})]\rangle$. To begin with, we define the right eigenoperators $\hat{r}_{\alpha}$
and the left eigenoperators $\hat{l}_{\alpha}$ of $\mathcal{L}^{\dagger}$
as \citep{Scarlatella2019} 
\begin{eqnarray}
	\mathcal{L}^{\dagger}(\hat{r}_{\alpha}) & = & \lambda_{\alpha}\hat{r}_{\alpha},\\
	\mathcal{L}(\hat{l}_{\alpha}) & = & \lambda_{\alpha}^{\ast}\hat{l}_{\alpha}.
\end{eqnarray}
The right and left eigenvectors satisfy the biorthogonal relation and the completeness relation:
\begin{equation}
	\text{Tr}(\hat{l}_{\alpha}^{\dagger}\hat{r}_{\beta})=\delta_{\alpha\beta},\sum_{\alpha}\hat{r}_{\alpha}\hat{l}_{\alpha}^{\dagger}=\mathbb{I}.
\end{equation}
In terms of the left and right eigenvectors, we can expand $\mathcal{L}^{\dagger}(\rho_{\bm{k}})$ as 
\begin{equation}
	\mathcal{L}^{\dagger}(\rho_{\bm{k}})=\sum_{\alpha}\lambda_{\alpha}\hat{r}_{\alpha}\text{Tr}(\hat{l}_{\alpha}^{\dagger}\rho_{\bm{k}}).
\end{equation}
Therefore, we obtain 
\begin{eqnarray}
	\langle[\rho_{-\bm{k}},\mathcal{L}^{\dagger}(\rho_{\bm{k}})]\rangle & = & \sum_{\alpha}\lambda_{\alpha}(\text{Tr}[\rho(t)\rho_{-\bm{k}}\hat{r}_{\alpha}]-\text{Tr}[\rho_{-\bm{k}}\rho(t)\hat{r}_{\alpha}])\text{Tr}(\hat{l}_{\alpha}^{\dagger}\rho_{\bm{k}})\nonumber \\
	& = & \sum_{\alpha}\lambda_{\alpha}\text{Tr}[[\rho(t),\rho_{-\bm{k}}]\hat{r}_{\alpha}]\text{Tr}(\hat{l}_{\alpha}^{\dagger}\rho_{\bm{k}}).
\end{eqnarray}
We further define a retarded Green's function as
\begin{align}
	\tilde{G}^{R}(\bm{k},t_{0},t):=-i\theta(t)\text{Tr}[\rho_{\bm{k}}e^{\mathcal{L}t}(e^{\mathcal{L}t_0}(\rho)\rho_{-\bm{k}})-\rho_{\bm{k}}e^{\mathcal{L}t}(\rho_{-\bm{k}}e^{\mathcal{L}t_0}(\rho))],
\end{align}
where $t_0$ is an arbitrary time. We can expand the Green's function as 
\begin{eqnarray}
	\tilde{G}^{R}(\bm{k},t_{0},t)
	& = & -i\theta(t)\sum_{\alpha}e^{\lambda_{\alpha}t}\text{Tr}\left\{[\rho(t_0),\rho_{-\bm{k}}]\hat{r}_{\alpha}\right\}\text{Tr}(\hat{l}_{\alpha}^{\dagger}\rho_{\bm{k}}).
\end{eqnarray}
Performing the Fourier transformation with respect to $t$, we obtain 
\begin{equation}
	\tilde{G}^{R}(\bm{k},t_{0},\omega)=\int dte^{i\omega t}\tilde{G}^{R}(\bm{k},t_{0},t)=\sum_{\alpha}\frac{1}{\omega-i\lambda_{\alpha}+i0^{+}}\text{Tr}[[\rho(t_0),\rho_{-\bm{k}}]\hat{r}_{\alpha}]\text{Tr}(\hat{l}_{\alpha}^{\dagger}\rho_{\bm{k}}).
\end{equation}
Integrating $\tilde{G}^R$ along a closed contour $C$ that contains all the poles $i\lambda_{\alpha}-i0^{+}$, we obtain
\begin{equation}
	\oint_{C}\frac{\omega d\omega}{2\pi}\tilde{G}^{R}(\bm{k},t_{0},\omega)=\sum_{\alpha}\lambda_{\alpha}\text{Tr}[[\rho(t_0),\rho_{-\bm{k}}]\hat{r}_{\alpha}]\text{Tr}(\hat{l}_{\alpha}^{\dagger}\rho_{\bm{k}})=2i\varepsilon_{\bm{k}}N(t_{0}).\label{eq:f}
\end{equation}
This is the $f$-sum rule in the open quantum system. We note that this $f$-sum rule holds even though the particle number is not conserved during the dynamics. This $f$-sum rule can be used to derive the relation
between the current-current correlation function and the number operator. To show this, we start from the dynamics of the local boson number $\langle\rho_{\bm{r}}(t)\rangle:=\text{Tr}[a_{\bm{r}}^{\dagger}a_{\bm{r}}\rho(t)]$ given by 
\begin{equation}
	\frac{\partial\langle\rho_{\bm{r}}(t)\rangle}{\partial t}=\text{Tr}[\mathcal{L}^{\dagger}\rho_{\bm{r}}\rho(t)]=i\langle[H,\rho_{\bm{r}}]\rangle+\frac{\gamma}{2}\int d^3\bm{r}\langle[2L_{\bm{r}}^{\dagger}\rho_{\bm{r}}L_{\bm{r}}-L_{\bm{r}}^{\dagger}L_{\bm{r}}\rho_{\bm{r}}-\rho_{\bm{r}}L_{\bm{r}}^{\dagger}L_{\bm{r}}]\rangle.\label{eq:continuous}
\end{equation}
After simplification, this dynamics can be expressed as 
\begin{equation}
	\frac{\partial\langle\rho_{\bm{r}}(t)\rangle}{\partial t}=-\nabla\cdot(\bm{j}_t):=-\nabla\cdot(\bm{j}_{c}+\bm{j}_{d}),
\end{equation}
where $\bm{j}_t:=\bm{j}_{c}+\bm{j}_{d}$ is the total current with
$\bm{j}_{c}$ being the current flow in the system and $\bm{j}_{d}$ being the current from the system to an environment induced by the two-body loss. They are given by 
\begin{eqnarray}
	\bm{j}_{c} & = & \frac{i}{2m}\langle[\nabla a^{\dagger}(\bm{r})a(\bm{r})-a^{\dagger}(\bm{r})\nabla a(\bm{r})]\rangle,
\end{eqnarray}
and $\nabla\cdot\bm{j}_d=2\gamma a^{\dagger}_{\bm{r}}a^{\dagger}_{\bm{r}}a_{\bm{r}}a_{\bm{r}}$, which is solved as
\begin{equation}
	\bm{j}_d(\bm{r})=2\gamma\int \frac{d\bm{r}'}{4\pi}\frac{\bm{r}-\bm{r}'}{|\bm{r}-\bm{r}'|^3} \langle a^{\dagger}_{\bm{r}'}a^{\dagger}_{\bm{r}'}a_{\bm{r}'}a_{\bm{r}'}\rangle.
\end{equation}
Fourier transforming Eq. \eqref{eq:continuous}, we obtain 
\begin{equation}
	\langle\mathcal{L}^{\dagger}(\rho_{\bm{k}}(\omega))\rangle=-i\omega\langle\rho_{\bm{k}}(\omega)\rangle=-i\bm{k}\cdot\bm{j}_t(\bm{k},\omega).
\end{equation}
It follows from Eq. \eqref{eq: f-sum-1} that
\begin{equation}\label{eq: f-sum-2}
	m\langle[\rho_{\bm{k}}(t),\mathcal{L}^{\dagger}(\rho_{-\bm{k}}(t))]\rangle=ik^{2}N(t),
\end{equation}
where the correlation function is defined on the basis of the Schwinger-Keldysh field theory as
\begin{equation}\label{eq:def}
	\langle[A(t),B(t))]\rangle:=\int D[a]D[a^{\dagger}](A_+(t)B_+(t)-A_-(t)B_-(t))e^{iS}=\text{Tr}[[A,B]e^{\mathcal{L}t}\rho].
\end{equation}
For convenience of the notations, we will use $\langle[A(t),B(t))]\rangle$ in the sense of Eq. \eqref{eq:def} in the following calculation. From the Fourier transformation of Eq. \eqref{eq: f-sum-2}, we have
\begin{equation}
	m\int dte^{i(\omega-\omega_1-\omega_{2})t}\frac{d\omega_{1}}{2\pi}\frac{d\omega_{2}}{2\pi}\langle[\rho_{\bm{k}}(\omega_{1}),\mathcal{L}^{\dagger}(\rho_{-\bm{k}}(\omega_{2}))]\rangle=ik^{2}N(\omega).\label{eq:f2}
\end{equation}
The left-hand side of Eq. \eqref{eq:f2} can be simplified as 
\begin{eqnarray}
	&  & m\int dt\frac{d\omega_{1}}{2\pi}\frac{d\omega_{2}}{2\pi}e^{i(\omega-\omega_{1}-\omega_{2})t}\langle[\rho_{\bm{k}}(\omega_{1}),\mathcal{L}^{\dagger}(\rho_{-\bm{k}}(\omega_{2}))]\rangle\nonumber \\
	& = & \frac{m}{2\pi}\int d\omega_{1}d\omega_{2}\delta(\omega-\omega_1-\omega_{2})\left\langle\left[\frac{-\bm{k}\cdot\bm{j}_t(\bm{k},\omega_{1})}{\omega_{1}},-i\bm{k}\cdot\bm{j}_t(-\bm{k},\omega_{2})\right]\right\rangle\nonumber \\
	& = & ik_{i}k_{j}m\int\frac{d\omega_{1}}{2\pi\omega_{1}}\langle[j_t^{i}(\bm{k},\omega_{1}),j_t^{j}(-\bm{k},\omega-\omega_{1})]\rangle,
\end{eqnarray}
and hence Eq. \eqref{eq: f-sum-2} can be rewritten as 
\begin{equation}
	N(\omega)=\frac{k_{i}k_{j}}{k^{2}}m\int\frac{d\omega_{1}}{2\pi\omega_{1}}\langle[j_t^{i}(\bm{k},\omega_{1}),j_t^{j}(-\bm{k},\omega-\omega_{1})]\rangle.\label{eq:N_omega}
\end{equation}
Furthermore, we define the total current-current correlation function as 
\begin{eqnarray}
	\gamma^{i,j}(\bm{k},\omega,t_{0}) & = & m\int dte^{i\omega t}\langle[j_t^{i}(\bm{k},t+t_{0}),j_t^{j}(-\bm{k},t_{0})]\rangle\nonumber \\
	& = & m\int dte^{i\omega(t+t_{0})}\langle[j_t^{i}(\bm{k},t+t_{0}),j_t^{j}(-\bm{k},t_{0})]\rangle e^{-i\omega t_{0}}\nonumber \\
	& = & m\langle[j_t^{i}(\bm{k},\omega),j_t^{j}(-\bm{k},t_{0})]\rangle e^{-i\omega t_{0}}. \label{eq:total_cc_corr}
\end{eqnarray}
By substituting this into \eqref{eq:N_omega}, we have 
\begin{eqnarray}
	N(t_{0}) & = & \frac{k_{i}k_{j}}{k^{2}}m\int\frac{d\omega_{1}}{2\pi\omega_{1}}\int\frac{d\omega}{2\pi}e^{-i\omega t_{0}}\langle[j_t^{i}(\bm{k},\omega_{1}),j_t^{j}(-\bm{k},\omega-\omega_{1})]\rangle\nonumber \\
	& = & \frac{k_{i}k_{j}}{k^{2}}m\int\frac{d\omega_{1}}{2\pi\omega_{1}}\int\frac{d\omega}{2\pi}e^{-i\omega_{1}t_{0}}\langle[j_t^{i}(\bm{k},\omega_{1}),j_t^{j}(-\bm{k},\omega-\omega_{1})]\rangle e^{-i(\omega-\omega_{1})t_{0}}\nonumber \\
	& = & m\frac{k_{i}k_{j}}{k^{2}}\int\frac{d\omega_{1}}{2\pi\omega_{1}}e^{-i\omega_{1}t_{0}}\langle[j_t^{i}(\bm{k},\omega_{1}),j_t^{j}(-\bm{k},t_{0})]\rangle\nonumber \\
	& = & \frac{k_{i}k_{j}}{k^{2}}\int\frac{d\omega_{1}}{2\pi\omega_{1}}\gamma^{i,j}(\bm{k},\omega_{1},t_{0}).
\end{eqnarray}
Since the longitudinal component of the correlation function is given
by $\gamma^{L}(\bm{k},\omega)=\frac{k_{i}k_{j}}{k^{2}}\gamma^{i,j}(\bm{k},\omega)$,
we can also express the $f$-sum rule as 
\begin{equation}
	\int\frac{d\omega}{2\pi\omega}\gamma^{L}(\bm{k},\omega,t_0)=N(t_0).\label{eq:total}
\end{equation}

In the following discussion of this section, we omit the subscript $c$ for the current operator $\bm{j}_{c}$ for simplicity.
Let us examine the normal fluid density, which is determined by the current response under an external perturbation $H\to H-\bm{u}\cdot\bm{j}$, where $\bm{u}$ is an external velocity field that represents the effect of a wall moving with velocity $\bm{u}$~\cite{Wen2007}. The current response is given by
\begin{equation}
	\langle J_{i}(t)\rangle=\rho_{n}^{i,j}u_{j},
\end{equation}
where we define the averaged current density as
\begin{equation}
	\langle J_{i}(t)\rangle=\frac{1}{V}\int d^{3}\bm{r}\langle j_{i}(\bm{r},t)\rangle.
\end{equation}
Since the two-body loss is uniform in space, we assume that the dissipative current does not influence the averaged current density. The only contribution originates from the closed current $\bm{j}$. It follows from the Lindbladian dynamics described by Eq. \eqref{eq:continuous} that the linear-response current density obeys
\begin{eqnarray}
	\frac{d}{dt}\langle J_{i}(t)\rangle & = & \frac{1}{V}\int d^{3}\bm{r}\frac{d}{dt}\langle j_{i}(\bm{r},t)\rangle\nonumber \\
	& = & \frac{1}{V}\int d^{3}\bm{r}\left\langle -i[j_{i},H]-\frac{\gamma}{2}\int d^3{\bm{r'}}\{j_{i},L_{\bm{r'}}^{\dagger}L_{\bm{r'}}\}+\gamma\int d^3\bm{r'}L_{\bm{r'}}^{\dagger}j_{i}L_{\bm{r'}}\right\rangle \nonumber \\
	& = & \frac{i}{V}m\int d^{3}\bm{r}\int d^{3}\bm{r'}\langle[j_{i}(\bm{r},t),j_{j}(\bm{r}',t)]\rangle u_{j}.
\end{eqnarray}
The averaged current density is thus given by 
\begin{eqnarray}
	\langle J_{i}(t)\rangle & = & \frac{i}{V}m\int d^{3}\bm{r}\int d^{3}\bm{r'}\int^{t}dt'\langle[j_{i}(\bm{r},t),j_{j}(\bm{r}',t')]\rangle u_{j}\nonumber \\
	& = & \frac{i}{V}m\int\frac{d\omega_{2}}{2\pi}\int d^{3}\bm{r}\int d^{3}\bm{r'}\int^{t}dt'e^{-i\omega_{2}t'}\langle[j_{i}(\bm{r},t),j_{j}(\bm{r}',\omega_{2})]\rangle u_{j}\nonumber \\
	& = & \frac{m}{V}\int\frac{d\omega_{2}}{2\pi\omega_{2}}\int d^{3}\bm{r}\int d^{3}\bm{r'}e^{-i\omega_{2}t}\langle[j_{i}(\bm{r},t),j_{j}(\bm{r}',\omega_{2})]\rangle u_{j},
\end{eqnarray}
where we cancel the constant term with the initial condition. Then the normal fluid density tensor is given by 
\begin{eqnarray}
	\rho_{n}^{i,j}(t) & = & \frac{m}{V^2}\int\frac{d\omega_{2}}{2\pi\omega_{2}}\int d^{3}\bm{r}\int d^{3}\bm{r'}e^{-i\omega_{2}t}\langle[j_{i}(\bm{r},t),j_{j}(\bm{r}',\omega_{2})]\rangle\nonumber \\
	& = & \lim_{\bm{k}\rightarrow0}m\int\frac{d\omega_{2}}{2\pi\omega_{2}}e^{-i\omega_{2}t}\langle[j_{i}(\bm{k},t),j_{j}(\bm{-k},\omega_{2})]\rangle\nonumber \\
	& = & \lim_{\bm{k}\rightarrow0}\int\frac{d\omega}{2\pi\omega}\tilde{\gamma}^{i,j}(\bm{k},\omega,t).
\end{eqnarray}
Here we define the current-current correlation function $\tilde{\gamma}$ for the current operator $\bm{j}_c$. To connect this with the tensor $\gamma^{i,j}$ in Eq. (\ref{eq:total_cc_corr}),
we first note that the dissipative current does not contribute to the correlation function since it is isotropic, i.e., $\int d^3\bm{r}\bm{j}_d(\bm{r})=0$. Hence, we can directly replace $\tilde{\gamma}$ with the total current-current correlator $\gamma$ in Eq. (\ref{eq:total_cc_corr}) as 
\begin{equation}
	\rho_{n}^{i,j}(t)=\lim_{\bm{k}\rightarrow0}\int\frac{d\omega}{2\pi\omega}\gamma^{i,j}(\bm{k},\omega,t).
\end{equation}
According to an analysis similar to the one applicable to a closed quantum system, we find that the normal fluid density corresponds to the transverse component of the
total current-current correlation function \cite{Ueda2010}. Together with Eq. \eqref{eq:total},
we can define the superfluid density as the difference between the
transverse and the longitudinal parts of the correlation function
tensor. We note that the derivation of the $f$-sum rule involves no approximation and can be considered as a general formula to calculate the normal fluid density and the superfluid density for an arbitrary interaction and dissipation. Here the $f$-sum rule is a consequence of the weak U$(1)$ symmetry of the system since under the weak $U(1)$ symmetry, the density matrix remains diagonal in the particle-number basis during time evolution. Therefore, the right-hand side of Eq. \eqref{eq:commutation} is only determined by the number of bosons.

\section{Derivation of the Quantum Depletion Density}

In the presence of dissipation, it is highly nontrivial whether quantum depletion contributes only to the superfluid or to the normal fluid as well. It is therefore important to carefully assess their individual contributions. In this section, we will address this issue in the context of open quantum systems.

We introduce some new concepts in the dissipative quantum many-body systems. Here we list them in the table 
below.
\begin{center}
	\begin{tabular}{|c|c|c|} \hline
		Concept & Definition & Corresponding equation \\ \hline
		Superfluid quantum depletion density $n_{sD}$& \makecell{The density of the quantum depletion part\\ engaging in the superfluid transport}& Eq. \eqref{eq: superfluid_2} \\ \hline
		Normal quantum depletion density $n_{nD}$& \makecell{The density of the quantum depletion part\\ engaging in the normal transport} & Eq. \eqref{eq: normalfluid_2} \\ \hline
	\end{tabular}
\end{center}
\subsection{The Case without a Dipolar Interaction}
In this section, we investigate the density of the quantum depletion, which is the density of the noncondensate part at absolute zero, of a dissipative molecular BEC. 
We begin by constructing the mean-field Lindbladian action. The Hamiltonian of a weakly interacting bosonic system is given by
\begin{eqnarray}
	H & = & \sum_{\bm{k}}\varepsilon_{\bm{k}}a_{\bm{k}}^{\dagger}a_{\bm{k}}+\frac{U_{R}}{2}\int d^3\bm{r}(a_{\bm{r}}^{\dagger})^{2}(a_{\bm{r}})^{2}\nonumber \\
	& = & \sum_{\bm{k}}\varepsilon_{\bm{k}}a_{\bm{k}}^{\dagger}a_{\bm{k}}+\frac{U_{R}}{2V}\sum_{\bm{k},\bm{q},\bm{p}}a_{\bm{k}}^{\dagger}a_{\bm{p}}^{\dagger}a_{\bm{p}-\bm{q}}a_{\bm{k}+\bm{q}},\label{Hamiltonian}
\end{eqnarray}
where $V$ is the volume of the system. For simplicity of discussions, we consider only a contact interaction here, and we will consider the dipolar-dipole interaction later. 
In the presence of dissipation, the dynamics of the system is described by the Lindblad equation: 
\begin{equation}
	\frac{d\rho}{dt}=\mathcal{L}\rho=-i[H,\rho]-\frac{\gamma}{2}\int d^3\bm{r}(\{L_{\bm{r}}^{\dagger}L_{\bm{r}},\rho\}-2L_{\bm{r}}\rho L_{\bm{r}}^{\dagger}).\label{Lindbland}
\end{equation}
Here we take the Lindblad operator for two-body loss as $L_{\bm{r}}=a_{\bm{r}}^{2}$.
We now apply the closed-time-contour path integral formalism to the Lindblad equation. The action is defined on the Schwinger-Keldysh contour \citep{Sieberer_2016} as 
\begin{equation}
	S=\int_{-\infty}^{\infty}dt\left[\sum_{\bm{k}}(a_{\bm{k}+}^{\dagger}i\partial_{t}a_{\bm{k}+}-a_{\bm{k}-}^{\dagger}i\partial_{t}a_{\bm{k}-})-H_{+}+H_{-}+\frac{i\gamma}{2}\int d^3\bm{r}(L_{\bm{r}+}^{\dagger}L_{\bm{r}+}+L_{\bm{r}-}^{\dagger}L_{\bm{r}-}-2L_{\bm{r}+}L_{\bm{r}-}^{\dagger})\right],
\end{equation}
where
\begin{equation}
	H_{\alpha}=\sum_{\bm{k}}\varepsilon_{\bm{k}}a_{\bm{k}\alpha}^{\dagger}a_{\bm{k}\alpha}+\frac{U_{R}}{2V}\sum_{\bm{k},\bm{q},\bm{p}}a_{\bm{k}\alpha}^{\dagger}a_{\bm{p}\alpha}^{\dagger}a_{\bm{p}-\bm{q},\alpha}a_{\bm{k}+\bm{q},\alpha}.
\end{equation} 
By applying Fourier transformation to the dissipation part, the Schwinger-Keldysh action is given by
\begin{eqnarray}
	S & = & \int_{-\infty}^{\infty}dt\Bigg[\sum_{\bm{k}}(a_{\bm{k}+}^{\dagger}(i\partial_{t}-\varepsilon_{\bm{k}})a_{\bm{k}+}-a_{\bm{k}-}^{\dagger}(i\partial_{t}-\varepsilon_{\bm{k}})a_{\bm{k}-})-\frac{U}{2V}\sum_{\bm{k},\bm{q},\bm{p}}a_{\bm{k}+}^{\dagger}a_{\bm{p}+}^{\dagger}a_{\bm{p}-\bm{q},+}a_{\bm{k}+\bm{q},+}\nonumber \\
	&  & +\frac{U^{\ast}}{2V}\sum_{\bm{k},\bm{q},\bm{p}}a_{\bm{k}-}^{\dagger}a_{\bm{p}-}^{\dagger}a_{\bm{p}-\bm{q},-}a_{\bm{k}+\bm{q},-}-i\frac{\gamma}{V}\sum_{\bm{k},\bm{q},\bm{p}}a_{\bm{k}-}^{\dagger}a_{\bm{p}-}^{\dagger}a_{\bm{p}-\bm{q},+}a_{\bm{k}+\bm{q},+}\Bigg],
\end{eqnarray}
where $U=U_{R}-i\gamma$. We employ the mean-field approximation to separate
the operators into the condensate part and noncondensate part. Assuming that most bosons in the system form a condensate, we have
\begin{equation}
	a_{0}^{\dagger}a_{0}\approx N,\sum_{\bm{k},\bm{k}\neq0}a_{\bm{k}}^{\dagger}a_{\bm{k}}\ll N.
\end{equation}
Here $N$ is the total number of bosons in the system. Since $N\gg1$,
we can replace the creation and annihilation operators of the condensate by c-numbers: 
\begin{equation}
	a_{0+}^{\dagger}\approx\sqrt{N}e^{i\theta_{+}},\ a_{0+}\approx\sqrt{N}e^{-i\theta_{+}},\ a_{0-}^{\dagger}\approx\sqrt{N}e^{i\theta_{-}},\ a_{0-}\approx\sqrt{N}e^{-i\theta_{-}}.
\end{equation}
From Eqs. \eqref{action-superfluid} and \eqref{eq:super_current_action}, we find that the difference between $\theta_{+}$ and $\theta_{-}$ only contributes to the dissipative current, i.e., loss of particles from the system to an environment. To calculate the response current from external perturbations, we here take $\theta_{+}=\theta_{-}=0$ for simplicity. Then the interaction terms become
\begin{equation}
	\sum_{\bm{k},\bm{q},\bm{p}}a_{\bm{k}+}^{\dagger}a_{\bm{p}+}^{\dagger}a_{\bm{p}-\bm{q},+}a_{\bm{k}+\bm{q},+}\approx N^{2}+N\sum_{\bm{k},\bm{k}\neq0}a_{-\bm{k},+}a_{\bm{k}+}+N\sum_{\bm{k},\bm{k}\neq0}a_{\bm{k}+}^{\dagger}a_{-\bm{k},+}^{\dagger}+4N\sum_{\bm{k},\bm{k}\neq0}a_{\bm{k}+}^{\dagger}a_{\bm{k}+},\label{mean1}
\end{equation}

\begin{equation}
	\sum_{\bm{k},\bm{q},\bm{p}}a_{\bm{k}-}^{\dagger}a_{\bm{p}-}^{\dagger}a_{\bm{p}-\bm{q},-}a_{\bm{k}+\bm{q},-}\approx N^{2}+N\sum_{\bm{k},\bm{k}\neq0}a_{-\bm{k},-}a_{\bm{k}-}+N\sum_{\bm{k},\bm{k}\neq0}a_{\bm{k}-}^{\dagger}a_{-\bm{k},-}^{\dagger}+4N\sum_{\bm{k},\bm{k}\neq0}a_{\bm{k}-}^{\dagger}a_{\bm{k}-},
\end{equation}
\begin{equation}
	\sum_{\bm{k},\bm{q},\bm{p}}a_{\bm{k}-}^{\dagger}a_{\bm{p}-}^{\dagger}a_{\bm{p}-\bm{q},+}a_{\bm{k}+\bm{q},+}\approx N^{2}+N\sum_{\bm{k},\bm{k}\neq0}a_{-\bm{k},+}a_{\bm{k}+}+N\sum_{\bm{k},\bm{k}\neq0}a_{\bm{k}-}^{\dagger}a_{-\bm{k},-}^{\dagger}+4N\sum_{\bm{k},\bm{k}\neq0}a_{\bm{k}-}^{\dagger}a_{\bm{k}+},\label{mean3}
\end{equation}
and the action can be simplified as
\begin{eqnarray}
	S & = & \int_{-\infty}^{\infty}dt\Bigg[\sum_{\bm{k}}(a_{\bm{k}+}^{\dagger}(i\partial_{t}-\varepsilon_{\bm{k}})a_{\bm{k}+}-a_{\bm{k}-}^{\dagger}(i\partial_{t}-\varepsilon_{\bm{k}})a_{\bm{k}-})-\frac{U_{R}}{2V}a_{0+}^{\dagger}a_{0+}^{\dagger}a_{0,+}a_{0,+}+\frac{U_{R}}{2V}a_{0-}^{\dagger}a_{0-}^{\dagger}a_{0,-}a_{0,-}\nonumber \\
	&  & -\frac{U^{\ast}n}{2}\sum_{\bm{k},\bm{k}\neq0}a_{-\bm{k},+}a_{\bm{k}+}-\frac{Un}{2}\sum_{\bm{k},\bm{k}\neq0}a_{\bm{k}+}^{\dagger}a_{-\bm{k},+}^{\dagger}-2Un\sum_{\bm{k},\bm{k}\neq0}a_{\bm{k}+}^{\dagger}a_{\bm{k}+}+\frac{U^{\ast}n}{2}\sum_{\bm{k},\bm{k}\neq0}a_{-\bm{k},-}a_{\bm{k}-}\nonumber \\
	&  & +\frac{Un}{2}\sum_{\bm{k},\bm{k}\neq0}a_{\bm{k}-}^{\dagger}a_{-\bm{k},-}^{\dagger}+2U^{\ast}n\sum_{\bm{k},\bm{k}\neq0}a_{\bm{k}-}^{\dagger}a_{\bm{k}-}-4i\gamma n\sum_{\bm{k},\bm{k}\neq0}a_{\bm{k}-}^{\dagger}a_{\bm{k}+}\Bigg]\\
	& = & \int_{-\infty}^{\infty}dt\left[\sum_{\bm{k}}(a_{\bm{k}+}^{\dagger}i\partial_{t}a_{\bm{k}+}-a_{\bm{k}-}^{\dagger}i\partial_{t}a_{\bm{k}-})-H_{+}+H_{-}-4i\gamma n\sum_{\bm{k},\bm{k}\neq0}a_{\bm{k}-}^{\dagger}a_{\bm{k}+}\right],\label{action}
\end{eqnarray}
where $n=N/V$ and
\begin{eqnarray}
	H_{\alpha} 
	& = & \frac{U_{R}n}{2}N+\sum_{\bm{k},\bm{k}\neq0}\left[(\varepsilon_{\bm{k}}+U_{R}n-2i\alpha\gamma n)a_{\bm{k}\alpha}^{\dagger}a_{\bm{k}\alpha}+\frac{U^{\ast}n}{2}a_{-\bm{k}\alpha}a_{\bm{k}\alpha}+\frac{Un}{2}a_{\bm{k}\alpha}^{\dagger}a_{-\bm{k}\alpha}^{\dagger}\right].\label{H}
\end{eqnarray}

We then move on to evaluate the quantum depletion density on the basis of the action \eqref{action}. Since the particle-number density $n(t)$ of the system decays with time and the quantum depletion part is also time-dependent, we consider the weak-dissipation case in which the system can reach a quasi-steady state in a time scale shorter than the inverse two-body loss rate. We first calculate the superfluid quantum depletion density $n_{sD}$, which is defined as the density of the quantum depletion part engaged in the superfluid transport. We first place bosons into a cylinder moving with velocity $\bm{v}$. Then the superfluid quantum depletion density $n_{sD}$ is defined from
\begin{equation}\label{eq: superfluid-current}
	\bm{j}_{sD}=mn_{sD}\bm{v},
\end{equation}
where $\bm{j}_{sD}$ is the mass current of bosons in the quantum depletion in the frame of reference of the cylinder. To obtain the superfluid quantum depletion density, we change the action \eqref{action} by $\bm{k}\to\bm{k}-m\bm{v}$ for the forward contour and $\bm{k}\to\bm{k}+m\bm{v}$ for the backward contour for the finite-momentum sector with $\bm{k}\neq0$~\cite{Wen2007,Mathematical_method_SF_1968} and the density of the superfluid quantum depletion part can be shown to be
\begin{equation}\label{eq: superfluid_2}
	n_{sD}=\frac{1}{m}\frac{\partial j_{sD}^{\alpha}}{\partial v^{\alpha}}.
\end{equation}

We next calculate the normal quantum depletion density $n_{nD}$, which is defined as the density of the quantum depletion part engaged into the normal transport. In this case, we change the Hamiltonians in Eq. \eqref{action} by $H_{+}\to H_{+}-\bm{v}\cdot\bm{j}_{+}$ and $H_{-}\to H_{-}+\bm{v}\cdot\bm{j}_{-}$. Then the response normal current for the quantum depletion part is given by
\begin{equation}
	\bm{j}_{nD}=mn_{nD}\bm{v}.
\end{equation}
and the density of the normal-fluid quantum depletion can be shown as
\begin{equation}\label{eq: normalfluid_2}
	n_{nD}=\frac{1}{m}\frac{\partial j_{nD}^{\alpha}}{\partial v^{\alpha}}.
\end{equation}
The relation between the superfluid current and the normal current is given by
\begin{equation}
	\bm{j}_{sD}=\bm{j}_D-\bm{j}_{nD},\ \bm{j}_D=\frac{m}{2V}\sum_{\bm{k}\neq 0}\langle a_{\bm{k}+}^{\dagger}a_{\bm{k}+}+a_{\bm{k}-}^{\dagger}a_{\bm{k}-}\rangle\bm{v}=mn_D\bm{v},
\end{equation}
where $n_D=n_{sD}+n_{nD}$ is the quantum depletion density. 
We begin by calculating the superfluid quantum depletion density. We introduce the perturbation $\bm{k}\rightarrow\bm{k}-m\bm{v}$ for the forward contour and $\bm{k}\rightarrow\bm{k}+m\bm{v}$ for the backward contour for the finite-momentum sector. Then the perturbed Schwinger-Keldysh action \eqref{action} is given by 
\begin{equation}
	S[\bm{v}]:=S-\bm{v}\cdot(\bm{j}_{+}[\bm{v}/2]+\bm{j}_{-}[\bm{v}/2]),
\end{equation}
where $\bm{j}_{\alpha}[\bm{v}]:=\bm{j}_{\alpha}- m\bm{v}n_{\alpha}$ representing the total current of the bosons relative to a reference frame with a velocity $\bm{v}$ and $n_{\alpha}:=\sum_{\bm{k},\bm{k}\neq0}a^{\dagger}_{\bm{k}\alpha}a_{\bm{k}\alpha}$. We extend the phase stiffness \cite{Coleman_2015} to nonequilibrium quasi-steady states as 
\begin{equation}
	Q_{ab}=-\frac{1}{2V}\frac{\partial^{2}F}{\partial v_{a}\partial v_{b}}\Big|_{v=0},\label{Qv}
\end{equation}
where we define $F[\bm{v}]$ as 
\begin{equation}
	F[\bm{v}]:=-i\log Z[\bm{v}],Z[\bm{v}]:= \int D[a_{\bm{k}+}(t),a_{-\bm{k}+}^{\dagger}(t),a_{\bm{k}-}(t),a_{-\bm{k},-}^{\dagger}(t)]e^{iS[\bm{v}]}.
\end{equation}
Then the response superfluid current is given by
\begin{equation}
	j_{sD,a}[\bm{v}]:=\langle j_{a}[\bm{v}]\rangle=-\frac{1}{2V}\frac{\partial F}{\partial v_{a}}
\end{equation}
since 
\begin{equation}
	-\frac{1}{V}\frac{\partial F}{\partial\bm{v}}=\frac{i}{V}\frac{1}{Z}\frac{\partial Z}{\partial\bm{v}}=\frac{1}{V}\frac{1}{Z}\int D\varphi e^{i[S(\varphi)-\bm{v}\cdot(\bm{j}_{+}+\bm{j}_{-})]}(\bm{j}_{+}[\bm{v}]+\bm{j}_{-}[\bm{v}])=\langle\hat{\bm{j}}_{+}[\bm{v}]+\hat{\bm{j}}_{-}[\bm{v}]\rangle=2\langle\hat{\bm{j}}[\bm{v}]\rangle,
\end{equation}
where we use $\bm{\hat{j}}=(\hat{\bm{j}}_{+}+\hat{\bm{j}}_{-})/2$
for the current operator and $\langle\bm{j}[\bm{v}]\rangle$ is the current in the frame of reference of the cylinder. From Eqs. \eqref{eq: superfluid-current} and \eqref{eq: superfluid_2}, we can see that the diagonal elements of the phase stiffness matrix $Q_{ab}$ is related to the superfluid quantum depletion density as 
\begin{equation}\label{eq: relation}
	Q_{ab}=mn_{sD}\delta_{ab}.
\end{equation}

Adding the perturbation $\bm{v}$ to the system, we find that the functional $Z[\bm{v}]$ is given by 
\begin{eqnarray}
	Z[\bm{v}] & = & \int D[a_{\bm{k}+}(t),a_{-\bm{k}+}^{\dagger}(t),a_{\bm{k}-}(t),a_{-\bm{k},-}^{\dagger}(t)]e^{iS[\bm{v}]}\nonumber \\
	& = & \int D[a_{\bm{k}+}(\omega),a_{-\bm{k}+}^{\dagger}(\omega),a_{\bm{k}-}(\omega),a_{-\bm{k},-}^{\dagger}(\omega)]e^{\frac{i}{2}\Psi_{\bm{k}}^{\dagger}(\omega)G^{-1}(\bm{k},\omega)\Psi_{\bm{k}}(\omega)},
\end{eqnarray}
where we define 
\begin{equation}
	\Psi_{\bm{k}}(\omega):=(a_{\bm{k}+}(\omega),a_{-\bm{k}+}^{\dagger}(\omega),a_{\bm{k}-}(\omega),a_{-\bm{k},-}^{\dagger}(\omega))^{T}
\end{equation}
and  
\begin{equation}
	G(\bm{k},\omega):=\left(\begin{array}{cccc}
		a_{1} & b & 0 & 0\\
		b^{\ast} & a_{2} & 0 & c\\
		c & 0 & -a_{1}^{\ast} & -b\\
		0 & 0 & -b^{\ast} & -a_{2}^{\ast}
	\end{array}\right)^{-1},\label{Green2}
\end{equation}
with $a_{1}=-(\varepsilon_{\bm{k}-m\bm{v}}+U_{R}n-2i\gamma n-\omega),b=-Un,a_{2}=-(\varepsilon_{\bm{k}+m\bm{v}}+U_{R}n-2i\gamma n+\omega),$ and $c=-4i\gamma n$.
In this case, we integrate all the bosonic degrees of freedom and
obtain the effective action: 
\begin{equation}
	Z=e^{iS_{\text{eff}}}=e^{iF},
\end{equation}
which leads to the form of the generating functional $F$ as 
\begin{equation}
	F=-i\sum_{\bm{k}}\text{Tr}\log[iG(\bm{k},\omega)].
\end{equation}
Next we calculate
(\ref{Qv}). The first derivative of $F[\bm{v}]$ gives the expression for the bosonic current:
\begin{equation}
	\langle j_{a}[\bm{v}]\rangle=\frac{-1}{2mV}\frac{\partial F}{\partial v_{a}}=-i\frac{1}{2V}\sum_{\bm{k},\bm{k}\neq0}\text{Tr}[(\sigma_{z}\otimes\sigma_{z})\nabla_{a}\mathcal{M}_{\bm{k}}G(\bm{k},\omega)],
\end{equation}
where
\begin{equation}
	\mathcal{M}_{\bm{k}}:=\left(\begin{array}{cccc}
		\varepsilon_{\bm{k}-m\bm{v}} & 0 & 0 & 0\\
		0 & \varepsilon_{\bm{k}+m\bm{v}}  & 0 & 0\\
		0 & 0 & \varepsilon_{\bm{k}-m\bm{v}}  & 0\\
		0 & 0 & 0 & \varepsilon_{\bm{k}+m\bm{v}} 
	\end{array}\right).
\end{equation}
In this case, the phase stiffness is given by
\begin{eqnarray}
	Q_{ab} & = & \frac{-1}{2V}\frac{\partial^{2}F}{\partial v_{a}\partial v_{b}}\bigg|_{v=0}\nonumber \\
	& = & -i\frac{m^{2}}{2V}\sum_{\bm{k},\bm{k}\neq0}\text{Tr}[(\sigma_{z}\otimes\sigma_{0})\nabla_{a}\nabla_{b}\varepsilon_{\bm{k}}G(\bm{k},\omega)]+i\frac{m^{2}}{2V}\sum_{\bm{k},\bm{k}\neq0}\nabla_{a}\varepsilon_{\bm{k}}\nabla_{b}\varepsilon_{\bm{k}}\text{Tr}[(\sigma_{z}\otimes\sigma_{z})G(\sigma_{z}\otimes\sigma_{z})G]. \label{eq:total_of_Q}
\end{eqnarray}
Here we apply the equality $\nabla_{b}G=-G\nabla_{b}G^{-1}G$. By integrating the first term by parts, we obtain 
\begin{equation}
	Q_{ab}=-i\frac{m^{2}}{2V}\sum_{\bm{k},\bm{k}\neq0}\nabla_{a}\varepsilon_{\bm{k}}\nabla_{b}\varepsilon_{\bm{k}}\left\{ \text{Tr}[(\sigma_{z}\otimes\sigma_{z})G(\sigma_{z}\otimes\sigma_{z})G]-\text{Tr}[(\sigma_{z}\otimes\sigma_{0})G(\sigma_{z}\otimes\sigma_{0})G]\right\} .\label{Qab}
\end{equation}
Here we transform the trace $\text{Tr}[AGAG]$ as
\begin{eqnarray}
	\text{Tr}[AGAG] & = & \text{Tr}[A[G,A]G]+\text{Tr}[A^{2}G^{2}]\nonumber \\
	& = & \text{Tr}[[G,A][G,A]]+\text{Tr}[[G,A]AG]+\text{Tr}[A^{2}G^{2}]\nonumber \\
	& = & \text{Tr}[[G,A][G,A]]+2\text{Tr}[A^{2}G^{2}]-\text{Tr}[AGAG],
\end{eqnarray}
which yields
\begin{equation}\label{eq: AGAG}
	\text{Tr}[AGAG]=\text{Tr}[A^{2}G^{2}]+\frac{1}{2}\text{Tr}[[G,A]^{2}].
\end{equation}
We apply Eq. \eqref{eq: AGAG} to the right-hand side of Eq. (\ref{Qab}) by replacing $A$ with $\sigma_{z}\otimes\sigma_{z}$ and $\sigma_{z}\otimes\sigma_{0}$. In both cases, we have $A^{2}=I$. Therefore, we simplify the formula as
\begin{equation}
	Q_{ab}=-i\frac{m^{2}}{4V}\sum_{\bm{k},\bm{k}\neq0}\nabla_{a}\varepsilon_{\bm{k}}\nabla_{b}\varepsilon_{\bm{k}}\left\{ \text{Tr}[[G,\sigma_{z}\otimes\sigma_{z}]^{2}]-\text{Tr}[[G,\sigma_{z}\otimes\sigma_{0}]^{2}]\right\} .\label{Qab2}
\end{equation}
We rewrite the Green's function as 
\begin{equation}
	G=\left(\begin{array}{cc}
		G_{11} & G_{12}\\
		G_{21} & G_{22}
	\end{array}\right).
\end{equation}
From Eq. (\ref{Green2}), we have
\begin{eqnarray}
	G_{11} & = & \frac{1}{|G^{-1}|}\left(\begin{array}{cc}
		a_{2}(a_{1}^{\ast}a_{2}^{\ast}-|b|^{2}) & -b(a_{1}^{\ast}a_{2}^{\ast}-|b|^{2})\\
		b^{\ast}(|b|^{2}+c^{2}-a_{1}^{\ast}a_{2}^{\ast}) & a_{1}(a_{1}^{\ast}a_{2}^{\ast}-|b|^{2})
	\end{array}\right),\\
	G_{12} & = & \frac{1}{|G^{-1}|}\left(\begin{array}{cc}
		|b|^{2}c & -a_{2}^{\ast}bc\\
		-a_{1}b^{\ast}c & a_{1}a_{2}^{\ast}c
	\end{array}\right),\\
	G_{21} & = & \frac{1}{|G^{-1}|}\left(\begin{array}{cc}
		a_{2}a_{1}^{\ast}c & -a_{1}^{\ast}bc\\
		-a_{2}b^{\ast}c & |b|^{2}c
	\end{array}\right),\\
	G_{22} & = & \frac{1}{|G^{-1}|}\left(\begin{array}{cc}
		-a_{1}^{\ast}(a_{1}a_{2}-|b|^{2}) & -b(|b|^{2}+c^{2}-a_{1}a_{2})\\
		b^{\ast}(a_{1}a_{2}-|b|^{2}) & -a_{2}^{\ast}(a_{1}a_{2}-|b|^{2})
	\end{array}\right)\,.
\end{eqnarray}
Here we define the determinant of the inverse of the Green's function as $|G^{-1}|$ given by
\begin{equation}
	|G^{-1}|=|a_{1}|^{2}|a_{2}|^{2}-a_{1}a_{2}|b|^{2}+|b|^{2}c^{2}-a_{1}^{\ast}a_{2}^{\ast}|b|^{2}+|b|^{4}.\label{deter}
\end{equation}
Then let us focus on the commutators in the phase stiffness $\left(\ref{Qab2}\right)$ which has two components 
\begin{eqnarray}\label{eq: first_trace}
	\text{Tr}[[G,\sigma_{z}\otimes\sigma_{z}]^{2}] & = & \text{Tr}\left[\left(\begin{array}{cc}
		G_{11} & G_{12}\\
		G_{21} & G_{22}
	\end{array}\right),\sigma_{z}\otimes\sigma_{z}\right]^{2}\nonumber \\
	& = & \text{Tr}[G_{11},\sigma_{z}]^{2}+\text{Tr}[G_{22},\sigma_{z}]^{2}-2\text{Tr}[\{G_{12},\sigma_{z}\}\{G_{21},\sigma_{z}\}],\\
	\text{Tr}[[G,\sigma_{z}\otimes\sigma_{0}]^{2}] & = & \text{Tr}\left[\left(\begin{array}{cc}
		G_{11} & G_{12}\\
		G_{21} & G_{22}
	\end{array}\right),\sigma_{z}\otimes\sigma_{0}\right]^{2}\nonumber \\
	& = & -4\text{Tr}[\{G_{12},G_{21}\}]\nonumber \\
	& = & -8\text{Tr}[G_{12}G_{21}],
\end{eqnarray}
where $\{A,B\}:=AB+BA$. Equation \eqref{eq: first_trace}
has the following four components:
\begin{eqnarray}
	\text{Tr}[G_{11},\sigma_{z}]^{2} & = & \frac{1}{|G^{-1}|^{2}}8|b|^{2}(a_{1}^{\ast}a_{2}^{\ast}-|b|^{2})(|b|^{2}+c^{2}-a_{1}^{\ast}a_{2}^{\ast}),\label{G11}\\
	\text{Tr}[G_{22},\sigma_{z}]^{2} & = & \frac{1}{|G^{-1}|^{2}}8|b|^{2}(a_{1}a_{2}-|b|^{2})(|b|^{2}+c^{2}-a_{1}a_{2}),\\
	\text{Tr}[\{G_{12},\sigma_{z}\}\{G_{21},\sigma_{z}\}] & = & \frac{1}{|G^{-1}|^{2}}4|b|^{2}c^{2}(a_{1}^{\ast}a_{2}+a_{1}a_{2}^{\ast}),\\
	\text{Tr}[G_{12}G_{21}] & = & \frac{1}{|G^{-1}|^{2}}|b|^{2}c^{2}[a_{2}a_{1}^{\ast}+|a_{2}|^{2}+|a_{1}|^{2}+a_{1}a_{2}^{\ast}].\label{G12}
\end{eqnarray}
Substituting Eqs. (\ref{G11})-(\ref{G12}) into Eq. \eqref{Qab2}, we obtain
\begin{align}
	Q_{ab}=&i\frac{4m^{2}}{V}\sum_{\bm{k},\bm{k}\neq0}\nabla_{a}\varepsilon_{\bm{k}}\nabla_{b}\varepsilon_{\bm{k}}\frac{-|b|^{2}}{2|G^{-1}|^{2}}\big[(a_{1}^{\ast}a_{2}^{\ast}-|b|^{2})(|b|^{2}+c^{2}-a_{1}^{\ast}a_{2}^{\ast})\nonumber \\
	&+\left(a_{1}a_{2}-|b|^{2}\right)(|b|^{2}+c^{2}-a_{1}a_{2})-c^{2}(|a_{2}|^{2}+|a_{1}|^{2})\big].
\end{align}
Due to the rotational symmetry of the system, the phase stiffness tensor $Q$ is diagonal and isotropic. Thus we can 
write $Q_{ab}$ as $Q\delta_{ab}$. In this case, we obtain 
\begin{align}
	Q=&\frac{4}{3}\int_{-\infty}^{\infty}\frac{d\omega}{2\pi}\int_{-\infty}^{\infty}k^{4}dk\frac{-|b|^{2}}{4\pi^{2}|G^{-1}|^{2}}\big[(a_{1}^{\ast}a_{2}^{\ast}-|b|^{2})(|b|^{2}+c^{2}-a_{1}^{\ast}a_{2}^{\ast})\nonumber \\
	&+\left(a_{1}a_{2}-|b|^{2}\right)(|b|^{2}+c^{2}-a_{1}a_{2})-c^{2}(|a_{2}|^{2}+|a_{1}|^{2})\big].\label{Qboson}
\end{align}
Equation \eqref{Qboson} gives a complete expression of the phase
stiffness for an arbitrary interaction strength $U_{R}$ and dissipation
strength $\gamma$ within the quasi-steady-state approximation. Since it is difficult to directly calculate the integral in Eq. \eqref{Qboson} in general, we consider two extreme cases: the weak-dissipation limit $U_{R}\gg\gamma$ and the weak-interaction limit $U_{R}\ll\gamma$. To consider these problems, we firstly calculate the determinant $|G^{-1}|$. With the help
of Eq. (\ref{deter}), we have 
\begin{eqnarray}
	|G^{-1}| & = & \varepsilon^{4}+2(4\gamma^{2}n^{2}-\omega^{2}-|U|^{2}n^{2})\varepsilon^{2}+(4\gamma^{2}n^{2}+\omega^{2})^{2}-2(4\gamma^{2}n^{2}-\omega^{2})|U|^{2}n^{2}+|U|^{4}n^{4}\nonumber \\
	& = & \omega^{4}-2k_{1}\omega^{2}+k_{2},
\end{eqnarray}
where $\varepsilon:=\varepsilon_{\bm{k}}+U_Rn$, $k_{1}:=(\varepsilon^{2}-4\gamma^{2}n^{2}-|U|^{2}n^{2})$, and $k_{2}:=(\varepsilon^{2}+4\gamma^{2}n^{2}-|U|^{2}n^{2})^{2}$.
The numerator of Eq. \eqref{Qboson} can be written as
\begin{eqnarray}
	&  & -[(\varepsilon+2i\gamma n)^{2}-\omega^{2}-|U|^{2}n^{2}]^{2}-[(\varepsilon-2i\gamma n)^{2}-\omega^{2}-|U|^{2}n^{2}]^{2}+32\gamma^{2}n^{2}(|U|^{2}n^{2}+8\gamma^{2}n^{2})\nonumber \\
	& = & -2[\omega^{4}-2k_{1}\omega^{2}+k_{1}^{2}-16\gamma^{2}n^{2}\varepsilon^{2}]+32\gamma^{2}n^{2}(|U|^{2}n^{2}+8\gamma^{2}n^{2}).
\end{eqnarray}

Let us first consider the case with $\gamma=0$ where there is no dissipation
and the system is closed. In such a case, the quantities $a_{1},a_{2},b,c$
become 
\begin{equation}
	a_{1}=\omega-\varepsilon_{\bm{k}}-U_{R}n,a_{2}=-\omega-\varepsilon_{\bm{k}}-U_{R}n,b=-U_{R}n,c=0.
\end{equation}
Then we substitute these quantities into the phase stiffness and obtain
\begin{equation}
	Q=\frac{4i}{3}\int_{-\infty}^{\infty}\frac{d\omega}{2\pi}\int_{-\infty}^{\infty}\frac{k^{4}}{2\pi^{2}}dk\frac{U_{R}^{2}n^{2}}{[-\omega^{2}+\varepsilon_{\bm{k}}^{2}+2\varepsilon_{\bm{k}}U_{R}n]^{2}}.\label{Qzero}
\end{equation}
Similarly, we can also calculate $Q'$ in the closed
system without using the Schwinger-Keldysh formalism and find its
relation with $Q$ as $Q'=Q$. This can be shown as follows.  

The Matsubara Green's function can be written as 
\begin{equation}
	\mathcal{G}=-\left(\begin{array}{cc}
		\varepsilon_{\bm{k}-m\bm{v}}+U_{R}n+i\omega_n & U_{R}n\\
		U_{R}n & \varepsilon_{\bm{k}+m\bm{v}}+U_{R}n-i\omega_n
	\end{array}\right)^{-1}=-(\varepsilon_{\bm{k}-m\bm{v}\sigma_{z}}+U_{R}n+U_{R}n\sigma_{x}+i\omega_n\sigma_{z})^{-1},
\end{equation}
where $\omega_n:=2\pi n/\beta$ with $n\in\mathbb{Z}$ is the bosonic Matsubara frequency and $\varepsilon_{\bm{k}-m\bm{v}\sigma_{z}}:=\left(\begin{array}{cc}
	\varepsilon_{\bm{k}-m\bm{v}} & 0\\
	0 & \varepsilon_{\bm{k}+m\bm{v}}
\end{array}\right)$. Then the first derivative of the free energy gives the expression for the bosonic current:
\begin{equation}
	\langle j_{a}\rangle=-\frac{1}{V}\frac{\partial F}{\partial v_{a}}=\frac{m}{\beta}\int \frac{d^3\bm{k}}{(2\pi)^3}\sum_n\text{Tr}[\sigma_{z}\nabla_{a}\varepsilon_{\bm{k}-m\bm{v}\sigma_{z}}\mathcal{G}(\bm{k},i\omega_n)].
\end{equation}
In this case the phase stiffness is given by 
\begin{eqnarray}
	Q'_{ab} & = & -\frac{1}{V}\frac{\partial^{2}F}{\partial v_{a}\partial v_{b}}|_{v=0}\nonumber \\
	& = & -\frac{m^{2}}{\beta}\int \frac{d^3\bm{k}}{(2\pi)^3}\sum_n\text{Tr}[\nabla_{ab}^{2}\varepsilon_{\bm{k}}\mathcal{G}]+\frac{m^{2}}{\beta}\int \frac{d^3\bm{k}}{(2\pi)^3}\sum_n\nabla_{a}\varepsilon_{\bm{k}}\nabla_{b}\varepsilon_{\bm{k}}\text{Tr}[\sigma_{z}\mathcal{G}\sigma_{z}\mathcal{G}]\\
	& = & -\frac{m^{2}}{\beta}\int \frac{d^3\bm{k}}{(2\pi)^3}\sum_n\nabla_{a}\varepsilon_{\bm{k}}\nabla_{b}\varepsilon_{\bm{k}}\text{Tr}[\sigma_{z}\mathcal{G}\sigma_{z}\mathcal{G}-\mathcal{G}^{2}]\nonumber \\
	& = & -\frac{m^{2}}{2\beta}\int \frac{d^3\bm{k}}{(2\pi)^3}\sum_n\nabla_{a}\varepsilon_{\bm{k}}\nabla_{b}\varepsilon_{\bm{k}}\text{Tr}[\sigma_{z},\mathcal{G}]^{2}.
\end{eqnarray}
An explicit form of the Green's function is
\begin{equation}
	\mathcal{G}=\frac{1}{(\omega^{2}_n+\varepsilon_{\bm{k}}^{2}+2\varepsilon_{\bm{k}}U_{R}n)}\left(\begin{array}{cc}
		-\varepsilon_{\bm{k}}-U_{R}n+i\omega_n & U_{R}n\\
		U_{R}n & -\varepsilon_{\bm{k}}-U_{R}n-i\omega_n
	\end{array}\right)
\end{equation}
and the commutator takes the form of 
\begin{equation}
	[\sigma_{z},\mathcal{G}]=\frac{1}{(\omega_n^{2}+\varepsilon_{\bm{k}}^{2}+2\varepsilon_{\bm{k}}U_{R}n)}\left(\begin{array}{cc}
		0 & -2U_{R}n\\
		2U_{R}n & 0
	\end{array}\right).
\end{equation}
Therefore,
\begin{equation}
	Q'_{ab}=\frac{4m^{2}}{\beta}\int \frac{d^3\bm{k}}{(2\pi)^3}\sum_n\nabla_{a}\varepsilon_{\bm{k}}\nabla_{b}\varepsilon_{\bm{k}}\frac{U_{R}^{2}n^{2}}{(\omega_n^{2}+\varepsilon_{\bm{k}}^{2}+2\varepsilon_{\bm{k}}U_{R}n)^{2}}.
\end{equation}
Since $\varepsilon_{\bm{k}}=\bm{k}^{2}/2m$, we have 
\begin{equation}
	Q'_{ab}=\frac{4m^{2}}{\beta}\sum_{n}\int\frac{d^{3}\bm{k}}{(2\pi)^{3}}\frac{k_{a}k_{b}}{m^{2}}\frac{U_{R}^{2}n^{2}}{(\omega_n^{2}+\varepsilon_{\bm{k}}^{2}+2\varepsilon_{\bm{k}}U_{R}n)^{2}}=\frac{4}{3\beta}\sum_{n}\int\frac{k^{4}dk}{2\pi^{2}}\frac{U_{R}^{2}n^{2}}{(\omega_n^{2}+\varepsilon_{\bm{k}}^{2}+2\varepsilon_{\bm{k}}U_{R}n)^{2}}\delta_{ab}.
\end{equation}
In the zero-temperature limit, defining $Q'_{ab}=:Q'\delta_{ab}$, we obtain
\begin{equation}\label{eq: closed_Q'}
	Q'=\frac{4}{3}\int \frac{d\omega'}{2\pi}\int\frac{k^{4}dk}{2\pi^{2}}\frac{U_{R}^{2}n^{2}}{(\omega'^{2}+\varepsilon_{\bm{k}}^{2}+2\varepsilon_{\bm{k}}U_{R}n)^{2}}.
\end{equation}
Equations (\ref{eq: closed_Q'}) and (\ref{Qzero}) can be transformed to each other via the Wick rotation: $\omega\to -i\omega'$. We therefore reach the conclusion 
\begin{equation}
	Q'=Q.
\end{equation}
The phase stiffness can be calculated as 
\begin{eqnarray}
	Q' & = & \frac{4}{3}\int\frac{d\omega'}{2\pi}\int\frac{k^{4}dk}{2\pi^{2}}\frac{U_{R}^{2}n^{2}}{(\omega'^{2}+\varepsilon_{\bm{k}}^{2}+2\varepsilon_{\bm{k}}U_{R}n)^{2}}\nonumber \\
	& = & \frac{2}{3}\int_{0}^{\infty}\frac{k^{4}dk}{2\pi^{2}}\frac{U_{R}^{2}n^{2}}{[\varepsilon_{\bm{k}}^{2}+2\varepsilon_{\bm{k}}U_{R}n]^{3/2}}\nonumber \\
	& = & \frac{1}{3\pi^{2}}\sqrt{n^{3}U_{R}^{3}}m^{5/2}\nonumber \\
	& \propto & (U_{R}n)^{3/2}.\label{QQ}
\end{eqnarray}
This is consistent with the quantum depletion density calculated by linear response theory
\citep{Ueda2010}:
\begin{equation}
	n_{D}=\frac{1}{3\pi^{2}}\sqrt{(nU_{R})^{3}m^{3}}.
\end{equation}

If we assume $U_{R}\gg\gamma\neq0$, i.e., the dissipation is very
weak but nonvanishing, we can calculate the phase stiffness from the trace
in Eq. \eqref{Qab} as
\begin{eqnarray}
	&  & \text{Tr}[(\sigma_{z}\otimes\sigma_{z})G(\sigma_{z}\otimes\sigma_{z})G]-\text{Tr}[(\sigma_{z}\otimes\sigma_{0})G(\sigma_{z}\otimes\sigma_{0})G]\nonumber \\
	& = & \frac{8U_{R}^{2}n^{2}}{(\varepsilon_{\bm{k}}(\varepsilon_{\bm{k}}+2U_{R}n)-\omega^{2})^{2}}+\frac{8(-6U_{R}^{2}n^{2}(\varepsilon_{\bm{k}}(\varepsilon_{\bm{k}}+2U_{R}n)+7\omega^{2})+(\varepsilon_{\bm{k}}(\varepsilon_{\bm{k}}+2U_{R}n)-\omega^{2})^{2})}{(\varepsilon_{\bm{k}}(\varepsilon_{\bm{k}}+2U_{R}n)-\omega^{2})^{4}}\gamma^{2}n^{2}.\label{eq:divergent}
\end{eqnarray}
Here we only expand the expression in \eqref{Qab} up to the second order in $\gamma n$. By taking the integration
over $\omega$, 
we have 
\begin{eqnarray}
	&  & -i\int d\omega\text{Tr}[(\sigma_{z}\otimes\sigma_{z})G(\sigma_{z}\otimes\sigma_{z})G]-\text{Tr}[(\sigma_{z}\otimes\sigma_{0})G(\sigma_{z}\otimes\sigma_{0})G]\nonumber \\
	& = & \frac{4\pi U_{R}^{2}n^{2}}{(\varepsilon_{\bm{k}}(\varepsilon_{\bm{k}}+2U_{R}n))^{3/2}}+2\pi\frac{(3(U_{R}n)^{2}+4\varepsilon_{\bm{k}}U_{R}n+2\varepsilon_{\bm{k}}^{2})\sqrt{\varepsilon_{\bm{k}}(\varepsilon_{\bm{k}}+2U_{R}n)}}{(\varepsilon_{\bm{k}}(\varepsilon_{\bm{k}}+2U_{R}n))^{3}}\gamma^{2}n^{2},
\end{eqnarray}
which gives the form of the phase stiffness as 
\begin{eqnarray}
	Q & = & \frac{m^{5/2}(U_{R}n)^{3/2}}{3\pi^{2}}\left[1+\frac{(\gamma n)^{2}}{2\sqrt{2}(U_{R}n)^{2}}\int_{0}^{\infty}dxx^{\frac{3}{2}}\frac{(3+4x+2x^{2})\sqrt{x(2+x)}}{x^{3}(2+x)^{3}}\right]\nonumber \\
	& \simeq & \frac{m^{5/2}}{3\pi^{2}}(U_{R}n)^{3/2}\left[1+\eta\left(\frac{\gamma}{U_{R}}\right)^{2}+O\left(\frac{\gamma}{U_R}\right)^3\right],\label{eq:integration}
\end{eqnarray}
where $\eta=(4+3\ln2)/8$. Hence, the superfluid quantum depletion density takes the form of
\begin{equation}
	n_{sD}=\frac{m^{3/2}}{3\pi^{2}}(U_{R}n)^{3/2}\left[1+\eta\left(\frac{\gamma}{U_{R}}\right)^{2}+O\left(\frac{\gamma}{U_R}\right)^3\right].
\end{equation}

In the other limit $U_{R}\to0$, we have 
\begin{equation}
	a_{1}=-(\varepsilon_{\bm{k}}-2i\gamma n-\omega),b=i\gamma n,a_{2}=-(\varepsilon_{\bm{k}}-2i\gamma n+\omega),c=-4i\gamma n.
\end{equation}
The numerator of the integrand in Eq. \eqref{Qboson} becomes 
\begin{equation}
	-2[\omega^{4}-2k_{1}\omega^{2}+k_{1}^{2}-16\gamma^{2}n^{2}\varepsilon_{\bm{k}}^{2}]+288\gamma^{4}n^{4},
\end{equation}
where $k_{1}=(\varepsilon_{\bm{k}}^{2}-5\gamma^{2}n^{2}),k_{2}=(\varepsilon_{\bm{k}}^{2}+3\gamma^{2}n^{2})^{2}$.
The denominator of the integrand in Eq.~(\ref{Qboson}) becomes 
\begin{equation}
	|G^{-1}|^{2}=(\omega^{4}-2k_{1}\omega^{2}+k_{2})^{2}.
\end{equation}
Hence, the phase stiffness can be rewritten as 
\begin{equation}
	Q=\frac{(2m)^{5/2}\gamma^{2}n^{2}}{3\pi^{2}}\int_{-\infty}^{\infty}d\omega\int_{0}^{\infty}\varepsilon_{\bm{k}}^{3/2}d\varepsilon_{\bm{k}}\frac{\omega^{4}-2k_{1}\omega^{2}+k_{1}^{2}-16\gamma^{2}n^{2}\varepsilon_{\bm{k}}^{2}-144\gamma^{4}n^{4}}{(\omega^{4}-2k_{1}\omega^{2}+k_{2})^{2}}\,. 
\end{equation}
By replacing the integrated variables as $\omega\rightarrow\gamma n\omega,\varepsilon_{\bm{k}}\rightarrow\gamma n\varepsilon_{\bm{k}}$,
we have 
\begin{equation}
	Q=\frac{(2m)^{5/2}}{3\pi^{2}}(\gamma n)^{3/2}\times A,\label{QU=00003D0}
\end{equation}
where
\begin{eqnarray*}
	A & = & \int_{-\infty}^{\infty}dx\int_{0}^{\infty}y^{3/2}dy\frac{x^{4}-2(y^{2}-5)x^{2}+(y^{2}-5)^{2}-16y^{2}-144}{(x^{4}-2(y^{2}-5)x^{2}+(y^{2}+3)^{2})^{2}}\\
	& = & \frac{1}{2^{11/2}\sqrt{\pi}}\Gamma\left(\frac{1}{4}\right)^{2}>0,
\end{eqnarray*}
where $\Gamma(x)$ is the Gamma function. From the expression (\ref{QU=00003D0}), we can see that when $U_{R}=0$,
the superfluid quantum depletion density is propotional to
$(\gamma n)^{3/2}$, which indicates that the quantum depletion is induced purely by dissipation. Furthermore, if we assume $0\neq U_{R}\ll\gamma$ and expand the phase stiffness \eqref{Qab} around $U_{R}=0$, we obtain 
\begin{eqnarray}
	&  & \text{Tr}[(\sigma_{z}\otimes\sigma_{z})G(\sigma_{z}\otimes\sigma_{z})G]-\text{Tr}[(\sigma_{z}\otimes\sigma_{0})G(\sigma_{z}\otimes\sigma_{0})G]\nonumber \\
	& = & -4\gamma^{2}n^{2}\left[\frac{1}{(\omega^{2}-4i\gamma n\omega-\varepsilon_{\bm{k}}^{2}-3\gamma^{2}n^{2})^{2}}+\frac{1}{(\omega^{2}+4i\gamma n\omega-\varepsilon_{\bm{k}}^{2}-3\gamma^{2}n^{2})^{2}}\right]\nonumber \\
	&  & +4\gamma nU_{R}n\left[\frac{4\varepsilon_{\bm{k}}}{(\omega^{2}-4i\gamma n\omega-\varepsilon_{\bm{k}}^{2}-3\gamma^{2}n^{2})^{3}}+\frac{4\varepsilon_{\bm{k}}}{(\omega^{2}+4i\gamma n\omega-\varepsilon_{\bm{k}}^{2}-3\gamma^{2}n^{2})^{2}}\right]\nonumber \\
	&  & +O(U_{R}^{2}n^{2}). \label{eq:multiplication_Matrix}
\end{eqnarray}
By substituting Eq.~(\ref{eq:multiplication_Matrix}) into Eq. \eqref{Qab} and integrating the result over
the variables $\omega$ and $\varepsilon_{\bm{k}}$, we obtain the phase stiffness as 
\begin{equation}
	Q=\frac{m^{5/2}(\gamma n)^{3/2}}{24\pi^{5/2}}\Gamma\left(\frac{1}{4}\right)^{2}\left(1+6\frac{U_{R}}{\gamma}\frac{\Gamma(3/4)^{2}}{\Gamma(1/4)^{2}}\right)+O(U_{R}^{2}/\gamma^{2}).
\end{equation}
Correspondingly, the superfluid quantum depletion density in the weak-interaction limit is given by
\begin{equation}
	n_{sD}=\frac{m^{3/2}(\gamma n)^{3/2}}{24\pi^{5/2}}\Gamma\left(\frac{1}{4}\right)^{2}\left(1+6\frac{U_{R}}{\gamma}\frac{\Gamma(3/4)^{2}}{\Gamma(1/4)^{2}}\right)+O(U_{R}^{2}/\gamma^{2}).
\end{equation}

\subsection{The Case with a Dipolar Interaction}
In the following, we also take the dipolar interaction into account to discuss application of our theory to a dissipative BEC of dipolar molecules. Following the
same procedure as above, we apply the Bogoliubov approximation and find the
Green's function as
\begin{equation}
	G (\bm{k}, \omega) = \left(\begin{array}{cccc}
		a_1 & b & 0 & 0\\
		b^{\ast} & a_2 & 0 & c\\
		c & 0 & - a_1^{\ast} & - b\\
		0 & 0 & - b^{\ast} & - a_2^{\ast}
	\end{array}\right)^{- 1},
\end{equation}
where $a_1 = - (\varepsilon_{\bm{k}- m\bm{v}} + (U_R + V_{d d}
(\bm{k})) n - 2 i \gamma n - \omega), b = - U n, a_2 = -
(\varepsilon_{\bm{k}+ m\bm{v}} + (U_R + V_{d d} (\bm{k})) n
- 2 i \gamma n + \omega)$, and $c = - 4 i \gamma n$. We can see that the effect of
dipolar interaction is just to replace the interaction strength $U_R$ with
$U_R + V_{d d} (\bm{k}) = U_R (1 - \varepsilon_{d d} + 3 \varepsilon_{d
	d} \cos^2 \theta_{\bm{k}})$, which is anisotropic in the momentum space.
For convenience, we introduce an effective interaction strength $\tilde{U}_R
(\bm{k}) := U_R + V_{d d} (\bm{k})$. Therefore, the phase
stiffness is given by
\begin{align}
	Q_{a b} = &i 4m^2\int\frac{d\omega}{2\pi}\frac{d^3\bm{k}}{(2\pi)^3} \frac{k_a k_b}{m^2} \frac{- | b |^2}{2 |
		G^{- 1} |^2} \big[ (a_1^{\ast} a_2^{\ast} - | b |^2) (| b |^2 + c^2 -
	a_1^{\ast} a_2^{\ast}) \nonumber \\
	&+ \left( a_1 a_2 - | b |^2 \right) (| b |^2 + c^2 -
	a_1 a_2) - c^2 (| a_2 |^2 + | a_1 |^2) \big], \label{eq:newphase}
\end{align}
where $a_1$ and $a_2$ include the effect of dipole-dipole interaction as defined above. We note that here the phase stiffness takes different values depending on the direction due to the anisotropic dipolar interaction. To proceed further, we average the phase stiffness over the direction of the applied perturbation. From Eq. \eqref{eq:newphase}, we find that
the nontrivial contribution only comes from the diagonal terms with $a = b$.
For simplicity, we define the direction of the subscript $a$ in Eq. \eqref{eq:newphase} as $(\cos \varphi \sin \zeta, \sin
\varphi \sin \zeta, \cos \zeta)$ and rewrite the phase stiffness as
\begin{equation}\label{eq: anisotropic_1}
	Q_{a a} = i 4m^2\int\frac{d\omega}{2\pi}\frac{d^3\bm{k}}{(2\pi)^3}\frac{q (k,\theta) k^2}{m^2} (\cos \theta \cos
	\zeta + \sin \theta \sin \zeta \cos (\phi - \varphi))^2,
\end{equation}
where
\begin{equation}
	q := \frac{- | b |^2}{2 | G^{- 1} |^2} \left[ (a_1^{\ast} a_2^{\ast} -
	| b |^2) (| b |^2 + c^2 - a_1^{\ast} a_2^{\ast}) + \left( a_1 a_2 - | b |^2
	\right) (| b |^2 + c^2 - a_1 a_2) - c^2 (| a_2 |^2 + | a_1 |^2) \right],
\end{equation}
and the direction of $\bm{k}$ is defined as $(\sin \theta \cos \phi,
\sin \theta \sin \phi, \cos \theta)$. Then by integrating Eq. \eqref{eq: anisotropic_1} over $\phi$, we have
\begin{equation}
	Q_{a a} = i 4m^2\int \frac{d \omega}{(2 \pi)^3} \int k^2 d k
	\sin \theta d \theta \frac{q (k,\theta) k^2}{m^2} \left[ (\cos \theta \cos \zeta)^2
	+ \frac{1}{2} (\sin \theta \sin \zeta)^2 \right],
\end{equation}
which depends only on the angle $\zeta$ in the direction of $a$. We then take the average over $\zeta$ to determine the averaged quantum depletion given by
\begin{eqnarray}\label{eq: average_Q}
	\bar{Q} & = & i \frac{4 m^2}{(2 \pi)^2 } \int \frac{d \omega}{2\pi}
	\int k^2 d k \sin \theta d \theta \frac{q (k,\theta) k^2}{m^2} \frac{2}{3} [(\cos
	\theta)^2 + (\sin \theta)^2] \nonumber\\
	& = & i \frac{8 m^2}{3 (2 \pi)^2 } \int \frac{d \omega}{2 \pi} \int
	k^2 d k \sin \theta d \theta \frac{q (k,\theta) k^2}{m^2} . 
\end{eqnarray}
After the integration over $k$, we are left with the expression of $\bar{Q}$ with
an average over $\theta$. We estimate the value of $\bar{Q}$ by
substituting $U_R$ with $U_R + V_{d d}$ and take the average over the angle
$\theta$. We consider the two limits: the weak-dissipation limit $U_R,
c_{d d} \gg \gamma$ and the weak-interaction limit $U_R, c_{d d} \ll \gamma$.
In the first case, the phase stiffness can be expressed as
\begin{align}
	\bar{Q} & = \frac{m^{5 / 2}}{3 \pi^2} \int \frac{1}{2} \sin \theta d \theta
	(\tilde{U}_R n)^{3 / 2}   \left[ 1 + \frac{(\gamma n)^2}{2 \sqrt{2}( \tilde{U}_R
		n)^2} \int_0^{\infty} d x \frac{(3 + 4 x + 2 x^2) \sqrt{x (2 + x)}}{x (2 +
		x)^3}+O\left[\left(\frac{\gamma}{U_R}\right)^3\right] \right] \nonumber\\
	& = \frac{m^{5 / 2}}{3 \pi^2} (U_R n)^{3 / 2} \left[ \frac{1}{8}
	\sqrt{1 + 2 \varepsilon_{d d}} \left[ 5 + \varepsilon_{d d} + \frac{3 (1 -
		\varepsilon_{d d})^2 \text{arcsinh} \sqrt{3 \varepsilon_{d d} / (1 -
			\varepsilon_{d d})}}{\sqrt{3 \varepsilon_{d d} (1 + 2 \varepsilon_{d d})}}
	\right]  \right. \nonumber\\
	&  + \left.\frac{\eta(\gamma n)^2}{\sqrt{U_R n}} \frac{1}{2 \sqrt{3
			\varepsilon_{d d}}} \log \left( 1 + \frac{2 \left( 3 \varepsilon_{d d} +
		\sqrt{3 \varepsilon_{d d} (1 + 2 \varepsilon_{d d})} \right)}{1 -
		\varepsilon_{d d}} \right)+O\left(\frac{\gamma}{U_R}\right)^3  \right] . \nonumber\\
	&=: \frac{m^{5 / 2}}{3 \pi^2} (U_R n)^{3 / 2}\left(h_1+h_2\frac{\eta(\gamma n)^2}{2\sqrt{U_R n}}+O\left(\frac{\gamma}{U_R}\right)^3\right),
\end{align}
where
\begin{align}
	h_1&:=\frac{1}{8}\sqrt{1 + 2 \varepsilon_{d d}} \left[ 5 + \varepsilon_{d d} + \frac{3 (1 -	\varepsilon_{d d})^2 \text{arcsinh} \sqrt{3 \varepsilon_{d d} / (1 -\varepsilon_{d d})}}{\sqrt{3 \varepsilon_{d d} (1 + 2 \varepsilon_{d d})}}	\right], \\
	h_2&:=\frac{1}{2 \sqrt{3\varepsilon_{d d}}} \log \left( 1 + \frac{2 \left( 3 \varepsilon_{d d} +\sqrt{3 \varepsilon_{d d} (1 + 2 \varepsilon_{d d})} \right)}{1 -\varepsilon_{d d}} \right).
\end{align}
We can prove that the two coeffients are enlarged by the ratio
$\varepsilon_{d d}$ between the dipolar and the contact interactions, which indicates that the dipolar interaction enhances the
repulsive interaction on average and hence increases the quantum depletion. Substituting the experimental data $\varepsilon_{dd}=0.833$ in Ref. \cite{Sebastian2023}, we obtain $h_1=1.204,h_2=1.305$.

In the weak-interaction limit where $U_R, c_{d d} \ll \gamma$, Eq. \eqref{eq: average_Q} becomes
\begin{equation}
	\bar{Q} = \int \frac{1}{2} \sin \theta d \theta \frac{m^{5 / 2} (\gamma n)^{3 /
			2}}{24 \pi^{5 / 2}} \Gamma \left( \frac{1}{4} \right)^2 \left[ 1 + 6
	\frac{\tilde{U}_R}{\gamma} \frac{\Gamma (3 / 4)^2}{\Gamma (1 / 4)^2}+O\left(\frac{U_R}{\gamma}\right)^2 \right]
	.
\end{equation}
After taking the average over the angle $\theta$, we find that this is
equivalent to the result with no dipolar interaction. This is because here the quantum depletion is mainly produced by dissipation and the effect of the dipolar interaction vanishes up to the first order in $\tilde{U}_R$.

Furthermore, we consider the normal quantum depletion density. We perturb the Hamiltonians in the Schwinger-Keldysh action by $H_{+}\to H_{+}-\bm{v}\cdot\bm{j}_{+}$ and $H_{-}\to H_{-}+\bm{v}\cdot\bm{j}_{-}$. The normal fluid density is determined from $\tilde{Q}_{ab} = \frac{-1}{2V}\frac{\partial^2 F}{\partial v_a \partial v_a}|_{v=0}=mn_{nD}\delta_{ab}$. Here $F$ is a functional in the presence of the source term. The normal fluid density tensor is equivalent to the second term in Eq. (\ref{eq:total_of_Q}), i.e.,
\begin{equation}
	\tilde{Q}_{ab} = i\frac{m^{2}}{2V}\int\frac{d\omega}{2\pi}\sum_{\bm{k},\bm{k}\neq0}\nabla_{a}\varepsilon_{\bm{k}}\nabla_{b}\varepsilon_{\bm{k}}\text{Tr}[(\sigma_{z}\otimes\sigma_{z})G(\sigma_{z}\otimes\sigma_{z})G],
\end{equation}
where the diagonal term is the normal fluid density. Using the explicit form of $G$, we obtain
\begin{align}
	&\text{Tr}[(\sigma_{z}\otimes\sigma_{z})G(\sigma_{z}\otimes\sigma_{z})G]\nonumber\\
	=&\frac{4\left\{\gamma^{2}n^{2}\left[\text{\ensuremath{\epsilon_{\bm{k}}}}^{2}(2\tilde{U}_Rn+\text{\ensuremath{\epsilon_{\bm{k}}}})^{2}-38\omega^{2}\text{\ensuremath{\epsilon_{\bm{k}}}}(2\tilde{U}_Rn+\text{\ensuremath{\epsilon_{\bm{k}}}})+5\omega^{4}\right]+\left(2\tilde{U}_Rn\text{\ensuremath{\epsilon_{\bm{k}}}}+\omega^{2}+\text{\ensuremath{\epsilon_{\bm{k}}}}^{2}\right)\left(2\tilde{U}_Rn\ensuremath{\epsilon_{\bm{k}}}-\omega^{2}+\text{\ensuremath{\epsilon_{\bm{k}}}}^{2}\right)^{2}\right\}}{\left(3\gamma^{2}n^{2}+4i\gamma n\omega+2\tilde{U}_Rn\text{\ensuremath{\epsilon_{\bm{k}}}}-\omega^{2}+\text{\ensuremath{\epsilon_{\bm{k}}}}^{2}\right)^{2}\left[(\gamma n-i\omega)(3\gamma n-i\omega)+\text{\ensuremath{\epsilon_{\bm{k}}}}(2\tilde{U}_Rn+\text{\ensuremath{\epsilon_{\bm{k}}}})\right]^{2}}\nonumber\\	
	+&\frac{4\left[-45\gamma^{6}n^{6}-\gamma^{4}n^{4}\left(21\text{\ensuremath{\epsilon_{\bm{k}}}}(2\tilde{U}_Rn+\text{\ensuremath{\epsilon_{\bm{k}}}})-23\omega^{2}\right)\right]}{\left(3\gamma^{2}n^{2}+4i\gamma n\omega+2\tilde{U}_Rn\text{\ensuremath{\epsilon_{\bm{k}}}}-\omega^{2}+\text{\ensuremath{\epsilon_{\bm{k}}}}^{2}\right)^{2}\left[(\gamma n-i\omega)(3\gamma n-i\omega)+\text{\ensuremath{\epsilon_{\bm{k}}}}(2\tilde{U}_Rn+\text{\ensuremath{\epsilon_{\bm{k}}}})\right]^{2}}.	
\end{align}
The above expression vanishes after intergration over $\omega$ from $-\infty$ to $\infty$ since the residues of all poles of the integrand in the upper-half complex plane of $\omega$ vanish. Thus, we can see the superfluid quantum depletion is just equal to the total quantum depletion and the normal fluid density is always zero, i.e., all the bosons belonging to the quantum depletion contribute to the superfluid transport.

\section{Derivation of The Spectral Function and the Excitation Spectrum}\label{sec: stability}
Here we derive an expression of the Green's function of the dissipative superfluid. Within the mean-field approximation, the correlation-function matrix in the momentum-energy space takes the form of 
\begin{eqnarray}
	G(\bm{k},\omega) & := & -i\left\langle \left(\begin{array}{c}
		a_{\bm{k}\bm{,+}}\\
		a_{-\bm{k}\bm{,+}}^{\dagger}\\
		a_{\bm{k},-}\\
		a_{-\bm{k},-}^{\dagger}
	\end{array}\right)\left(\begin{array}{cccc}
		a_{\bm{k},+}^{\dagger} & a_{-\bm{k},+} & a_{\bm{k},-}^{\dagger} & a_{-\bm{k},-}\end{array}\right)\right\rangle \nonumber \\
	& = & \left(\begin{array}{cccc}
		-\frac{\varepsilon_{\bm{k}}+U_{R}n-2i\gamma n-\omega}{2} & -\frac{Un}{2}\\
		-\frac{U^{\ast}n}{2} & -\frac{\varepsilon_{\bm{k}}+U_{R}n-2i\gamma n+\omega}{2} &  & -2i\gamma n\\
		-2i\gamma n &  & \frac{\varepsilon_{\bm{k}}+U_{R}n+2i\gamma n-\omega}{2} & \frac{Un}{2}\\
		&  & \frac{U^{\ast}n}{2} & \frac{\varepsilon_{\bm{k}}+U_{R}n+2i\gamma n+\omega}{2}
	\end{array}\right)^{-1}
\end{eqnarray}
(see Eq. \eqref{Green2}). By directly calculating the inverse of the matrix, we have the following Green's
functions \citep{Kamenev_2011}: 
\begin{eqnarray}
	G^{T}(\bm{k},\omega) & := & -i\langle a_{\bm{k},+}a_{\bm{k},+}^{\dagger}\rangle\nonumber \\
	& = & \frac{2(\omega+U_{R}n-2i\gamma n+\varepsilon_{\bm{k}})\left[\omega^{2}-\varepsilon_{\bm{k}}^{2}-2\varepsilon_{\bm{k}}(U_{R}n+2i\gamma n)+\gamma n(5\gamma n-4iU_{R}n)\right]}{\omega^{4}+2\left[5\gamma^{2}n^{2}-\varepsilon_{\bm{k}}(\varepsilon_{\bm{k}}+2U_{R}n)\right]\omega^{2}+[\varepsilon_{\bm{k}}(\varepsilon_{\bm{k}}+2U_{R}n)+3\gamma^{2}n^{2}]^{2}},\\
	G^{<}(\bm{k},\omega) & := & -i\langle a_{\bm{k},+}a_{\bm{k},-}^{\dagger}\rangle\nonumber \\
	& = & \frac{-8i\gamma n(U_{R}^{2}n^{2}+\gamma^{2}n^{2})}{\omega^{4}+2\left[5\gamma^{2}n^{2}-\varepsilon_{\bm{k}}(\varepsilon_{\bm{k}}+2U_{R}n)\right]\omega^{2}+[\varepsilon_{\bm{k}}(\varepsilon_{\bm{k}}+2U_{R}n)+3\gamma^{2}n^{2}]^{2}},\\
	G^{>}(\bm{k},\omega) & := & -i\langle a_{\bm{k},-}a_{\bm{k},+}^{\dagger}\rangle\nonumber \\
	& = & \frac{-8i\gamma n\left[(\omega+\varepsilon_{\bm{k}}+U_{R}n)^{2}+4\gamma^{2}n^{2}\right]}{\omega^{4}+2\left[5\gamma^{2}n^{2}-\varepsilon_{\bm{k}}(\varepsilon_{\bm{k}}+2U_{R}n)\right]\omega^{2}+[\varepsilon_{\bm{k}}(\varepsilon_{\bm{k}}+2U_{R}n)+3\gamma^{2}n^{2}]^{2}},\\
	G^{\tilde{T}}(\bm{k},\omega) & := & -i\langle a_{\bm{k},-}a_{\bm{k},-}^{\dagger}\rangle\nonumber \\
	& = & \frac{2(\omega+U_{R}n+2i\gamma n+\varepsilon_{\bm{k}})\left[-\omega^{2}+\varepsilon_{\bm{k}}^{2}+2\varepsilon_{\bm{k}}(U_{R}n-2i\gamma n)-\gamma n(5\gamma n+4iU_{R}n)\right]}{\omega^{4}+2\left[5\gamma^{2}n^{2}-\varepsilon_{\bm{k}}(\varepsilon_{\bm{k}}+2U_{R}n)\right]\omega^{2}+[\varepsilon_{\bm{k}}(\varepsilon_{\bm{k}}+2U_{R}n)+3\gamma^{2}n^{2}]^{2}}.
\end{eqnarray}
To obtain the spectral function, we first consider the retarded Green's function $G^R$, which is given by the relation 
\begin{eqnarray}
	\left(\begin{array}{cc}
		G^{K} & G^{R}\\
		G^{A} & 0
	\end{array}\right) & = & \left(\begin{array}{cc}
		1 & 1\\
		1 & -1
	\end{array}\right)\left(\begin{array}{cc}
		G^{T} & G^{<}\\
		G^{>} & G^{\tilde{T}}
	\end{array}\right)\left(\begin{array}{cc}
		1 & 1\\
		1 & -1
	\end{array}\right).
\end{eqnarray}
Therefore, the Keldysh, retarded, and advanced Green's functions are
given by 
\begin{eqnarray}
	G^{K}(\bm{k},\omega) & = & \frac{-8i\gamma n\left[(\omega+\varepsilon_{\bm{k}}+U_{R}n)^{2}+U_{R}^{2}n^{2}+5\gamma^{2}n^{2}\right]}{\omega^{4}+2(5\gamma^{2}n^{2}-\varepsilon_{\bm{k}}(\varepsilon_{\bm{k}}+2U_{R}n))\omega^{2}+[\varepsilon_{\bm{k}}(\varepsilon_{\bm{k}}+2U_{R}n)+3\gamma^{2}n^{2}]^{2}},\\
	G^{R}(\bm{k},\omega) & = & \frac{2(\omega+\varepsilon_{\bm{k}}+U_{R}n+2i\gamma n)}{\omega^{2}+4i\gamma n\omega-\varepsilon_{\bm{k}}^{2}-2U_{R}n\varepsilon_{\bm{k}}-3\gamma^{2}n^{2}},\\
	G^{A}(\bm{k},\omega) & = & \frac{2(\omega+\varepsilon_{\bm{k}}+U_{R}n-2i\gamma n)}{\omega^{2}-4i\gamma n\omega-\varepsilon_{\bm{k}}^{2}-2U_{R}n\varepsilon_{\bm{k}}-3\gamma^{2}n^{2}}=(G^{R})^{\ast}.
\end{eqnarray}
From the Green's functions, we obtain the spectral function
as \citep{Coleman_2015}
\begin{eqnarray}
	A(\bm{k},\omega) & = & \frac{i}{2\pi}(G^{<}+G^{>})\nonumber \\ \label{eq:spectral}
	& = & \frac{1}{\pi}\frac{4\gamma n\left[(\omega+\varepsilon_{\bm{k}}+U_{R}n)^{2}+U_{R}^{2}n^{2}+5\gamma^{2}n^{2}\right]}{\omega^{4}+2(5\gamma^{2}n^{2}-\varepsilon_{\bm{k}}(\varepsilon_{\bm{k}}+2U_{R}n))\omega^{2}+[\varepsilon_{\bm{k}}(\varepsilon_{\bm{k}}+2U_{R}n)+3\gamma^{2}n^{2}]^{2}}.
\end{eqnarray}
We note that when $U_{R}=0$, the spectral function becomes 
\begin{eqnarray}
	A(\bm{k},\omega) & = & \frac{1}{\pi}\frac{4\gamma n\left[(\omega+\varepsilon_{\bm{k}})^{2}+5\gamma^{2}n^{2}\right]}{\omega^{4}+2(5\gamma^{2}n^{2}-\varepsilon_{\bm{k}}^{2})\omega^{2}+[\varepsilon_{\bm{k}}^{2}+3\gamma^{2}n^{2}]^{2}}.
\end{eqnarray}
In the other limit $\gamma=0$, one can prove that Eq. \eqref{eq:spectral} is proportional to $\delta(\omega^2-\varepsilon_{\bm{k}}(\varepsilon_{\bm{k}}+2U_Rn))$, which reproduces the spectral function of a closed quantum system.

The excitation spectrum of the dissipative superfluid is given by the poles
of the spectral function as 
\begin{equation}
	\omega^{4}+2(5\gamma^{2}n^{2}-\varepsilon_{\bm{k}}(\varepsilon_{\bm{k}}+2U_{R}n))\omega^{2}+[\varepsilon_{\bm{k}}(\varepsilon_{\bm{k}}+2U_{R}n)+3\gamma^{2}n^{2}]^{2}=0.
\end{equation}
This equation can be factorized as 
\begin{equation}
	(\omega^{2}+4i\gamma n\omega-\varepsilon_{\bm{k}}^{2}-2U_{R}n\varepsilon_{\bm{k}}-3\gamma^{2}n^{2})(\omega^{2}-4i\gamma n\omega-\varepsilon_{\bm{k}}^{2}-2U_{R}n\varepsilon_{\bm{k}}-3\gamma^{2}n^{2})=0.\label{eq:poles}
\end{equation}
The solutions to Eq. \eqref{eq:poles} can be expressed as 
\begin{eqnarray}\label{eq: omega_12}
	\omega_{1,2} & = & -2i\gamma n\pm\sqrt{\varepsilon_{\bm{k}}(\varepsilon_{\bm{k}}+2U_{R}n)-\gamma^{2}n^{2}},\\ \label{eq: omega_34}
	\omega_{3,4} & = & 2i\gamma n\pm\sqrt{\varepsilon_{\bm{k}}(\varepsilon_{\bm{k}}+2U_{R}n)-\gamma^{2}n^{2}}.
\end{eqnarray}
For those momenta that satisfy $\varepsilon_{\bm{k}}(\varepsilon_{\bm{k}}+2U_{R}n)<\gamma^{2}n^{2}$,
the real parts of all the spectra vanish. For those momenta that satisfy 
$\varepsilon_{\bm{k}}(\varepsilon_{\bm{k}}+2U_{R}n)>\gamma^{2}n^{2}$,
the real parts of $\omega_{1(2)}$ are equal to those of $\omega_{3(4)}$ since $\omega_{1}=\omega_{3}^{\ast}$ and $\omega_{2}=\omega_{4}^{\ast}$.
Hence, we reach the conclusion that 
\begin{equation}
	\text{Re}[\omega_{1}]=\text{Re}[\omega_{3}]=-\text{Re}[\omega_{2}]=-\text{Re}[\omega_{4}].
\end{equation}
There is only one nontrivial real part in the four spectra. 
The poles of the retarded Green's
function give the spectra $\omega_{1,2}$. These two spectra indeed coincide
with the spectra of the observables $A_{\bm{k}}$ in Ref. \cite{Ce2022}. We can understand the roles of the spectra from another framework of the Schwinger-Keldysh action. We first transform the action into another basis. By defining the retarded or advanced operators $a_{\bm{k},R}=\frac{1}{2}(a_{\bm{k},+}+a_{\bm{k},-})$
and $a_{\bm{k},A}=a_{\bm{k},+}-a_{\bm{k},-}$ \citep{Kamenev_2011},
we have 
\begin{equation}
	a_{\bm{k}+}^{\dagger}i\partial_{t}a_{\bm{k}+}-a_{\bm{k}-}^{\dagger}i\partial_{t}a_{\bm{k}-}=a_{\bm{k},A}^{\dagger}i\partial_{t}a_{\bm{k},R}-a_{\bm{k},R}^{\dagger}i\partial_{t}a_{\bm{k},A},
\end{equation}
\begin{eqnarray}
	-H_{+}+H_{-}-4i\gamma n\sum_{\bm{k},\bm{k}\neq0}a_{\bm{k}-}^{\dagger}a_{\bm{k}+} & = & \sum_{\bm{k},\bm{k}\neq0}-(\varepsilon_{\bm{k}}+U_{R}n-2i\gamma n)a_{\bm{k},A}^{\dagger}a_{\bm{k},R}-(\varepsilon_{\bm{k}}+U_{R}n+2i\gamma n)a_{\bm{k},R}^{\dagger}a_{\bm{k},A}\nonumber \\
	&  & -\frac{U^{\ast}n}{2}(a_{-\bm{k},R}a_{\bm{k},A}+a_{-\bm{k},A}a_{\bm{k},R})-\frac{Un}{2}(a_{\bm{k},R}^{\dagger}a_{-\bm{k},A}^{\dagger}+a_{\bm{k},A}^{\dagger}a_{-\bm{k},R}^{\dagger})\nonumber \\
	&  &+2i\gamma n (a_{\bm{k},A}^\dagger a_{\bm{k},A}+a_{-\bm{k},A}^\dagger a_{-\bm{k},A} )
	.
\end{eqnarray}
Hence, the action can be reorganized as 
\begin{equation}
	S=\frac{1}{2}\sum_{\bm{k},\bm{k}\neq0}\int\frac{d\omega}{2\pi}\left(\begin{array}{cccc}
		a_{\bm{k},R}^{\dagger} & a_{-\bm{k},R} & a_{\bm{k},A}^{\dagger} & a_{-\bm{k},A}\end{array}\right)\left(\begin{array}{cc}
		O_{2\times2} & G\\
		G^{\dagger} & 2i\gamma n I_{2\times2}
	\end{array}\right)\left(\begin{array}{c}
		a_{\bm{k},R}\\
		a_{-\bm{k},R}^{\dagger}\\
		a_{\bm{k},A}\\
		a_{-\bm{k},A}^{\dagger}
	\end{array}\right),\label{eq:action4}
\end{equation}
where 
\begin{equation}\label{eq: G_ex}
	G=\left(\begin{array}{cc}
		\omega-(\varepsilon_{\bm{k}}+U_{R}n+2i\gamma n) & -Un\\
		-U^{\ast}n & -\omega-(\varepsilon_{\bm{k}}+U_{R}n-2i\gamma n)
	\end{array}\right).
\end{equation}
The conditions of poles of the Green's function are given by 
\begin{equation}
	\det(G)=0\Rightarrow(\omega-(\varepsilon_{\bm{k}}+U_{R}n+2i\gamma n))(\omega+(\varepsilon_{\bm{k}}+U_{R}n-2i\gamma n))+|U|^{2}n^{2}=0,
\end{equation}
\begin{equation}
	\det(G^{\dagger})=0\Rightarrow(\omega-(\varepsilon_{\bm{k}}+U_{R}n-2i\gamma n))(\omega+(\varepsilon_{\bm{k}}+U_{R}n+2i\gamma n))+|U|^{2}n^{2}=0.
\end{equation}
These two equations can be rewritten as 
\begin{eqnarray}
	\omega^{2}-4i\gamma n\omega-\varepsilon_{\bm{k}}^{2}-2U_{R}n\varepsilon_{\bm{k}}-3\gamma^{2}n^{2} & = & 0,\\
	\omega^{2}+4i\gamma n\omega-\varepsilon_{\bm{k}}^{2}-2U_{R}n\varepsilon_{\bm{k}}-3\gamma^{2}n^{2} & = & 0.
\end{eqnarray}
Hence, $\omega_{1,2}$ are the poles of the retarded Green's function, which coincide with those obtained from the Gross-Pitaevskii equation~\cite{liu2022weakly}, and $\omega_{3,4}$ are the poles of the advanced Green's function. 

To examine the elementary excitations in the system, we need to diagonalize
the action \eqref{eq:action4}. We first review a similar transformation in a closed quantum system. The mean-field Hamiltonian takes the form of 
\begin{equation}\label{eq: Hamiltonian}
	H=\frac{1}{2}\left(\begin{array}{cc}
		a_{\bm{k}}^{\dagger} & a_{-\bm{k}}\end{array}\right)\left(\begin{array}{cc}
		\varepsilon_{\bm{k}}+Un & Un\\
		Un & \varepsilon_{\bm{k}}+Un
	\end{array}\right)\left(\begin{array}{c}
		a_{\bm{k}}\\
		a_{-\bm{k}}^{\dagger}
	\end{array}\right).
\end{equation}
To derive the matrix for the Bogoliubov transformation, we first transform
the matrix as 
\begin{equation}
	\left(\begin{array}{cc}
		\varepsilon_{\bm{k}}+Un & Un\\
		Un & \varepsilon_{\bm{k}}+Un
	\end{array}\right)\rightarrow\left(\begin{array}{cc}
		\varepsilon_{\bm{k}}+Un & Un\\
		Un & \varepsilon_{\bm{k}}+Un
	\end{array}\right)\left(\begin{array}{cc}
		1 & 0\\
		0 & -1
	\end{array}\right)=\left(\begin{array}{cc}
		\varepsilon_{\bm{k}}+Un & -Un\\
		Un & -(\varepsilon_{\bm{k}}+Un)
	\end{array}\right)\label{eq:needed_diagonalization}
\end{equation}
since the action in the frequency space is given by
\begin{equation}
	S=\int d\omega \frac{1}{2}\left(\begin{array}{cc}
		a_{\bm{k}}^{\dagger} & a_{-\bm{k}}\end{array}\right)\left(\begin{array}{cc}
		\omega-(\varepsilon_{\bm{k}}+Un) & -Un\\
		-Un & -\omega-(\varepsilon_{\bm{k}}+Un)
	\end{array}\right)\left(\begin{array}{c}
		a_{\bm{k}}\\
		a_{-\bm{k}}^{\dagger}
	\end{array}\right).
\end{equation}
Therefore, the matrix $\left(\begin{array}{cc}
	1 & 0\\
	0 & -1
\end{array}\right)$ in Eq. \eqref{eq:needed_diagonalization} transforms the matrix $\left(\begin{array}{cc}
	\omega & 0\\
	0 & -\omega
\end{array}\right)$ in the action into $\left(\begin{array}{cc}
	\omega & 0\\
	0 & \omega
\end{array}\right)$, which enables us to use a similarity transformation to diagonalize the Hamiltonian \eqref{eq: Hamiltonian}. The eigenvalues and eigenvectors of the matrix are 
\begin{align}
	E_{1}=\sqrt{\varepsilon_{\bm{k}}(\varepsilon_{\bm{k}}+2Un)} &,\  v_{1}=\left(1,1+x^{2}-x\sqrt{x^{2}+2}\right)^{T},\\
	E_{2}=-\sqrt{\varepsilon_{\bm{k}}(\varepsilon_{\bm{k}}+2Un)} & ,\  v_{2}=\left(1+x^{2}-x\sqrt{x^{2}+2},1\right)^{T},
\end{align}
where $x=\sqrt{\varepsilon_{\bm{k}}/Un}$. Hence, the similarity transformation that diagonalizes the matrix 
(\ref{eq:needed_diagonalization})
can be written as 
\begin{equation}
	M=\left(\begin{array}{cc}
		1 & 1+x^{2}-x\sqrt{x^{2}+2}\\
		1+x^{2}-x\sqrt{x^{2}+2} & 1
	\end{array}\right)\equiv\left(\begin{array}{cc}
		1 & \alpha\\
		\alpha & 1
	\end{array}\right),
\end{equation}
where $\alpha=1+x^{2}-x\sqrt{x^{2}+2}$.
This similarity transformation can be shown to be equivalent to the Bogoliubov transformation by renormalizing the matrix as $\tilde{M} := M/\sqrt{1-\alpha^2}$ such that $\det(\tilde{M})=1$. Thus we obtain the Bogoliubov transformation matrix \citep{Ueda2010}
\begin{equation}
	\tilde{M}=\left(\begin{array}{cc}
		\frac{1}{\sqrt{1-\alpha^{2}}} & \frac{\alpha}{\sqrt{1-\alpha^{2}}}\\
		\frac{\alpha}{\sqrt{1-\alpha^{2}}} & \frac{1}{\sqrt{1-\alpha^{2}}}
	\end{array}\right).\label{Bogoliubov}
\end{equation}
Similarly, in open quantum systems, we need to diagonalize the
matrix 
\begin{equation}
	K=\left(\begin{array}{cc}
		\varepsilon_{\bm{k}}+U_{R}n+2i\gamma n & Un\\
		U^{\ast}n & \varepsilon_{\bm{k}}+U_{R}n-2i\gamma n
	\end{array}\right)
\end{equation}
from the matrix $G$ in Eq. \eqref{eq: G_ex}. We follow the same procedure as above and transform
$K$ as $K\sigma_{z}$ for diagonalization. The problem is equivalent
to the diagonalization of the matrix 
\begin{equation}
	\left(\begin{array}{cc}
		\varepsilon_{\bm{k}}+U_{R}n+2i\gamma n & -Un\\
		U^{\ast}n & -(\varepsilon_{\bm{k}}+U_{R}n-2i\gamma n)
	\end{array}\right).
\end{equation}
The eigenvectors are given by 
\begin{eqnarray}
	v_{1} & = & \left(1,\frac{U_{R}n+\varepsilon_{\bm{k}}+\sqrt{\varepsilon_{\bm{k}}(\varepsilon_{\bm{k}}+2U_{R}n)-\gamma^{2}n^{2}}}{Un}\right)^{T}\nonumber \\
	& \equiv & \left(1,\bar{\alpha}\right)^{T},\\
	v_{2} & = & \left(\frac{U_{R}n+\varepsilon_{\bm{k}}-\sqrt{\varepsilon_{\bm{k}}(\varepsilon_{\bm{k}}+2U_{R}n)-\gamma^{2}n^{2}}}{U^{\ast}n},1\right)^{T}\nonumber \\
	& \equiv & (\alpha,1)^{T},
\end{eqnarray}
with
\begin{equation}
	\alpha=\frac{U_{R}n+\varepsilon_{\bm{k}}-\sqrt{\varepsilon_{\bm{k}}(\varepsilon_{\bm{k}}+2U_{R}n)-\gamma^{2}n^{2}}}{U^{\ast}n},\bar{\alpha}=\frac{U_{R}n+\varepsilon_{\bm{k}}-\sqrt{\varepsilon_{\bm{k}}(\varepsilon_{\bm{k}}+2U_{R}n)-\gamma^{2}n^{2}}}{Un}.
\end{equation}
We note that $\alpha^{\ast}=\bar{\alpha}$ only holds when $\varepsilon_{\bm{k}}(\varepsilon_{\bm{k}}+2U_{R}n)-\gamma^{2}n^{2}\geqslant0$
, where the spectra in Eqs. \eqref{eq: omega_12} and \eqref{eq: omega_34} have a nonzero real part. Therefore, the diagonalization matrix is given by 
\begin{equation}
	M=\left(\begin{array}{cc}
		1 & \alpha\\
		\bar{\alpha} & 1
	\end{array}\right).
\end{equation}
After renormalization $\tilde{M}:= M/\sqrt{1-\alpha\bar{\alpha}}$, we obtain the Bogoliubov matrix for the retarded operators
\begin{equation}
	\tilde{M}=\left(\begin{array}{cc}
		\frac{1}{\sqrt{1-\alpha\bar{\alpha}}} & \frac{\alpha}{\sqrt{1-\alpha\bar{\alpha}}}\\
		\frac{\bar{\alpha}}{\sqrt{1-\alpha\bar{\alpha}}} & \frac{1}{\sqrt{1-\alpha\bar{\alpha}}}
	\end{array}\right),
	\left(\begin{array}{cc}\bar{b}_{\bm{k}, R} &
		b_{- \bm{k}, R}\end{array}\right) = \left(\begin{array}{cc}a^{\dagger}_{\bm{k}, R} &
		a_{- \bm{k}, R}\end{array}\right)\left(\begin{array}{cc}
		\frac{1}{\sqrt{1 - \alpha \bar{\alpha}}} & \frac{\alpha}{\sqrt{1 - \alpha
				\bar{\alpha}}}\\
		\frac{\bar{\alpha}}{\sqrt{1 - \alpha \bar{\alpha}}} & \frac{1}{\sqrt{1 -
				\alpha \bar{\alpha}}}
	\end{array}\right),
\end{equation}
and the Bogoliubov matrix for the advanced operators 
\begin{equation}
	\tilde{M}^{-1}=\left(\begin{array}{cc}
		\frac{1}{\sqrt{1-\alpha\bar{\alpha}}} & -\frac{\alpha}{\sqrt{1-\alpha\bar{\alpha}}}\\
		-\frac{\bar{\alpha}}{\sqrt{1-\alpha\bar{\alpha}}} & \frac{1}{\sqrt{1-\alpha\bar{\alpha}}}
	\end{array}\right),\left(\begin{array}{c}
		b_{\bm{k},A}\\
		\bar{b}_{-\bm{k},A}
	\end{array}\right)=\left(\begin{array}{cc}
		\frac{1}{\sqrt{1-\alpha\bar{\alpha}}} & \frac{\alpha}{\sqrt{1-\alpha\bar{\alpha}}}\\
		\frac{\bar{\alpha}}{\sqrt{1-\alpha\bar{\alpha}}} & \frac{1}{\sqrt{1-\alpha\bar{\alpha}}}
	\end{array}\right)\left(\begin{array}{c}
		a_{\bm{k},A}\\
		a_{-\bm{k},A}^{\dagger}
	\end{array}\right).
\end{equation}
Hence, the action becomes 
\begin{equation}
	S=\frac{1}{2}\sum_{\bm{k},\bm{k}\neq0}\int\frac{d\omega}{2\pi}\left(\begin{array}{cccc}
		b^{\dagger}_{\bm{k},R} & b_{-\bm{k},R} & b^{\dagger}_{\bm{k},A} & b_{-\bm{k},A}\end{array}\right)\left(\begin{array}{cc}
		O_{2\times2} & H'\\
		(H')^{\dagger} & 2i\gamma n \tilde{M}^{-2}
	\end{array}\right)\left(\begin{array}{c}
		b_{\bm{k},R}\\
		b^{\dagger}_{-\bm{k},R}\\
		b_{\bm{k},A}\\
		b^{\dagger}_{-\bm{k},A}
	\end{array}\right),
\end{equation}
where 
\begin{equation}
	H'=i\partial_{t}\sigma_{z}-\left(\begin{array}{cc}
		2i\gamma n-\sqrt{\varepsilon_{\bm{k}}(\varepsilon_{\bm{k}}+2U_{R}n)-\gamma^{2}n^{2}} & 0\\
		0 & -2i\gamma n-\sqrt{\varepsilon_{\bm{k}}(\varepsilon_{\bm{k}}+2U_{R}n)-\gamma^{2}n^{2}}
	\end{array}\right).
\end{equation}
In the expressions above, we can see $\bar{b}_{\bm{k}}=b_{\bm{k}}^{\dagger}$ only holds for $\varepsilon_{\bm{k}}(\varepsilon_{\bm{k}}+2U_Rn)>\gamma^2n^2$.
	
	

\begin{figure}[t]
		\includegraphics[width=0.6\columnwidth]{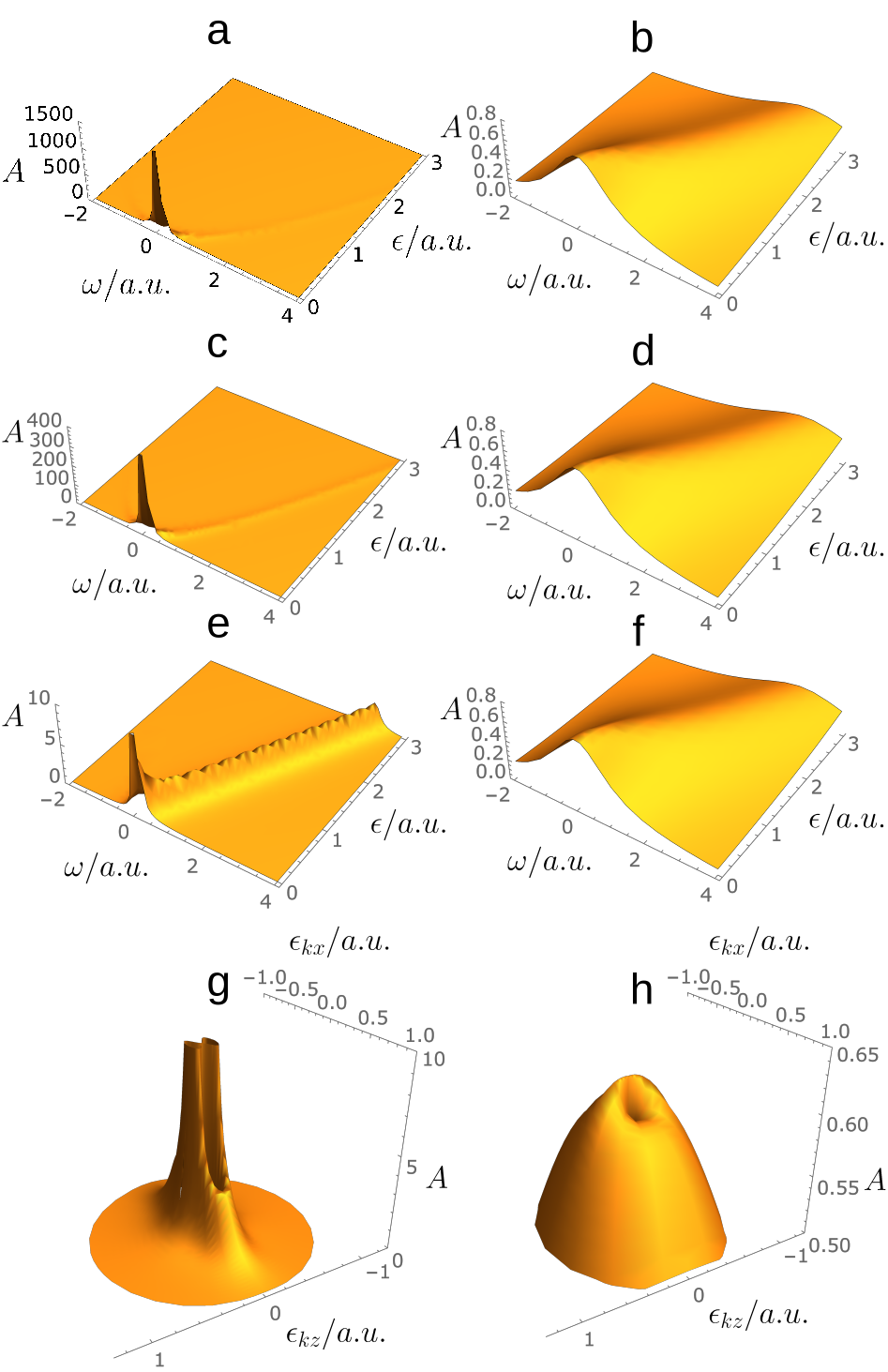}
		
		\caption{Spectral function $A(\omega)$ of a dissipative molecular
			BEC as a function of the  kinetic energy
			$\epsilon=\frac{|\bm{k}|^{2}}{2m}$ and frequency $\omega$ under different regimes and different
			directions. Here we choose $\epsilon_{dd}=0.833$ and set the polarization of dipole moments along the $z$-axis. Figures a,c,e,g
			show the weak-dissipation regime where $U_{R}n=1.0$ a.u., and
			$\gamma n=0.1$ a.u.. Figures b,d,f,h show the weak-interaction
			regime where $U_{R}n=0.1$ a.u., and $\gamma n=1.0$ a.u.. In Figs. a-f, the directions are $\theta_{\bm{k}}=0,\pi/4,\pi/2$ from top to bottom.
			In Figs. g and h, we fix $\omega=0.5$ a.u.., where $\epsilon_{kx}:=k_x^2/(2m)$ and $\epsilon_{kz}:=k_z^2/(2m)$.}
		
		\label{spectral_dipoledipole}
	\end{figure}

	Furthermore, we calculate the excitation spectrum and the spectral function in the presence of the dipole-dipole interaction. Recall that the Fourier
	transform of the dipole-dipole interaction is \cite{Ueda2010}
	\begin{equation}
		V_{dd}(\bm{k})=\begin{cases}
			-\frac{8\pi}{3}c_{dd}(1-3\cos^{2}\theta_{\bm{k}}) & \bm{k}\neq0;\\
			-\frac{8\pi}{3}c_{dd} & \bm{k}=0.
		\end{cases}
	\end{equation}
	The mean-field approximation for the dipole-dipole potential is \-
	\begin{equation}
		\begin{aligned} & \frac{1}{2}\sum_{\bm{k}_1,\bm{k}_2,\bm{k}_3,\bm{k}_4} a_{\bm{k}_{1}}^{\dagger}a_{\bm{k}_{2}}^{\dagger}a_{\bm{k}_{3}}a_{\bm{k}_{4}}V_{dd}(\bm{k}_{1}-\bm{k}_{4})\delta_{\bm{k}_{1}+\bm{k}_{2}-\bm{k}_{3}-\bm{k}_{4}}\\
			\approx & \frac{1}{2}N^{2}V_{dd}(0)+\frac{N}{2}\sum_{\bm{k},\bm{k}\neq0}a_{\bm{k}}a_{-\bm{k}}V_{dd}(\bm{k})+\frac{N}{2}\sum_{\bm{k},\bm{k}\neq0}a_{\bm{k}}^{\dagger}a_{-\bm{k}}^{\dagger}V_{dd}(\bm{k})+N\sum_{\bm{k},\bm{k}\neq0}a_{\bm{k}}^{\dagger}a_{\bm{k}}[V_{dd}(0)+V_{dd}(\bm{k})].
		\end{aligned}
	\end{equation}
	Thus, the effective interaction in the Keldysh contour can be written
	as 
	\begin{equation}
		S=\int_{-\infty}^{\infty}dt\left[\sum_{\bm{k},\bm{k}\neq0}(a_{\bm{k}+}^{\dagger}i\partial_{t}a_{\bm{k}+}-a_{\bm{k}-}^{\dagger}i\partial_{t}a_{\bm{k}-})-H_{+}+H_{-}-4i\gamma n\sum_{\bm{k},\bm{k}\neq0}a_{\bm{k}-}^{\dagger}a_{\bm{k}+}\right],
	\end{equation}
	where
	\begin{align}
		H_{\alpha}=&\frac{(U_{R}+V_{dd}(0))n}{2}N+\sum_{\bm{k},\bm{k}\neq0}\Bigg[(\varepsilon_{\bm{k}}+(U_{R}+V_{dd}(\bm{k}))n-2i\alpha\gamma n)a_{\bm{k}\alpha}^{\dagger}a_{\bm{k}\alpha}\nonumber\\
		&\left.+\frac{(U^{\ast}+V_{dd}(\bm{k}))n}{2}a_{-\bm{k}\alpha}a_{\bm{k}\alpha}+\frac{(U+V_{dd}(\bm{k}))n}{2}a_{\bm{k}\alpha}^{\dagger}a_{-\bm{k}\alpha}^{\dagger}\right].
	\end{align}
	Following the same calculations as above with $U_{R}$ replaced by $\widetilde{U}_{R}=U_{R}+V_{dd}(\bm{k})$, we obtain the spectral function as
	\begin{equation}
		A(\bm{k},\omega)=\frac{1}{\pi}\frac{4\gamma n\left[(\omega+\varepsilon_{\bm{k}}+(U_{R}+V_{dd}(\bm{k}))n)^{2}+((U_{R}+V_{dd}(\bm{k}))n)^{2}+5\gamma^{2}n^{2}\right]}{\omega^{4}+2[5\gamma^{2}n^{2}-\varepsilon_{\bm{k}}(\varepsilon_{\bm{k}}+2(U_{R}+V_{dd}(\bm{k}))n)]\omega^{2}+[\varepsilon_{\bm{k}}(\varepsilon_{\bm{k}}+2(U_{R}+V_{dd}(\bm{k}))n)+3\gamma^{2}n^{2}]^{2}}.\label{eq:spectral_dipoledipole}
	\end{equation}
	The excitation spectra and the associated Bogoliubov transformation are given by 
	\begin{equation}
		\begin{aligned}\label{eq:Bogoliubov_spectrum}
			\omega_{1,2} & =-2i\gamma n\pm\sqrt{\varepsilon_{\bm{k}}(\varepsilon_{\bm{k}}+2(U_{R}+V_{dd}(\bm{k}))n)-\gamma^{2}n^{2}},\\
			\omega_{3,4} & =2i\gamma n\pm\sqrt{\varepsilon_{\bm{k}}(\varepsilon_{\bm{k}}+2(U_{R}+V_{dd}(\bm{k}))n)-\gamma^{2}n^{2}},\\
			\left(\begin{array}{c}
				b_{\bm{k},A}\\
				\bar{b}_{-\bm{k},A}
			\end{array}\right) & =\left(\begin{array}{cc}
				\frac{1}{\sqrt{1-\alpha\bar{\alpha}}} & \frac{\alpha}{\sqrt{1-\alpha\bar{\alpha}}}\\
				\frac{\bar{\alpha}}{\sqrt{1-\alpha\bar{\alpha}}} & \frac{1}{\sqrt{1-\alpha\bar{\alpha}}}
			\end{array}\right)\left(\begin{array}{c}
				a_{\bm{k},A}\\
				a_{-\bm{k},A}^{\dagger}
			\end{array}\right),
		\end{aligned}
	\end{equation}
	where
	\begin{align}
		\alpha&=((U_{R}+V_{dd}(\bm{k}))n+\varepsilon_{\bm{k}}-\sqrt{\varepsilon_{\bm{k}}(\varepsilon_{\bm{k}}+2(U_{R}+V_{dd}(\bm{k}))n-\gamma^{2}n^{2})})/((U^{*}+V_{dd}(\bm{k}))n),\nonumber\\
		\bar{\alpha}&=((U_{R}+V_{dd}(\bm{k}))n+\varepsilon_{\bm{k}}-\sqrt{\varepsilon_{\bm{k}}(\varepsilon_{\bm{k}}+2(U_{R}+V_{dd}(\bm{k}))n-\gamma^{2}n^{2})})/((U+V_{dd}(\bm{k}))n).
	\end{align}
	The dipole-dipole interaction only modifies the strength of the repulsive interaction. 
	In Fig. \ref{spectral_dipoledipole}, we show the spectral function for an experimental value of $\varepsilon_{dd}=0.833$~\cite{Sebastian2023} along different directions for the weak-interaction limit where $\gamma\gg U_{R}$ and the weak-dissipation limit where $U_{R}\gg\gamma$. Figures \ref{spectral_dipoledipole}a, c, and e show that spectral functions strongly depend on the measured direction in the weak-dissipation limit, while Figs. \ref{spectral_dipoledipole}b, d and f show that the system is dominated by the dissipation and the spectral function is nearly independent of the polarization direction in the weak-interaction limit. 
	
\begin{figure}
		\includegraphics[width=0.6\columnwidth]{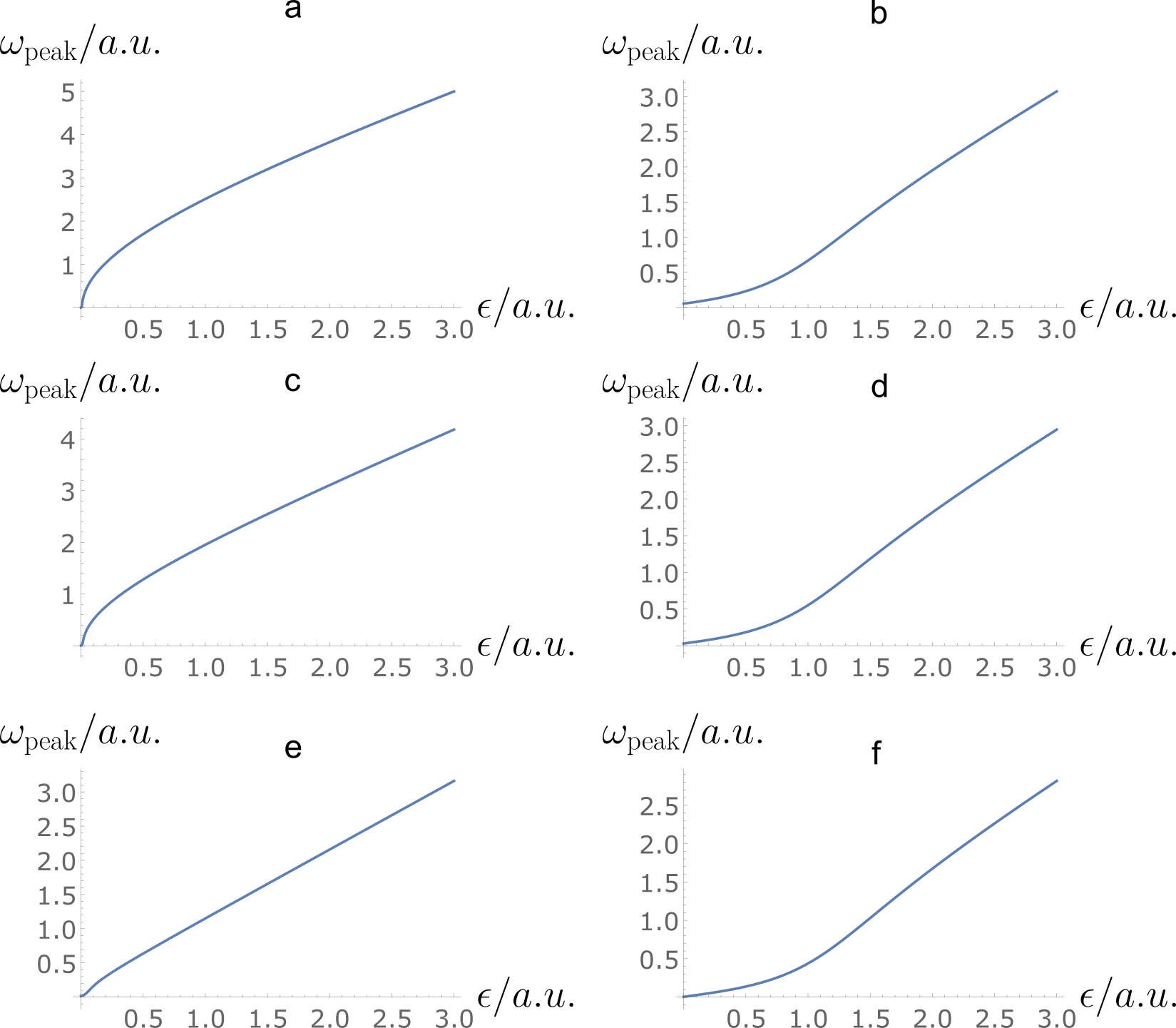}
		
		\caption{Peak frequency of the spectral function (\ref{eq:spectral_dipoledipole}) of a dissipative
			BEC as a function of the  kinetic energy
			$\epsilon=\frac{|\bm{k}|^{2}}{2m}$ in different regimes and different
			directions. Here we choose $\varepsilon_{dd}=0.833.$ Figures a,c and e show the weak-dissipation regime where $U_{R}n=1.0$ a.u. and  $\gamma n=0.1$ a.u.. Figures b,d and f show the weak-interaction
			regime where $U_{R}n=0.1$ a.u. and $\gamma n=1.0$ a.u.. In a-f, the directions are $\theta_{\bm{k}}=0,\pi/4,\pi/2$ from top to bottom.
			In the regime $\tilde{U}_{R}\gg\gamma$, $\omega_{\mathrm{peak}}$
			exhibits behavior initially following a square-root dependence and subsequently
			showing a crossover to a linear dependence on $\epsilon$. In the
			weak-interaction regime, the peak is less sensitive to the direction
			and shows a bending at $\epsilon=1.0$ a.u. where the real part of the
			spectrum increases. }
		
		\label{fig1}
	\end{figure}

	Additionally, we also plot the peak frequency $\omega_{\mathrm{peak}}=\mathrm{argmax}_\omega A(\bm{k},\omega)$ of the spectral function as a function of the kinetic energy $\epsilon=|\bm{k}|^{2}/2m$ in Fig. \ref{fig1}. In the weak-interaction limit,
	the peak frequency is nearly independent of the direction $\theta_{\bm{k}}$
	since the system is dominated by the dissipation. In the weak-dissipation regime, the peak frequency
	is more sensitive to the direction and $\omega_{\mathrm{peak}}$ exhibits the well-known spectrum of an interacting BEC \citep{Schmitt2015} in a closed system which confirms our calculation. 
	
	From the figures, we see that there always exists one peak in both cases for a given momentum and the strength of interaction and that the peak is broadened by the dissipation. This broadening indicates a finite lifetime of quasiparticles in the system. This spectral function shows distinct behavior in the weak-dissipation and weak-interaction limits. Since the spectral function can be measured in the state-of-the-art experimental platform \citep{Brown2020,PhysRevB.97.125117,Bismut2012}, it can be used to experimentally test our results.

	We examine the stability of the dissipative BEC with dipolar interactions. A BEC is unstable when one of the imaginary part of the spectra $\omega_{1,2}$ is positive. For the spectra \eqref{eq:Bogoliubov_spectrum}, they most likely become unstable for the angle $\theta_{\bm{k}}=\pi/2$, where the spectra can be rewritten as
	\begin{equation}\label{eq:omega12}
		\omega_{1,2}=-2i\gamma n\pm\sqrt{\varepsilon_{\bm{k}}(\varepsilon_{\bm{k}}+2(U_{R}-V_{dd})n)-\gamma^{2}n^{2}},
	\end{equation}
	where $V_{dd}:=8\pi c_{dd}/3$. The system becomes unstable if the imaginary part of $\omega_1$ is positive. From Eq. \eqref{eq:omega12}, we can see that the system becomes unstable when
	\begin{equation} 
		V_{dd}>\sqrt{3}\gamma+U_R \label{eq:stability_result}
	\end{equation} 
	for $U_R>0$. Thus, the dissipation helps stablize the condensate since the dissipation leads to an effective repulsive interaction. 
	
	We note that in the long-wavelength limit, the spectrum $\omega_{1}$ can be rewritten as
	\begin{equation}
		\omega_{1}\simeq-i\gamma n-iD(\bm{k})k^2,
	\end{equation}
	where $D(\bm{k}):=\tilde{U}_R/(2\gamma m)=(U_R-V_{dd}(\bm{k}))/(2\gamma m)$ is the diffusion coefficient. The dispersion relation $-iD(\bm{k})k^2$ represents that the propagation of excitation is diffusive. Hence, in contrast to BECs in closed systems, the speed of sound is ill-defined in the dissipative condensate.
	
\begin{figure}
		\includegraphics[width=\columnwidth]{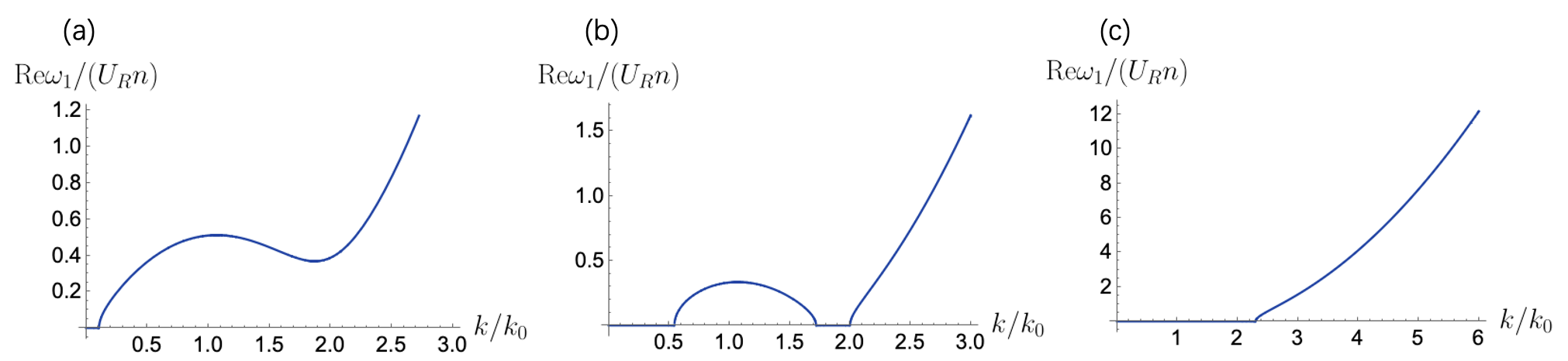}
		
		\caption{The real part of the Liouvillian spectrum $\omega_1$ in two-dimensional systems. Here we define $k_0:=\sqrt{2mU_Rn}$ as the characteristic wave number and take the parameters $V=0.98\sqrt{U_R/mn}$ for all figures and (a): $\gamma/U_R=0.1$, (b): $\gamma/U_R=0.4$, (c): $\gamma/U_R=0.6$. The critical value above which the roton excitations disappear is given by $\gamma/U_R=0.1456$.}
		
		\label{fig2}
	\end{figure}

	In a two-dimensional dipolar BEC, there exist roton-like excitations if the dipoles are perpendicular to the plane~\cite{Boudjem2013}. Here we place the dipoles on the $x$-$y$ plane. The Fourier transformation of the dipole-dipole interaction is given by
	\begin{equation}
		V_{d d} (\bm{k}) = \int d^3 \bm{r}e^{-i\bm{k} \cdot
			\bm{r}} V_{d d} (\bm{r}) .
	\end{equation}
	When the dipoles are polarized along the $z$-axis, the interaction in the momentum space is given by
	\begin{equation}
		V_{d d} (\bm{k}) = \int d^3 \bm{r}e^{-i\bm{k} \cdot
			\bm{r}} \frac{c_{d d}}{r^3} = -2 \pi c_{d d}k,
	\end{equation}
	which leads to the roton-like excitations in the system. In the closed quantum system, the condition for an extremum of the excitation energy is given by
	\begin{equation}
		\frac{d}{d k} \sqrt{\varepsilon_{\bm{k}} (\varepsilon_{\bm{k}} +
			2 \tilde{U}_R n)} = 0 \Rightarrow k = \frac{1}{2} \left( 3 m n V \pm \sqrt{9
			(m n V)^2 - 8 U_R n m} \right),
	\end{equation}
	where $V := 2 \pi c_{d d}$ is the strength of the dipole-dipole
	interaction and $\tilde{U}_R:=U_R-Vk$ is the total interaction~\cite{Boudjem2013}. Thus, the roton-like excitation happens at
	\begin{equation}
		k_c = \left( 3 m n V +
		\sqrt{9 (m n V)^2 - 8 U_R n m} \right) / 2.
	\end{equation}
	Now we consider the open quantum system. The Liouvillian spectrum in the two-dimensional system is given by
	\begin{equation}
		\omega_1=-2i\gamma n\pm\sqrt{\varepsilon_{\bm{k}}(\varepsilon_{\bm{k}}+2(U_{R}-Vk)n)-\gamma^{2}n^{2}},
	\end{equation}
	where the real part of the excitation spectrum is $\text{Re}[\omega_1] = 0$ if
	$\varepsilon_{\bm{k}} (\varepsilon_{\bm{k}} + 2 \tilde{U}_R n)
	\leqslant \gamma^2 n^2$ and $\text{Re}[\omega_1] = \sqrt{\varepsilon_{\bm{k}}
		(\varepsilon_{\bm{k}} + 2 \tilde{U}_R n) - \gamma^2 n^2}$ if
	$\varepsilon_{\bm{k}} (\varepsilon_{\bm{k}} + 2 \tilde{U}_R n) >
	\gamma^2 n^2$. By taking the derivative of the real part of the excitation spectrum with respect to $k$, we can see that the roton-like excitation happens if $\varepsilon_{\bm{k}_c} (\varepsilon_{\bm{k}_c} + 2
	\tilde{U}_R n) > \gamma^2 n^2$. Hence, substituting the critical momentum
	in the excitation spectrum, we can rewrite the condition as
	\begin{equation}\label{eq:loss-roton}
		\gamma^2 n^2 < \frac{1}{32} \left( 3 \sqrt{m} V n + \sqrt{9 m (n
			V)^2 - 8 U_R n} \right)^2 \left( 4 U_R n - 3 m (n V)^2 - V n \sqrt{9 (m n
			V)^2 - 8 U_R n m} \right).
	\end{equation}
	If the loss rate satisfies Eq. \eqref{eq:loss-roton}, the roton-like excitation takes place. Otherwise, roton-like excitations do not exist. The Liouvillian spectrum of both the cases are shown in Fig. \ref{fig2}.
	
\end{document}